\documentclass[onecolumn,a4paper,journal]{IEEEtran}
\usepackage[utf8]{inputenc}
\usepackage[T1]{fontenc}
\usepackage[caption=false,font=footnotesize,labelfont=sf,textfont=sf]{subfig}
\usepackage{float}
\usepackage{array}
\usepackage[unicode]{hyperref}
\usepackage{url}
\usepackage{balance}
\usepackage{ifthen}
\usepackage{cite}
\usepackage[cmex10]{amsmath} 
\usepackage{amsfonts}
\usepackage{booktabs}
\usepackage[cal = pxtx, scr = dutchcal]{mathalpha}
\usepackage[ruled,noline,linesnumbered,noend]{algorithm2e}
\usepackage{multirow}
\usepackage[pdftex]{graphicx}
\graphicspath{{./figures/}}
\DeclareGraphicsExtensions{.pdf,.jpeg,.png}
\usepackage{tikz}
\usepackage{pgfplots}
\pgfplotsset{compat=1.18}
\definecolor{darkred}{RGB}{202,0,32}
\definecolor{darkblue}{RGB}{5,113,176}
\definecolor{lightred}{RGB}{244,165,130}
\definecolor{lightblue}{RGB}{146,197,222}
\pgfplotscreateplotcyclelist{waterfall-floor}{%
red,mark=.\\%
red,only marks, mark=x\\%
orange,mark=.\\%
orange,only marks, mark=x\\%
green,mark=.\\%
green,only marks, mark=x\\%
blue,mark=.\\%
blue,only marks, mark=x\\%
violet,mark=.\\%
violet,only marks, mark=x\\%
}

\pgfplotscreateplotcyclelist{waterfall-floor-mc}{%
red,only marks, mark=x\\%
orange,only marks, mark=x\\%
green,only marks, mark=x\\%
blue,only marks, mark=x\\%
violet,only marks, mark=x\\%
}
\usepackage{orcidlink}

\usepackage{amsthm}
\newtheorem{theorem}{Theorem}
\newtheorem{proposition}{Proposition}
\newtheorem{assumption}{Assumption}

\usetikzlibrary{automata, positioning}
\usepackage{relsize}

\newcommand{\flips}[1]{\ensuremath{\mathcal{F}_t = {#1}}}
\newcommand{\flipsy}[1]{\ensuremath{\mathcal{F}_{t\mid y} = {#1}}}
\newcommand{\weight}[1]{\ensuremath{\mathrm{wt}(#1)}}
\newcommand{\fliplltoone}{\pi_{\small
\begin{smallmatrix}
 l-1 \to l\\
\mathtt{flip\ to}\ 1\\
\end{smallmatrix}}
}

\newcommand{\fliplltozero}{\pi_{\small
\begin{smallmatrix}
 l-1 \to l\\
\mathtt{flip\ to}\ 0\\
\end{smallmatrix}}
}
\newcommand{\vectmat}[1]{\ensuremath{\mathtt{#1}}}
\newcommand{\sevent}[1]{\ensuremath{\mathtt{E}_{(#1)}}}

\newcommand{\eventij}[1]{\ensuremath{\mathtt{E}_{(i,j), #1}}}
\newcommand{\prob}[1]{\ensuremath{\Pr(#1)}}
\newcommand{\condprob}[2]{\ensuremath{\Pr\left(#1\, \mid \, #2\right)}}
\newcommand{\interprob}[2]{\ensuremath{\Pr\left(#1\, \cap \, #2\right)}}
\newcommand{\punsatz}{\ensuremath{p_{\mathtt{unsat}|0}}}
\newcommand{\punsato}{\ensuremath{p_{\mathtt{unsat}|1}}}
\newcommand{\upc}{\ensuremath{\mathtt{upc}_j}}
\newcommand{\setofsat}{\ensuremath{\mathrm{Supp}(\vectmat{h}_{:,j})_{\mathtt{sat}}}}
\newcommand{\setofunsat}{\ensuremath{\mathrm{Supp}(\vectmat{h}_{:,j})_{\mathtt{unsat}}}}
\newcommand{\colsupport}{\ensuremath{\mathrm{Supp}(\vectmat{h}_{:,j})}}
\newcommand{\discre}{\ensuremath{\mathrm{d}}}
\newcommand{\discplus}{\ensuremath{\mathrm{d}_+}}
\newcommand{\discminus}{\ensuremath{\mathrm{d}_-}}
\newcommand{\support}[1]{\ensuremath{\mathrm{Supp}(#1)}}
\newcommand{\jset}[2]{\ensuremath{\mathbf{J}_{#1,#2}}}
\newcommand{\bindist}[3]{\ensuremath{\text{\sc Bin}\left(#1,\,#2,\,#3\right)}}
\newcommand{\berndist}[1]{\ensuremath{\text{\sc Bernoulli}\left(#1\right)}}
\newcommand{\hygdist}[4]{\ensuremath{\text{\sc Hyg}\left(#1,\,#2,\,#3,\,#4\right)}}
\newcommand{\pflipz}[1]{\ensuremath{p_{\mathtt{flip}|0,#1}}}
\newcommand{\pnoflipo}[1]{\ensuremath{p_{\neg\mathtt{flip}|1,#1}}}

\newcommand{\pflipzz}{\ensuremath{p_{\mathtt{flip}|00}}}
\newcommand{\pflipzo}{\ensuremath{p_{\mathtt{flip}|01}}}
\newcommand{\pflipoz}{\ensuremath{p_{\mathtt{flip}|10}}}
\newcommand{\pflipoo}{\ensuremath{p_{\mathtt{flip}|11}}}

\newcommand{\chichg}{\ensuremath{\chi_{\updownarrow \mathtt{odd}}(f,\,\epsilon_{01})}}
\newcommand{\chimaint}{\ensuremath{\chi_{\leftrightarrow \mathtt{odd}}(f,\,\epsilon_{11})}}

\newcommand{\mathalert}[1]{\ensuremath{{\color{blue}#1}}}
\newtheorem{lemma}{Lemma}

\begin{document}
%
%
\title{Estimating the Decoding Failure Rate of Binary Regular Codes Using
Iterative Decoding}
%
%
\author{
Alessandro Annechini, Alessandro Barenghi,
Gerardo Pelosi~\IEEEmembership{Member,~IEEE}
\thanks{
A.~Annechini, A. Barenghi and G. Pelosi are with Politecnico di Milano,
Department of Electronics, Information and Bioengineering (DEIB), P.zza
Leonardo da Vinci, 32. 20133 Milan, Italy
(e-mail: alessandro.annechini@mail.polimi.it,
alessandro.barenghi@polimi.it,
gerardo.pelosi@polimi.it).
}
\thanks{This paper was presented in part at the IEEE International Symposium on Information Theory (ISIT) 2024 ~\cite{DBLP:conf/isit/AnnechiniBP24}
Manuscript received $month$ $day$, 2026; revised $month$ $day$, 2026.
}
}
%
%
\markboth{
IEEE TRANSACTIONS ON INFORMATION THEORY,~VOL.~XXX, NO.~XXX, YYYY~2026}%
{A. Annechini, A. Barenghi, G. Pelosi,
Estimating the DFR of Regular Codes Using Iterative Decoding%
}
%
%
\IEEEpubid{0000--0000~\copyright~2026 IEEE}
%
%
\maketitle
%
%
\begin{abstract}
Over the years, providing closed-form estimates of the decoding failure rate of
iterative decoders for low- and moderate-density binary parity-check codes has
attracted significant interest in the research community.
Recently, interest in this topic has increased due to the use of iterative
decoders in post-quantum cryptosystems, where the desired decoding failure
rates (DFRs) are less than or equal to $2^{-128}$ and impossible to estimate
via Monte Carlo simulations.
In this work, we propose a new technique that provides accurate DFR estimates
for a two-iteration (parallel) bit-flipping decoder that can also be used for
cryptographic purposes.
We successfully estimate the bit-flipping probabilities at the second decoder
iteration and provide an estimate of the syndrome weight distribution before and
after the first iteration as a function of the code parameters and error weight
(i.e., the number of non-null bits).
We validate our results numerically by comparing the modeled and simulated
syndrome weights, the incorrectly guessed error bit distribution at the end of
the first iteration, and the DFR after two iterations in both the floor and
waterfall regimes for simulatable codes.
Finally, we apply our method to estimate the DFR of the Low-dEnsity parity-check
coDe-bAsed cryptographic system (LEDAcrypt), a post-quantum
key encapsulation method that employs a two-iteration bit-flipping decoder.
We show that the DFR estimate resulting from the chosen code parameters can be
improved by a factor larger than $2^{70}$ with respect to previous estimation
techniques, when $128$-bit security is required.
This allows for a $20$\% reduction in public key and ciphertext sizes at no
security loss. Furthermore, we note that our results can be applied to the
post-quantum cryptosystem known as Bit Flipping Key Encapsulation (BIKE) by
replacing the current ``BIKE-flip decoder'' with the two-iteration decoder
(adjusting the parameters as necessary to yield the same security guarantees as
the original specification) and consequently endowing BIKE with the property of
indistinguishability under an adaptive chosen-ciphertext attack (IND-CCA$2$),
provably.

\end{abstract}
%
%
\begin{IEEEkeywords}
\noindent Regular binary codes, Decoding failure rate estimation,
Parallel bit-flipping decoder, LDPC, Code-based cryptography,
Post-quantum cryptography
\end{IEEEkeywords}
%
%
\section{Introduction}\label{sec:intro}
\noindent \IEEEPARstart{R}{egular}
binary codes are a large family of codes
characterized by admitting a parity-check matrix in which the row and column
weights (i.e., the number of non-zero bits) are the same for all rows and columns.
Among these codes, those with relatively sparse parity-check matrices have
garnered significant interest for engineering purposes since
Robert Gallager's observation in~\cite{DBLP:journals/tit/Gallager62} that they
can be decoded using a fixed-point iteration method.
Low Density Parity Check (LDPC) codes were first introduced in~\cite{DBLP:journals/tit/Gallager62}
as a class of sparse, regular codes with a polynomial-time iterative decoding
algorithm that exhibits excellent error correction performance.
It is acknowledged that, whilst iterative decoders can be applied to dense
parity-check matrices in principle, the termination of their computation is
not guaranteed, and their expected running time is exponential in the size of
the code.
Indeed, despite the regularity property, the iterative decoding algorithm
does not perform better than it would on a code with a fully random parity-check
matrix if a maximum likelihood decoding is desired.
The decoding of such codes is known to be an NP-hard task~\cite{DBLP:conf/isit/Santhi08},
as the one of decoding random codes~\cite{DBLP:journals/tit/BerlekampMT78}.
As a consequence of the effectiveness of the iterative decoding on sparse
regular codes, they are currently the ones most interesting for engineering
purposes; in particular, they find practical applications from communications to
memory storage and cryptography.

The desirable features of sparse regular codes come at the cost of a significant
difficulty in providing closed-form bounds for their Decoding Failure Rate (DFR).
While the increased availability of computing power allows to estimate the
said DFR via Monte Carlo simulations (i.e., sampling random error vectors with
a given number of non-zero components, trying to decode them and counting
the number of failures), reliably estimating very low failure rates (e.g.,
$10^{-12}$ and below) still requires considerable time.
One context where closed form estimates for the DFR of binary codes, belonging
to the category of either Low or Moderate Density Parity Check (LDPC or MDPC)
codes, are remarkably important is the design of post-quantum cryptosystems.
Indeed, post-quantum cryptosystems such as BIKE~\cite{BIKE} and LEDACrypt~\cite{LEDA}
employ a Quasi-Cyclic (QC) LDPC/MDPC code as private key.
The difference between LDPC and MDPC codes is the Hamming weight of
a parity-check matrix row, which is $\mathcal{O}(\log(n))$ for LDPC codes, and
$\mathcal{O}(\sqrt{n\log(n)})$ for MDPC codes, where $n$ is the code length.
In both cases, whenever a decoding failure takes place during the decryption of
a ciphertext, information regarding the private key is leaked to an (active)
attacker~\cite{DBLP:journals/tit/GuoJW19}.
To attain security against active attackers (formally captured by the notion of
\emph{indistinguishability against adaptive chosen ciphertext attacks}, a.k.a.
IND-CCA$2$) both cryptosystems require the DFR of the employed codes to be below
$2^{-128}$, for a decoder of choice.
To this end, BIKE relies on an extrapolation based on Monte Carlo simulations of
the iterative decoder behavior at
higher values of DFR~\cite{DBLP:conf/pqcrypto/SendrierV19}, while LEDACrypt employs a two-iterations bit flipping
decoder for which it combines a first-iteration
model~\cite{DBLP:conf/icete/BaldiBCPS20,DBLP:conf/icete/BaldiBCPS20a}
with a conservative, code-specific, upper bound for the error correction
capability of the second
iteration~\cite{DBLP:conf/isit/Tillich18,DBLP:journals/tcom/SantiniBBC20}.

Cryptographic-grade low DFR values cannot be estimated via numerical simulation,
and have therefore led to a further significant amount of research in providing
closed form bounds for the DFR of a decoder, explicitly encouraged also by
NIST~\cite{MoodyPres}.
J. Chaulet~\cite{DBLP:phd/hal/Chaulet17} provided an estimate of the
distribution of the syndrome weights (before the first iteration of the
iterative decoder acts on it) with a good fit on the average value, which our
work improves, and a model for the probability of parity-check equations to
be unsatisfied at the first decoder iteration from which our contribution takes
inspiration.
The work by Vasseur~\cite{DBLP:phd/hal/Vasseur21} proposes a different analytical
approach from the one we employ, to estimate the decoding failure rate of a
sequential decoder (step-by-step), in order to employ the estimates as proxies
for the ones of parallel decoders.
The observations on the syndrome weight involve the fact that error vectors
with remarkably low syndrome weight tend to be decoded with lower probability.
In an affine line of work, the authors of~\cite{DBLP:phd/hal/Vasseur21,
DBLP:journals/iacr/Vasseur21} observe that errors vectors having regularities
such as runs of non-zero bits, or non-zero bits placed at regular intervals,
are less frequently decoded by QC-MDPC iterative decoders.
The authors of~\cite{DBLP:conf/cbc/BaldiBCPS21} and~\cite{DBLP:conf/pqcrypto/ArpinBHLPR22}
observed that, while the aforementioned error vectors are indeed
harder to decode, they appear to represent a relatively low fraction of the
overall non-decodable errors.
 \IEEEpubidadjcol
Going in a different direction, J-P. Tillich~\cite{DBLP:conf/isit/Tillich18}
provided a code-specific technique to determine the maximum weight of an error
which is guaranteed to be corrected by an iterative decoder operating on a
QC-LDPC/QC-MDPC code, and proved that the DFR falls exponentially quickly
when the code length is increased, while keeping the weight of the parity
check rows $\mathcal{O}(\sqrt{n\log(n)})$.

\smallskip
\noindent \textbf{Contribution.} In this work, we describe how to
accurately estimate the DFR of a two-iterations (parallel)
bit flipping decoder for $(v,w)$-regular codes, such as the LDPC/MDPC
code used in~\cite{LEDA}.
In doing so, we provide a closed form derivation of the syndrome weight
distribution, providing a better fit with respect to the model
in~\cite{DBLP:phd/hal/Chaulet17}.
Furthermore, we propose a technique to assess the bit values in the
error vector estimate processed by bit-flipping decoders that do not match
the actual error vector values after the first iteration, improving the
accuracy with respect to the results presented in~\cite{DBLP:conf/icete/BaldiBCPS20}.
Finally, we propose a technique, relying on the previous two results, to derive
the amount of mismatches between the error estimate and the actual error
elaborated by the decoder after the second iteration, deriving the
two-iterations DFR.
In our derivation, we provide the first quantitative analysis of the
approximation technique employed in several studies.
This technique considers the outcome of different parity-check 
equations
as independent,
in order to simplify the characterization
of the flips performed by the decoder.

We report the results of extensive numerical simulations validating
the goodness of fit of our model~\footnote{
code available at \url{https://crypto.deib.polimi.it/DFR_codebase.zip}},
both for the final DFR values obtained, and the intermediate quantities employed.
Our results show a very good match of the DFR curve both in the waterfall
and the floor regions for a variety of parity-check matrix and error densities.

Finally, we apply our results to obtain a reliable estimate of the
DFR of a randomly picked ($v,w$)-regular LDPC/MDPC code, providing notable
tightness improvements on the previous method of~\cite{DBLP:journals/tcom/SantiniBBC20}.
In particular, our technique allows to improve the tightness of the DFR bounds
currently employed in the post-quantum key encapsulation method
LEDACrypt~\cite{LEDA} by a factor greater than $2^{70}$ (when $128$-bit security
is required), allowing a significant reduction in the required keysize and
ciphertext size at no security loss.
Furthermore, we note that the analyzed two-iterations decoder is applicable
also to the post-quantum key encapsulation mechanism known as BIKE,
where swapping it with the current ``BIKE-flip decoder'' (and adjusting the parameters)
would provide the required strong security guarantee of indistinguishability
against adaptive chosen ciphertext attacks (i.e., the IND-CCA$2$ property).
We note that our results improve the tightness of the estimates from 
LEDACrypt~\cite{LEDA}, which did not consider explicitly the effects of
the quasi-cyclicity of the employed codes (which are also employed in BIKE).
A recent work~\cite{cryptoeprint:2025/2218} shows that the impact of
considering the quasi cyclicity does not impact the syndrome weight distribution
computation technique reported in our work, which matches the one 
in~\cite{DBLP:conf/isit/AnnechiniBP24}.
In the same work~\cite{cryptoeprint:2025/2218}, the authors note that ``\emph{for
BIKE the probability of the events where the regular model differs from
the QC-MDPC syndrome distribution is too low to be of concern}''.

%
\section{Preliminaries}\label{sec:background}
\noindent\textbf{Notation.}
In the following, vectors are denoted as lowercase letters with a computer modern typewriter font
(e.g., $\vectmat{v}$).
$\support{\vectmat{v}}$ denotes the set of positions in $\{0,\ldots,n$$-$$1\}$ of
the non null coordinates of $\vectmat{v}$.
$\weight{\vectmat{v}}$ denotes the Hamming weight of $\vectmat{v}$, i.e., the count of
its non null components. Transposition is denoted as $\vectmat{v}^{\texttt{T}}$.
Similarly, a matrix $\vectmat{M}_{n_1 \times n_2}$ has $n_1$ rows and $n_2$ columns, while
$\vectmat{m}_{i,:}$ and $\vectmat{m}_{:,j}$ denote the $i$-th row and the $j$-th column
of $\vectmat{M}$, with $0 \leq i \leq n_1$$-$$1$, $0 \leq j \leq n_2$$-$$1$, respectively.
Random variables are denoted with calligraphic-faced uppercase
 roman letters (e.g., $\mathcal{X}$).
We denote sets with uppercase boldface letters, e.g., $\mathbf{C}$.
Given a binomially distributed random
variable $\mathcal{X}$ we denote the formula of its probability mass function (p.m.f.)
$\Pr(\mathcal{X} = k)$ as $\bindist{n}{k}{p}$, where $k$ is the number of successes out
of $n$ trials with success probability $p$,
i.e., $\bindist{n}{k}{p}=\binom{n}{k}p^k(1-p)^{n-k}$.
In the case of a random variable $\mathcal{Y}$ having a Bernoulli p.m.f.,
i.e., with a binomial distribution where a single trial with success probability
$p$ is conducted, we denote its formula as $\berndist{p}$,  while
the formula of the distribution of a hypergeometric random variable $\mathcal{Z} \sim \Pr(\mathcal{Z} = k)$,
is denoted as $\hygdist{N}{K}{n}{k}$, i.e.,
$\hygdist{N}{K}{n}{k}=\frac{\binom{K}{k} \binom{N-K}{n-k} }{\binom{N}{n}}$, where
$N$ is the population size, $K$ is the number of success states in the population,
$n$ is the number of draws (i.e., quantity drawn in each trial), and
$k$ is the number of observed successes.
\bigskip

A binary linear code $\mathbf{C}$ of length $n$ is a
$k$-dimensional ($0 < k < n$) vector subspace of $\mathbb{F}_2^n$, where
$\mathbb{F}_2$ is the field with two elements ($\{0, 1\}$),
described via a rank-$k$ generator matrix $\vectmat{G} \in \mathbb{F}_2^{k\times n}$ such
that $\mathbf{C} = \{\vectmat{c}\in \mathbb{F}_2^n\, \mid \, \exists\, \vectmat{m}
\in \mathbb{F}_2^k,\, \vectmat{c} = \vectmat{m}\, \vectmat{G}\}$,
and is also referred to as a code $\mathbf{C}[n, k]$.
The \emph{rate} of $\mathbf{C}$ is defined as $k/n$, where $k$ is
the dimension of $\mathbf{C}$, as a (row-)vector space over $\mathbb{F}_2$.
Being a linear subspace of dimension $k$, the code $\mathbf{C}$ can be described as the
kernel of a matrix $\vectmat{H} \in \mathbb{F}_2^{(n-k)\times n}$ so that
$\mathbf{C} = \{\vectmat{c} \in
\mathbb{F}_2^n\ |\ \vectmat{H}\,\vectmat{c}^{\mathtt{T}} = 0_{(n-k) \times 1}\}$, and
$\vectmat{G}\vectmat{H}^{\mathtt{T}} = 0 \in \mathbb{F}_2^{k\times (n-k)}$.
The matrix $\vectmat{H}$ is called the parity-check matrix of the code $\mathbf{C}$ and,
in general, any choice of $\vectmat{H}$ whose rows form a basis of the dual space
$\mathbf{C}_{\bot} = \{\vectmat{x} \in \mathbb{F}_2^n\
|\, \forall\ \vectmat{c} \in \mathbf{C},\, \ \vectmat{x} \cdot \vectmat{c}^{\mathtt{T}}
= 0\}$ describes the same code.
We will denote elements of vectors and matrices over $\mathbb{F}_2$ as 
asserted, when their value is $1$, and clear, when their value is $0$.
The parity-check matrix has also an interpretation as an adjacency matrix of a
bipartite graph, a.k.a. the {\em factor graph} of the code, defined as follows.
On the left side there are $n$ vertices, called {\em variable nodes}, one for
each position of the sequence of values defining a codeword.
On the right side there are $r$$=$$n$$-$$k$ vertices, called
{\em factor nodes} or {\em check nodes}, one for each row of the parity-check matrix.
The value ($1$ or $0$) of each column in a row of $\vectmat{H}$ indicates whether the
corresponding variable node is connected to the check node (value $1$) or not
(value $0$). If all nodes on the same side have the same degree, the graph is
categorized as {\em left-regular}, {\em right-regular}, or {\em biregular}
if every vertex has the same degree.
This work focuses on codes that admit a parity-check matrix where all rows
and columns have a fixed constant number of asserted bits, referred to as
$(v,w)$-regular codes, where $w$ and $v$ denote the number of $1$'s lying in
any row and any column of the parity-check
matrix, respectively, or equivalently with a factor graph exhibiting all check
nodes with degree $w$ and all variable nodes with degree $v$.
In the following, the number of asserted bits in either a
row or a column of a binary matrix, as well as in any binary vector will be
referred to as the (Hamming) \emph{weight} of the row, column, or vector at hand,
respectively.
\begin{algorithm}[!t]
  \DontPrintSemicolon
  {\small
  \KwIn{$\widetilde{\vectmat{c}} \in \mathbb{F}_2^n$: $1\times n$ error-affected binary codeword,\newline
  $\vectmat{s} \in \mathbb{F}_2^r$: $r\times 1$ syndrome,\newline
  $\vectmat{H} \in \mathbb{F}_2^{r \times n}$: $r\times n$ parity-check matrix,\newline
  $\mathtt{iterMax}$: max number of permitted iterations.
  }
  \KwOut{$\bar{\vectmat{e}} = [\bar{e}_0, \ldots, \bar{e}_{n-1}] \in \mathbb{F}_2^n$: $1 \times n$ binary error vector estimate,\newline
  $\mathtt{decodeOk}$: value indicating success, $1$, or fail, $0$.
  }
  $\mathtt{iter} \gets 0$\;
  $\bar{\vectmat{e}} \gets \mathtt{0}_{1 \times (n-1)}$ \tcp*[l]{\scriptsize bit vector }
  \While{($\vectmat{s} \neq 0_{r\times 1}$ \textbf{\em and} $\mathtt{iter} < \mathtt{iterMax}$)}{
        \For{$j$ \textbf{\em from} $0$ \KwTo $n-1$}{
           $\upc \gets \langle \vectmat{s}, \vectmat{h}_{:,j}\rangle$\tcp*[l]{\scriptsize integer products and additions: $\sum_i (s_i \cdot h_{i,j})$ }
        }
     $\mathtt{th} \gets \text{\sc ThresholdChoice}(\mathtt{iter},\vectmat{s})$\;
     \For{$j$ \textbf{\em from} $0$ \KwTo $n-1$}{
        \If{($\mathtt{upc}_j \geq \mathtt{th}$)}{
                 $\bar{e}_j \gets \bar{e}_j\oplus 1$  \tcp*[l]{\scriptsize modulo-2 addition}
                 $\vectmat{s} \gets \vectmat{s} \oplus \vectmat{h}_{:, j}$  \tcp*[l]{\scriptsize component-wise modulo-2 addition}
        }
     }
     $\mathtt{iter} \gets \mathtt{iter} + 1$\;
  }
  \If{($\vectmat{s} = 0_{r\times 1}$)}{$\mathtt{decodeOk} \gets 1$\;}\Else{$\mathtt{decodeOk} \gets 0$\;}
  \KwRet{$\bar{\vectmat{e}}, \mathtt{decodeOk}$}\;
  }
  \caption{{\sc Bit Flipping Algorithm}\label{algo:bitflipping}}
\end{algorithm}

Low Density Parity Check (LDPC) codes, originally studied
by Gallager~\cite{DBLP:journals/tit/Gallager62} and denoted as Ensemble A
in~\cite{DBLP:journals/tit/LitsynS02}, are $(v,w)$-regular codes that
admit a sparse parity-check matrix and sparse factor graph, and are hence amenable
to polynomial time decoding algorithms, exhibiting column and row weights
in the range of $O(\log(n))$.
The main reason for the interest in these codes is due to their
ability to provide highly reliable communications at code rates that
are extremely close to channel capacity.
Increasing moderately the number of asserted elements in each row of the
parity-check matrix up to values in the range of $O(\sqrt{n\log{(n)}})$,
the codes are also known as Moderate Density Parity Check Codes
(MDPC)~\cite{DBLP:journals/corr/abs-0911-3262, DBLP:conf/isit/MisoczkiTSB13}.

In the following section we are going to focus on modeling the
statistical error correction properties of the iterative (parallel) bit flipping
decoding algorithm proposed by Gallager's~\cite{DBLP:journals/tit/Gallager62},
when applied to a generic $(v,w)$-regular binary code.
Algorithm~\ref{algo:bitflipping} reports an operative description of the
decoding approach, while Figure~\ref{fig:1:notation} depicts the state of
the algorithm during one iteration of its decoding process.

The iterative decoding process takes as input the parity-check matrix
$\vectmat{H} = [h_{i,j}]$, $i$$\in$$\{0,\ldots,r$$-$$1\}$, $j$$\in$$\{0, \ldots,
n$$-$$1\}$ of a code, and the value of a syndrome, $\vectmat{s}_{r\times 1}$, of an error
affected codeword, $\widetilde{\vectmat{c}}_{1 \times n}$.
The error-affected codeword is assumed to be
obtained as $\widetilde{\vectmat{c}} = \vectmat{c} \oplus \vectmat{e}$, where
$\vectmat{c} = [c_0, c_1, \ldots, c_{n-1}] \in \mathbb{F}_2^n$ is a  $1 \times n$
legit codeword, i.e., $\vectmat{c}\in\mathbf{C}$,
$\vectmat{e} = [e_0, e_1, \ldots, e_{n-1}] \in
\mathbb{F}_2^n$ is an unknown $1 \times n$ error vector with weight $t$, while $\oplus$
denotes the component-wise modulo-$2$ addition between the binary values in
$\vectmat{c}$ and $\vectmat{e}$.
As a consequence, the $r \times 1 $ syndrome value,
$\vectmat{s} = [s_0, s_1 \ldots, s_{r-1}] \in \mathbb{F}_2^r$,
taken as input of the decoding process
described by Algorithm~\ref{algo:bitflipping} matches the following equalities:
$\vectmat{s}$$=$$\vectmat{H}\,
\widetilde{\vectmat{c}}$$=$$\vectmat{H} (\vectmat{c}\oplus \vectmat{e})^{\mathtt{T}} =
\vectmat{H}\vectmat{e}^{\mathtt{T}}$.

Given a parity-check matrix $\vectmat{H} \in \mathbb{F}_2^{r \times k}$,
a syndrome $\vectmat{s} \in \mathbb{F}^n$, and an error vector $\vectmat{e} \in \mathbb{F}^n$, a
{\em parity-check equation} is defined as follows.
Its right-hand side is a syndrome bit $s_i$, $i$$\in$$\{0,\ldots,r$$-$$1\}$, 
whilst its left-hand side is the sum of $n$ unknown error bits 
$e_j$, $j$$\in$$\{0, \ldots, n$$-$$1\}$, each of which is multiplied by the 
binary coefficient $h_{i,j}$ in the corresponding position of the $i$-th 
row in the parity-check matrix.
Following the example depicted in Figure~\ref{fig:1:notation}, we have
that the first row of the depicted parity-check matrix is associated to equation
$\sum_{i=0}^{13} h_{0,i} e_i = h_{0,0} e_0 \oplus h_{0,3} e_3 \oplus h_{0,9} e_9
\oplus h_{0,11} e_{11} \oplus = s_0$,
which, plugging in the depicted known values for the elements of $\vectmat{H}$ and $\vectmat{s}$,
becomes $e_0 \oplus e_3 \oplus e_9 \oplus e_{11} \oplus = 1$.
We deem a parity-check equation {\em satisfied} when $s_i = 0$, and {\em unsatisfied}
otherwise: rows corresponding to unsatisfied parity-check equations are
highlighted in cyan in the running example of Figure~\ref{fig:1:notation}, 
and the previously considered equation is unsatisfied.
We consider an element of the error vector $e_j$, corresponding to a position
$0 \leq j < n$, to be \emph{involved} in a parity-check equation $i$, $0 \leq i < r$,
if the corresponding coefficient in the parity-check equation, $h_{i,j}$,
is asserted. The reason behind the nomenclature is that an involved error
element will indeed influence the value of the known term of the equation (i.e., of the
syndrome bit).

Algorithm~\ref{algo:bitflipping} attempts to find the value of $\vectmat{e}$ by
iteratively refining a binary vector $\bar{\vectmat{e}}$ containing the current
best estimate for $\vectmat{e}$.
After each iteration, the algorithm updates the value of the syndrome
to be used for the next iteration so that the equality
$\vectmat{H}(\bar{\vectmat{e}}\oplus \vectmat{e})^{\mathtt{T}}=\vectmat{s}$
holds, terminating either as soon as the syndrome
has a null value, or after a predetermined maximum number of iterations is
reached.
We note that, while $\vectmat{s}=0_{r \times 1}$ does not necessarily imply that
$\bar{\vectmat{e}}=\vectmat{e}$,
it is commonplace to employ a test on the nullity of the syndrome to determine
if a correct decoding action took place (lines $12$--$15$ of
Algorithm~\ref{algo:bitflipping}), as working in a error density regime where
this is not achieved with certainty leads to a poor error correction
performance~\cite{DBLP:conf/isit/Tillich18}.
From now on, we will denote the vector $\bar{\vectmat{e}}\oplus \vectmat{e}$ as
the vector of discrepancies between the value of
the error $\vectmat{e}$ (unknown to the decoder) and the value of its estimate
$\bar{\vectmat{e}}$
(known to the decoder).
We will indicate the asserted bits in $\vectmat{e}$ (i.e., the bit positions
$j$ where $e_j = 1$) as {\em incorrect}, while the asserted bits in
$\bar{\vectmat{e}}\oplus \vectmat{e}$ (i.e., the bit positions
$j$ where $\bar{e}_j \neq e_j $) will be denoted as {\em discrepancies} or
{\em incorrectly guessed} bits.

\begin{figure}[!t]
     \resizebox{!}{0.16\textwidth}{
        \begin{tikzpicture}[scale=1.0,node distance=2cm]
\matrix (PCM) [nodes={draw=black,minimum size=5mm}]  {
     \node[color=cyan]{1};& \node[color=cyan]{0};& \node[color=cyan]{0};& \node[color=cyan]{1};& \node[color=cyan]{0};& \node[color=cyan]{0};& \node[color=cyan]{0};&   \node[color=cyan]{0};& \node[color=cyan]{0};& \node[color=cyan]{1};& \node[color=cyan]{0};& \node[color=cyan]{1};& \node[color=cyan]{0};& \node[color=cyan]{0};\\
    \node{0};& \node{1};& \node{0};& \node{0};& \node{1};& \node{0};& \node{0};&   \node{0};& \node{0};& \node{0};& \node{1};& \node{0};& \node{1};& \node{0};\\
    \node[color=cyan]{0};& \node[color=cyan]{0};& \node[color=cyan]{1};& \node[color=cyan]{0};& \node[color=cyan]{0};& \node[color=cyan]{1};& \node[color=cyan]{0};&   \node[color=cyan]{0};& \node[color=cyan]{0};& \node[color=cyan]{0};& \node[color=cyan]{0};& \node[color=cyan]{1};& \node[color=cyan]{0};& \node[color=cyan]{1};\\
    \node{0};& \node{0};& \node{0};& \node{1};& \node{0};& \node{0};& \node{1};&   \node{1};& \node{0};& \node{0};& \node{0};& \node{0};& \node{1};& \node{0};\\
    \node{1};& \node{0};& \node{0};& \node{0};& \node{1};& \node{0};& \node{0};&   \node{0};& \node{1};& \node{0};& \node{0};& \node{0};& \node{0};& \node{1};\\
    \node[color=cyan]{0};& \node[color=cyan]{1};& \node[color=cyan]{0};& \node[color=cyan]{0};& \node[color=cyan]{0};& \node[color=cyan]{1};& \node[color=cyan]{0};&   \node[color=cyan]{1};& \node[color=cyan]{0};& \node[color=cyan]{1};& \node[color=cyan]{0};& \node[color=cyan]{0};& \node[color=cyan]{0};& \node[color=cyan]{0};\\
    \node[color=cyan]{0};& \node[color=cyan]{0};& \node[color=cyan]{1};& \node[color=cyan]{0};& \node[color=cyan]{0};& \node[color=cyan]{0};& \node[color=cyan]{1};&   \node[color=cyan]{0};& \node[color=cyan]{1};& \node[color=cyan]{0};& \node[color=cyan]{1};& \node[color=cyan]{0};& \node[color=cyan]{0};& \node[color=cyan]{0};\\  };
   
  \matrix (synd) [right of=PCM,xshift=2.5cm,nodes={draw=black,minimum size=5mm}]  {
\node[color=cyan]{1};\\
\node{0};\\
\node[color=cyan]{1};\\
\node{0};\\
\node{0};\\
\node[color=cyan]{1};\\
\node[color=cyan]{1};\\
};

  \matrix (err) [below of=PCM,yshift=-0.5cm,nodes={draw=black,minimum size=5mm}]  {
    \node {?};& \node{?};&  \node{?};& \node{?};& \node{?};&    \node{?};& \node{?};& \node{?};& \node{?};& \node{?};& \node{?};& \node{?};& \node{?};& \node{?};\\   
  } ;
  
  \matrix (errbar) [below of=err,yshift=1.3cm,nodes={draw={black},minimum size=5mm}]  {
    \node {0};& \node{0};&  \node{0};& \node{0};&\node{0};& \node{0};& \node{0};& \node{0};& \node{0};& \node{0};& \node{0};& \node{0};& \node{0};& \node{0};\\
  };
	
	\matrix (discr) [below of=errbar,yshift=1.3cm,nodes={draw={black},minimum size=5mm}]  {
    \node {?};& \node{?};&  \node{?};& \node{?};&\node{?};& \node{?};& \node{?};& \node{?};& \node{?};& \node{?};& \node{?};& \node{?};& \node{?};& \node{?};\\
  };
  
\node (PCMlab)   [left of=PCM,   xshift=-3cm, align=left] {\LARGE $\vectmat{H}$};
\node (errlab)   [left of=err,   xshift=-3cm, align=left] {\LARGE $\vectmat{e}$};
\node (synd lab) [right of=synd, xshift=-1.0cm] {\LARGE $\vectmat{s}$};
\node (errbarlab) [left of=errbar,xshift=-3cm, align=left] {\LARGE $\bar{\vectmat{e}}_{(0)}$};
\node (discrlab) [left of=discr,xshift=-3.6cm, align=left] {\LARGE $\vectmat{e} \oplus \bar{\vectmat{e}}_{(0)}$};
\end{tikzpicture}
     }
     \resizebox{!}{0.16\textwidth}{
        \begin{tikzpicture}[scale=1.5,node distance=2cm]
\matrix (PCM) [nodes={draw=black,minimum size=5mm}]  {
    \node[color=cyan]{1};& \node[color=cyan]{0};& \node[color=cyan]{0};& \node[color=cyan]{1};& \node[color=cyan]{0};& \node[color=cyan]{0};& \node[color=cyan]{0};&   \node[color=cyan]{0};& \node[color=cyan]{0};& \node[color=cyan]{1};& \node[color=cyan]{0};& \node[color=cyan]{1};& \node[color=cyan]{0};& \node[color=cyan]{0};\\
    \node{0};& \node{1};& \node{0};& \node{0};& \node{1};& \node{0};& \node{0};&   \node{0};& \node{0};& \node{0};& \node{1};& \node{0};& \node{1};& \node{0};\\
    \node[color=cyan]{0};& \node[color=cyan]{0};& \node[color=cyan]{1};& \node[color=cyan]{0};& \node[color=cyan]{0};& \node[color=cyan]{1};& \node[color=cyan]{0};&   \node[color=cyan]{0};& \node[color=cyan]{0};& \node[color=cyan]{0};& \node[color=cyan]{0};& \node[color=cyan]{1};& \node[color=cyan]{0};& \node[color=cyan]{1};\\
    \node{0};& \node{0};& \node{0};& \node{1};& \node{0};& \node{0};& \node{1};&   \node{1};& \node{0};& \node{0};& \node{0};& \node{0};& \node{1};& \node{0};\\
    \node{1};& \node{0};& \node{0};& \node{0};& \node{1};& \node{0};& \node{0};&   \node{0};& \node{1};& \node{0};& \node{0};& \node{0};& \node{0};& \node{1};\\
    \node[color=cyan]{0};& \node[color=cyan]{1};& \node[color=cyan]{0};& \node[color=cyan]{0};& \node[color=cyan]{0};& \node[color=cyan]{1};& \node[color=cyan]{0};&   \node[color=cyan]{1};& \node[color=cyan]{0};& \node[color=cyan]{1};& \node[color=cyan]{0};& \node[color=cyan]{0};& \node[color=cyan]{0};& \node[color=cyan]{0};\\
    \node[color=cyan]{0};& \node[color=cyan]{0};& \node[color=cyan]{1};& \node[color=cyan]{0};& \node[color=cyan]{0};& \node[color=cyan]{0};& \node[color=cyan]{1};&   \node[color=cyan]{0};& \node[color=cyan]{1};& \node[color=cyan]{0};& \node[color=cyan]{1};& \node[color=cyan]{0};& \node[color=cyan]{0};& \node[color=cyan]{0};\\
  };
   
 \matrix (synd) [right of=PCM,xshift=2.5cm,nodes={draw=black,minimum size=5mm}]  {
\node[color=cyan]{1};\\
\node{0};\\
\node[color=cyan]{1};\\
\node{0};\\
\node{0};\\
\node[color=cyan]{1};\\
\node[color=cyan]{1};\\
};

  \matrix (errbar) [below of=PCM,yshift=-0.5cm,nodes={draw=black,minimum size=5mm}]  {
    \node {0};& \node{0};&  \node{0};& \node{0};&\node{0};& \node{0};& \node{0};& \node{0};& \node{0};& \node{0};& \node{0};& \node{0};& \node{0};& \node{0};\\
  } ;
  
  \matrix (upc) [below of=errbar,yshift=1cm,nodes={draw={black},minimum size=5mm}]  {
    \node {1};& \node{1};&  \node{2};& \node{1};&\node{0};& \node{2};& \node{1};& \node{1};& \node{1};& \node{2};& \node{1};& \node{2};& \node{0};& \node{1};\\
  };
  
  \node (upc)       [left of=upc, xshift=-2.3cm, align=left] {\LARGE $\mathtt{upc}$};
  \node (PCMlab)    [left of=PCM,   xshift=-2.1cm, align=left] {\LARGE $\vectmat{H}$};
  \node (synd lab)  [right of=synd, xshift=-1.0cm] {\LARGE $\vectmat{s}$};
  \node (errbarlab) [left of=errbar,xshift=-2.2cm, align=left] {\LARGE $\bar{\vectmat{e}}_{(0)}$};
\end{tikzpicture}
     }
     \resizebox{!}{0.16\textwidth}{
        \begin{tikzpicture}[scale=1.5,node distance=2cm]
\matrix (PCM) [nodes={draw=black,minimum size=3mm}]  {
    \node{1};& \node{0};& \node[color=cyan]{0};& \node{1};& \node{0};& \node[color=cyan]{0};& \node{0};&   \node{0};& \node{0};& \node[color=cyan]{1};& \node{0};& \node[color=cyan]{1};& \node{0};& \node{0};\\
    \node{0};& \node{1};& \node[color=cyan]{0};& \node{0};& \node{1};& \node[color=cyan]{0};& \node{0};&   \node{0};& \node{0};& \node[color=cyan]{0};& \node{1};& \node[color=cyan]{0};& \node{1};& \node{0};\\
    \node{0};& \node{0};& \node[color=cyan]{1};& \node{0};& \node{0};& \node[color=cyan]{1};& \node{0};&   \node{0};& \node{0};& \node[color=cyan]{0};& \node{0};& \node[color=cyan]{1};& \node{0};& \node{1};\\
    \node{0};& \node{0};& \node[color=cyan]{0};& \node{1};& \node{0};& \node[color=cyan]{0};& \node{1};&   \node{1};& \node{0};& \node[color=cyan]{0};& \node{0};& \node[color=cyan]{0};& \node{1};& \node{0};\\
    \node{1};& \node{0};& \node[color=cyan]{0};& \node{0};& \node{1};& \node[color=cyan]{0};& \node{0};&   \node{0};& \node{1};& \node[color=cyan]{0};& \node{0};& \node[color=cyan]{0};& \node{0};& \node{1};\\
    \node{0};& \node{1};& \node[color=cyan]{0};& \node{0};& \node{0};& \node[color=cyan]{1};& \node{0};&   \node{1};& \node{0};& \node[color=cyan]{1};& \node{0};& \node[color=cyan]{0};& \node{0};& \node{0};\\
    \node{0};& \node{0};& \node[color=cyan]{1};& \node{0};& \node{0};& \node[color=cyan]{0};& \node{1};&   \node{0};& \node{1};& \node[color=cyan]{0};& \node{1};& \node[color=cyan]{0};& \node{0};& \node{0};\\
  };
   
  \matrix (synd) [right of=PCM,xshift=2.5cm,nodes={draw=black,minimum size=3mm}]  {
\node[color=cyan]{1};\\
\node[color=cyan]{0};\\
\node[color=cyan]{0};\\
\node[color=cyan]{0};\\
\node[color=cyan]{0};\\
\node[color=cyan]{1};\\
\node[color=cyan]{0};\\
};
  
  \matrix (errbar) [below of=PCM,yshift=-0.5cm,nodes={draw=black,minimum size=3mm}]  {
    \node {0};& \node{0};&  \node[color=cyan]{1};& \node{0};&\node{0};& \node[color=cyan]{1};& \node{0};& \node{0};& \node{0};& \node[color=cyan]{1};& \node{0};& \node[color=cyan]{1};& \node{0};& \node{0};\\
  } ;
  
  \node (PCMlab)   [left of=PCM,   xshift=-2.1cm, align=left] {\LARGE $\vectmat{H}$};
  \node (synd lab) [right of=synd, xshift=-1.0cm] {\LARGE $\vectmat{s}_{(1)}$};
  \node (errbarlab) [left of=errbar,xshift=-2.1cm, align=left] {\LARGE $\bar{\vectmat{e}}_{(1)}$};
\end{tikzpicture}
     }
     \caption{Bit flipping algorithm variables at the beginning of the decoding procedure
     (i.e., when the number of iterations equals zero, $\mathtt{iter} = 0$) for a toy code $\mathbf{C}[n=14, k=7]$, where
     the parity-check matrix has column weight $v=2$ and row weight $w=4$, with a error weight $t=2$.
     Rows corresponding to unsatisfied parity-check equations, and their corresponding syndrome bits are highlighted
     in cyan.
     The unknown error vector is denoted as $\vectmat{e} = [{e}_0, \ldots,{e}_{n-1}]$, the error vector estimate
     is denoted as $\bar{\vectmat{e}} = [\bar{e}_0, \ldots, \bar{e}_{n-1}]$, while
     the sequence of discrepancies between the bits of the actual error vector and the corresponding
     bits in the error vector estimate just before the decoding computation is denoted as
     $\vectmat{e} \oplus \bar{\vectmat{e}}_{(0)}$.
     At the end of each iteration of the decoding procedure the equality
     $\vectmat{s}$$=$$\vectmat{H}$$(\vectmat{e}$$\oplus$$\bar{\vectmat{e}}_{(\mathtt{iter})})^{\mathtt{T}}$ holds.
     \label{fig:1:notation}}
\end{figure}

The iterative correction procedure starts by initializing the value of the
error estimate $\bar{\vectmat{e}}$ to the null vector $0_{1\times n}$ (line $2$, and
central subfigure in Figure~\ref{fig:1:notation}).
Each algorithm iteration is split up in three phases. In the first phase 
(lines $4$-$5$), it computes the inner product between the bit vector $\vectmat{s}$ and
the bits in each column of $\vectmat{H}$, considering them as integers, obtaining a
quantity known as the ``unsatisfied parity-check [equation count]'' (upc)
bound to the $j$-th bit position in the error vector and stores such a value
in a variable $\mathtt{upc}_j$, $j \in \{0,\ldots,n\}$.
Indeed, $\mathtt{upc}_j = \langle \vectmat{s}, \vectmat{h}_{:,j}\rangle = \sum_{i=0}^{r-1} s_i\cdot h_{i,j}$
(where $\vectmat{h}_{:,j}$ denotes the $j$-th column of $\vectmat{H}$, and $s_i\cdot h_{i,j}$ is used as an integer value) is the number of
unsatisfied parity-check equations in which the $j$-th element of the error
vector $e$ is involved.
Figure~\ref{fig:1:notation}, center, reports the computed values of $\mathtt{upc}$ during
the first iteration of the decoding algorithm.
In the second phase, a threshold $\mathtt{th}$$\in$$\left\{\lceil 
\frac{v+1}{2}\rceil,\ldots, v\right\}$
is computed as a function of the current value of the syndrome and the current 
iteration count.
In the third phase (lines $7$-$10$), the algorithm evaluates
for each $j$$\in$$\{0,\ldots,n$$-$$1\}$, if $\mathtt{upc}_j$ is greater than the
threshold ${\mathtt{th}}$, and in the affirmative case it flips the value in
$\bar{e}_j$ (i.e., $\bar{e}_j \leftarrow \bar{e}_j \oplus 1$)
and updates the syndrome by adding to it the $j$-th column of $\vectmat{H}$
(i.e., $\vectmat{s} \leftarrow \vectmat{s} \oplus \vectmat{h}_{:,j}$).
To avoid ambiguities, whenever referring to the value of a variable of the
decoding algorithm after a specific iteration has finished acting on it, we 
suffix the variable itself with the iteration number.
Figure~\ref{fig:1:notation}, right, reports the value of the error estimate
$\bar{\vectmat{e}}$ after the first iteration is complete, that is $\bar{\vectmat{e}}_{(1)}$,
considering a threshold value equal to $2$, i.e., flipping all the bits in 
$\bar{\vectmat{e}}_{(1)}$ in a position $j$ where the corresponding value of $\mathtt{upc}_j$
is greater or equal to $2$. The columns of $\vectmat{H}$ corresponding to such positions are also
highlighted in cyan, and are the ones added to $\vectmat{s}_{(0)}$ to obtain the depicted
value of $\vectmat{s}_{(1)}$.

%
%
\section{Decoding Failure Rate Model}
Our approach to modeling the DFR of the iterative decoder starts by
characterizing the statistical distribution of the weight of the syndrome
in input to the decoder, $\weight{\vectmat{s}_{(0)}}$ (Section~\ref{subsec:synwt}).
Once this is done, we
proceed assuming that such a weight is known for the remainder of the computations.
To obtain the value of the modeled DFR, we employ the law of total probability
to combine the syndrome-weight dependent DFR estimates together, weighting them
according to the syndrome weight distribution.
The (syndrome-weight dependent) two-iterations DFR estimate is obtained in
two steps.

The first step (described in Section~\ref{subsec:first}) computes the
statistical distribution of the upc values during the first iteration of
the decoder, and derives from it the distribution of the amount of
correctly flipped bits ($\discminus$) in the error estimate $\bar{\vectmat{e}}$, and
the distribution of the amount of incorrectly flipped bits ($\discplus$)
in the error estimate $\bar{\vectmat{e}}$.

The second step (described in Section~\ref{subsec:second}) starts from the
distributions of $\discplus$ and $\discminus$ and logically partitions the positions
of bit values of the error vector into four classes, according to whether
the bit values in those positions were correctly estimated or not during the
first iteration.
The two-iterations DFR estimate is obtained by computing separately the contribution
of the aforementioned four classes.


An instrumental working assumption, which will be used in our derivations starting
from Section~\ref{subsec:first}, is the following:
\begin{assumption}\label{assumption:indep}
    Let $\mathcal{S}_0$, $\mathcal{S}_1$, $\dots$, $\mathcal{S}_{v-1}$ be the random variables modeling
    the outcome of $v$ parity-check equations (among $\vectmat{H}\vectmat{e}^{\mathtt{T}} = \vectmat{s}$)
    indexed by the position values
    in the support of a given column of the parity-check matrix $\vectmat{H}$,
    $\vectmat{h}_{:,j}$, $0 \leq j \leq n-1, \weight{\vectmat{h}_{:,j}} = v$.
    Then $\mathcal{S}_i \sim \berndist{p_i}$, for some probability $p_i$
    ($0 \leq i \leq v-1$),
    independently from the outcome of any other parity-check.
\end{assumption}
We will employ this assumption to model the distribution of the $\mathtt{upc}$ counters
as the sum of Bernoulli random variables. This assumption is used is several other works
on regular LDPC and MDPC codes~\cite{DBLP:conf/isit/Tillich18,DBLP:conf/pqcrypto/SendrierV19,DBLP:journals/iacr/ArpinLPRTV25}.
\newline
In Section~\ref{subsec:assumption}, we will discuss the impact of this
approximation on the estimation of the decoding failure probability,
providing evidence that this assumption is not only reasonable, but
also slightly pessimistic in terms of the estimated DFR.

\subsection{Modeling Syndrome Weight Distribution}\label{subsec:synwt}

\begin{figure}[t]
    \centering
    \scalebox{0.7}{
\begin{tikzpicture}[
    >=stealth,
    node distance = 3.2cm and 6cm,
    every state/.style={circle, thick, draw, minimum size=20mm, font=\small},
    every edge/.append style={thick}
]

\node[state] (t0) {$\mathcal{W}_{\ell-1}=0$};
\node[state, right of=t0] (t1) {$\mathcal{W}_{\ell-1}=1$};
\node[state, right of=t1, draw=none] (t2) {\large $\cdots$};
\node[state, right of=t2] (tx) {$\mathcal{W}_{\ell-1}=x$};
\node[state, right of=tx, draw=none] (t3) {\large $\cdots$};
\node[state, right of=t3] (tr1) {$\mathcal{W}_{\ell-1}=r-1$};
\node[state, right of=tr1] (tr) {$\mathcal{W}_{\ell-1}=r$};

\node[state, below=1cm of t0] (b0) {$\mathcal{W}_{\ell}=0$};
\node[state, right of=b0] (b1) {$\mathcal{W}_{\ell}=1$};
\node[state, right of=b1, draw=none] (b2) {\large $\cdots$};
\node[state, right of=b2] (by) {$\mathcal{W}_{\ell}=y$};
\node[state, right of=by, draw=none] (b3) {\large $\cdots$};
\node[state, right of=b3] (br1) {$\mathcal{W}_{\ell}=r-1$};
\node[state, right of=br1] (br) {$\mathcal{W}_{\ell}=r$};

\path[->]
  (tx) edge node[right]{$p_{x,y,\ell}$} (by)
  (t0) edge node[right, yshift=-0.2cm, xshift=0.4cm]{$p_{0,1,\ell}$} (b1)
  (t1) edge node[left, yshift=-0.2cm, xshift=-0.4cm]{$p_{1,0,\ell}$} (b0)
  (tr1) edge node[right, yshift=-0.2cm, xshift=0.4cm]{$p_{r-1,r,\ell}$} (br)
  (tr) edge  node[left, yshift=-0.2cm, xshift=-0.4cm]{$p_{r,r-1,\ell}$} (br1)
  (t1) edge (b2)
  (tr1) edge (b3)
  ;
\end{tikzpicture}
}
    \caption{Graphical representation of the $\ell$-th step of the non-homogeneous
    Markov chain defined in Section~\ref{subsec:synwt}.
    Each state $\mathcal{W}_{\ell-1} = x$, $0 \leq x \leq r$, corresponds to each
    possible value of the Hamming weight of the syndrome at step $\ell-1$
    (i.e., $(\vectmat{e}, \vectmat{s})_{\ell-1}$),
    while the states $\mathcal{W}_{\ell} = y$, $0 \leq y \leq r$,
    correspond to the possible values of the Hamming weight of the syndrome at
    step $\ell$ (i.e., $(\vectmat{e}, \vectmat{s})_{\ell-1}$).
    The transition probability from state $\mathcal{W}_{\ell-1} = x$ to $\mathcal{W}_{\ell} = y$ is
    $p_{x,y,\ell} = \Pr(\mathcal{W}_{\ell} = y \mid \mathcal{W}_{\ell-1} = x)$.}
    \label{fig:markovchain}
\end{figure}

In this section, we derive the probability
that the syndrome $\vectmat{s}_{(0)}$ at the
beginning of the decoding process has a Hamming
weight $\weight{\vectmat{s}_{(0)}} = y$.
To this end, we define a non-homogeneous
Markov chain modeling a sequence of
random variables $\mathcal{W}_\ell$,
$0 \leq \ell \leq t$, in such a way that the
distribution of $\mathcal{W}_t$
matches the distribution of
$\weight{\vectmat{s}_{(0)}}$, and we show how the probability $\Pr\left(\mathcal{W}_t=y\right)$
can be computed
starting from the (known) distribution of $\mathcal{W}_0$,
i.e., the first random variable of the sequence.

Let us denote as $(\vectmat{e}, \vectmat{s})_\ell$, $\ell$$\in$$\{0,1,2 \ldots, t\}$,
a pair of values representing an error vector $\vectmat{e}$ and its corresponding syndrome
$\vectmat{s}=\vectmat{H}\vectmat{e}^{\mathtt{T}}$ both indexed by the weight of the error vector $\ell$.
We consider the syndrome we want to model and its corresponding error vector
with weight $t$ as the last pair in the sequence $(\vectmat{e}, \vectmat{s})_0,
( \vectmat{e}, \vectmat{s})_1, (\vectmat{e}, \vectmat{s})_2, \ldots, (\vectmat{e}, \vectmat{s})_t$,
where $(\vectmat{e}, \vectmat{s})_0$ contains a null
error vector and its null syndrome, while $(\vectmat{e}, \vectmat{s})_{\ell}$, $\ell \geq 1$,
denotes a pair with an error vector that includes the same asserted bits of the
error vector in $(\vectmat{e}, \vectmat{s})_{\ell-1}$ plus an additional single asserted bit that is
uniformly randomly placed in one out of the $n$$-$$(\ell-1)$ available positions.
Analogously, the syndrome value in $(\vectmat{e}, \vectmat{s})_{\ell}$, $\ell \geq 1$, is
assumed to differ from the one in $(\vectmat{e}, \vectmat{s})_{\ell-1}$ due to the
bitwise addition of the column of the parity-check matrix $\vectmat{H}$ selected by the
said additional asserted bit. The vector $\vectmat{s}_{(0)}$ in input to the decoder can
be seen as the syndrome in the final pair $(\vectmat{e}, \vectmat{s})_{t}$, computed as the sum of
the $t$ columns of the parity-check matrix $\vectmat{H}$ indexed by asserted bits in $\vectmat{e}$.

We model the Hamming weight of each syndrome in the aforementioned sequence as an
instance of a discrete random variable $\mathcal{W}_\ell$ bound to a probability
mass function $\Pr(\mathcal{W}_\ell = y)$, with $\ell \in \{0,\ldots, t\}$,
$y$$\in$$\{0, \ldots,r\}$, which can be represented operatively as an array with $r$$+$$1$ elements
$\mathbf{\mathtt{wp}}_{(\ell)} = [\mathrm{wp}_{(\ell),0},\ldots,
\mathrm{wp}_{(\ell),x},\ldots, \mathrm{wp}_{(\ell),r}]$. Starting from the distribution
of the weight of the syndrome of a null error vector, $\Pr(\mathcal{W}_0 = y) = \mathtt{wp}_{(0),y}$
where $\mathbf{\mathtt{wp}}_{(0)} = [1, 0, \ldots, 0]$, we define a non-homogeneous Markov
chain with $r$$+$$1$ states, where each step models the change in the distribution of the
Hamming weight of the syndrome (defined as $\mathcal{W}_\ell$ after $\ell$ steps) brought by
the addition of one column of $\vectmat{H}$. Such a Markov chain is uniquely defined by
$\mathbf{\mathtt{wp}}_{(0)}$ and the
transition matrices $\mathbf{\mathtt{P}}_{(\ell)} =
[p_{x,y,\ell}]_{\scriptstyle x,y \in \{0,\ldots,r\}}$.
Specifically, the distribution of each random variable $\mathcal{W}_\ell$ is
derived through the following vector-matrix multiplication
$\mathbf{\mathtt{wp}}_{(\ell)} =
\mathbf{\mathtt{wp}}_{(\ell-1)}\cdot \mathbf{\mathtt{P}}_{(\ell)}$, with
$\ell$$\in$$\{1,\ldots,t\}$, where each transition probability
$p_{x,y,\ell} = \Pr(\mathcal{W}_{\ell} = y | \mathcal{W}_{\ell-1} = x)$ is a function of
the initial and final weight of the syndrome as well as of the step $\ell$
considered along the chain.

Figure~\ref{fig:markovchain} depicts the state change in the $\ell$-th step of the Markov chain.
After $\ell-1$ steps, the random variable $\mathcal{W}_{\ell-1}$ models the Hamming weight of
the syndrome $\vectmat{s}$ in $(\vectmat{e}, \vectmat{s})_{\ell-1}$,
with $\weight{\vectmat{e}} = \ell-1$, bound to the p.m.f.
$\Pr(\mathcal{W}_{\ell-1}=x) = \mathbf{\mathrm{wp}}_{(\ell-1),x}$.
During the $\ell$-th step, one column of the parity
check matrix is added to the syndrome, causing a change in its Hamming weight.
The term $p_{x,y,\ell}$ indicates the transition probability from state
$\mathcal{W}_{\ell-1} = x$ to state $\mathcal{W}_{\ell} = y$. Since the probability
distribution of $\mathcal{W}_{0}$ is known, computing $p_{x,y,\ell}$ for $0 \leq x,y \leq r$
and $0 \leq \ell \leq t$ is sufficient to derive the p.m.f. of $\mathcal{W}_{t}$, that is
the distribution of the Hamming weight of the
syndrome $\vectmat{s}_{(0)}$ in input to the decoder.

During each step, the addition of a column of $\vectmat{H}$ causes $v$ bits to flip in the syndrome.
In the following we model the number of flips induced on any syndrome bit, subsequently
we derive the probabilities of flipping up a clear bit and flipping down an asserted bit
of a syndrome along the aforementioned chain, and finally compute the statistical distribution
of the weight of a syndrome and the transition probabilities $p_{x,y,\ell}$.
We denote as $\mathcal{F}_\ell \in \{0,\ldots,\min(w,\ell)\}$ the discrete random variable
modeling the intersection between the support of a parity-check in $\vectmat{H}$
(i.e. of a row $\vectmat{h}_{i,:}$ of $\vectmat{H}$, $0\leq i < r$)
ad the support of an error vector $e$ with Hamming weight $\weight{\vectmat{e}}=\ell$.
We point out that $\mathcal{F}_\ell$ corresponds
to the number of flips applied to a syndrome bit of an error with weight $\ell$,
during the computation of the syndrome itself by means of sequential additions of $\ell$ columns of $\vectmat{H}$.
The range of such variable is justified by the fact that the maximum
number of flips happening to a syndrome bit is limited by the smaller value between
the total number of erroneous bits, $\ell$, and the number of
parity-check equation terms, $w$.
The p.m.f.  of $\mathcal{F}_\ell$ follows a hypergeometric
distribution $\phi_\ell(f)$$=$$\Pr(\mathcal{F}_\ell = f) =
\hygdist{n}{w}{\ell}{f} =
\frac{\binom{w}{f}\binom{n-w}{\ell-f}}{\binom{n}{\ell}}$.\\

We begin by modeling the probability of flipping a syndrome bit during step $\ell$, given its value at step $\ell-1$.
\begin{proposition}\label{prop:fliplltoone}
    The probability that any given syndrome bit that was not asserted at step $\ell$$-$$1$ is flipped up to $1$ at step
    $\ell$ is:
$$
\fliplltoone(\ell)=
\left({\displaystyle \sum_{f=0, \mathtt{even}}^{\min(\ell-1, w)}
\left(\frac{w-f}{n-(\ell-1)} \cdot \phi_{\ell-1}(f)\right)}\right)
\mathlarger{\mathlarger{\mathlarger{\mathlarger{\mathlarger{/}}}}}
\left({\displaystyle \sum_{f'=0, \mathtt{even}}^{\min(\ell-1, w)} \phi_{\ell-1}(f'))}\right)
$$
\end{proposition}
\begin{proof}
    We can rewrite the probability $\fliplltoone(\ell)$ in terms of $\mathcal{F}_{\ell}$.
    Indeed, if a syndrome bit is not asserted at the step $\ell-1$, then its value of $\mathcal{F}_{\ell-1}$
    is be even. Moreover, if $\mathcal{F}_{\ell-1} = f$ and the bit is flipped, then $\mathcal{F}_{\ell} = f+1$.
    Therefore:
$$
\fliplltoone(\ell) =
\sum_{f=0, \mathtt{even}}^{\min(\ell-1, w)}
\Pr(\mathcal{F}_{\ell} = f+1 | \mathcal{F}_{\ell-1} = f)
\Pr(\mathcal{F}_{\ell-1} = f | \mathcal{F}_{\ell-1}\text{ even})
$$
For odd values of $f$, we have $\Pr(\mathcal{F}_{\ell-1} = f | \mathcal{F}_{\ell-1}\text{ even}) = 0$.
For even values of $f$, we obtain $\Pr(\mathcal{F}_{\ell-1} = f | \mathcal{F}_{\ell-1}\text{ even})$ as:
$$
\Pr(\mathcal{F}_{\ell-1} = f | \mathcal{F}_{\ell-1}\text{ even}) =
\frac{
    \Pr(\mathcal{F}_{\ell-1} = f)
}{
    {\displaystyle \sum_{f'=0, \mathtt{even}}^{\min(\ell-1, w)}}
    \Pr(\mathcal{F}_{\ell-1} = f')
} =
\frac{
    \phi_{\ell-1}(f)
}{
    {\displaystyle \sum_{f'=0, \mathtt{even}}^{\min(\ell-1, w)}}
    \phi_{\ell-1}(f')
}
$$
We now recall that $(\vectmat{e}, \vectmat{s})_{\ell-1}$ and $\vectmat{e}, \vectmat{s})_{\ell}$
only differ by one erroneous bit, introduced at step $\ell$.
The probability that $\mathcal{F}_{\ell}=f+1$, given $\mathcal{F}_{\ell-1}=f$,
corresponds to the probability that the erroneous bit introduced in step $\ell$ is also included
in the parity-check. Since the total number of unselected bits at step $\ell-1$ is $n-(\ell-1)$, out
of which $w-f$ are within the parity-check, the probability of flipping the syndrome bit in such case is:
$$
\Pr(\mathcal{F}_{\ell} = f+1 | \mathcal{F}_{\ell-1} = f) =
\frac{
w-f
}{
n-(\ell-1)
}
$$
Combining the expressions above yields the stated formula for $\fliplltoone(\ell)$.
\end{proof}

\begin{proposition}\label{prop:fliplltozero}
    The probability that any given syndrome bit that was asserted at step $\ell$$-$$1$ is flipped down to $0$ at step
    $\ell$ is:
$$
\fliplltozero(\ell) =
\left({\displaystyle \sum_{f=1, \mathtt{odd}}^{\min(\ell-1, w)}
\left(\frac{w-f}{n-(\ell-1)} \cdot \phi_{\ell-1}(f)\right)}\right)
\mathlarger{\mathlarger{\mathlarger{\mathlarger{\mathlarger{/}}}}}
\left({\displaystyle \sum_{f'=1, \mathtt{odd}}^{\min(\ell-1, w)} \phi_{\ell-1}(f'))}\right)
$$
\end{proposition}
\begin{proof}
The proof is analogous to the one for $\fliplltoone(\ell)$.
The only difference lies in the assumption on the parity of $\mathcal{F}_{\ell-1}$, which is now assumed
to be odd given that the parity-check is unsatisfied (i.e. the corresponding syndrome bit is asserted).
\end{proof}

Observe that $\fliplltozero(\ell)$ is undefined for $\ell=1$ since $\phi_0(0) = 1$,
leading to the denominator of $\fliplltoone(1)$ being equal to zero.
Nonetheless, its value can  be easily derived: no parity-check is unsatisfied at
step $0$, meaning that $\fliplltozero(1)$ is null.

We now analyze the change of the syndrome weight from step $\ell$$-$$1$ to step $\ell$,
to derive the p.m.f.  $p_{x,y,\ell}$$=$$\Pr(\mathcal{W}_\ell = y |
\mathcal{W}_{\ell-1} = x)$, with $x,y \in \{0,\ldots, r\}$.
Since the column weight of the parity-check matrix is $v$, the overall amount
of flips taking place among all $r$ bits of the syndrome when an additional
error bit is asserted is $v$.
As a consequence, the weight $y$ of the syndrome at step $\ell$ is obtained from
flipping up $a$ clear bits out of $r$$-$$x$, and flipping down
$v$$-$$a$ asserted bits out of $x$, for all admissible
values of $a$, i.e., $a$$\in$$\left\{ \max\{0,v-x)\}, \ldots,
\min\{r-x, v\}\right\}$.
The weight of the syndrome after the $\ell$-th step is completed is thus
$y = x+a-(v-a)$, or equivalently it holds that $r-y = r-x-a+(v-a)$,
from which we derive $a = \frac{y-x+v}{2}$. We now derive the probability of transitioning from a
syndrome Hamming weight $\mathcal{W}_{\ell-1} = x$ to a syndrome Hamming weight $\mathcal{W}_{\ell} = y$.

\begin{proposition}
Let $p_{x,y,\ell}$ be the probability that $\mathcal{W}_{\ell} = y$, given $\mathcal{W}_{\ell-1} = x$.
Then:
$$
p_{x,y,\ell} =
\begin{cases}
1, & \text{if } \ell=1,\, x=0,\, y=v \\
\frac{
\omega_{\mathtt{up}}\left(\frac{y-x+v}{2}\right)
\cdot
\omega_{\mathtt{down}}\left(\frac{y-x+v}{2}\right)
}{
\sum_{a=\max\{0,v-x\}}^{\min\{r-x, v\}}
\omega_{\mathtt{up}}\left(a\right) \cdot \omega_{\mathtt{down}}\left(a\right)
}, & \text{if }
\ell \geq 2, y \equiv_2 x+v, \max(0, x-v) \leq y \leq \min(x+v, r) \\
0, & \text{otherwise}
\end{cases}
$$
\text{where:}
$$
\omega_{\mathtt{up}}\left(a\right) := \bindist{r-x}{a}{\fliplltoone(\ell)} \quad \text{and} \quad
\omega_{\mathtt{down}}\left(a\right) := \bindist{x}{v-a}{\fliplltozero(\ell)}
$$
\end{proposition}
\begin{proof}
Considering $\mathcal{W}_{\ell-1} = x$, during step $\ell$ a certain number $a$ of satisfied
parity-checks will become unsatisfied due to the addition of a column of $\vectmat{H}$ to the syndrome,
while $v-a$ unsatisfied checks will become satisfied.

Proposition~\ref{prop:fliplltoone} defines $\fliplltoone(\ell)$ as the probability that any
satisfied parity-check at step $\ell-1$ becomes unsatisfied at step $\ell$, while
Proposition~\ref{prop:fliplltozero} defines $\fliplltozero(\ell)$ as the probability
that any unsatisfied parity-check at step $\ell-1$ becomes satisfied at step $\ell$.
Considering one specific value of $a$, the probability that $a$ satisfied parity-checks become unsatisfied is:
$\omega_{\mathtt{up}}\left(a\right) := \bindist{r-x}{a}{\fliplltoone(\ell)}$.
Likewise, the probability that $v-a$ unsatisfied parity-checks become satisfied is:
$\omega_{\mathtt{down}}\left(a\right) := \bindist{x}{v-a}{\fliplltozero(\ell)}$.
Considering all the admissible values for $a$, the probability of moving
to any admissible syndrome weight $y$ at step $\ell$ is computed as:
$$
\displaystyle
\sum_{a=\max\{0,v-x\}}^{\min\{r-x, v\}}
\left(
 \omega_{\mathtt{up}}\left(a\right) \cdot \omega_{\mathtt{down}}\left(a\right)
\right).
$$
Thus, for any admissible value of $a$, the probability $\Pr(\mathcal{W}_\ell = y | \mathcal{W}_{\ell-1} = x)$,
where $y = x - v + 2a$, is:
$$
\rho(x, y, \ell):=\frac{
\omega_{\mathtt{up}}\left(\frac{y-x+v}{2}\right)
\cdot
\omega_{\mathtt{down}}\left(\frac{y-x+v}{2}\right)
}{
\sum_{a=\max\{0,v-x\}}^{\min\{r-x, v\}}
\omega_{\mathtt{up}}\left(a\right) \cdot \omega_{\mathtt{down}}\left(a\right)
}
$$

\noindent The transition probability
$p_{x,y,\ell}$$=$$\Pr(\mathcal{W}_\ell = y | \mathcal{W}_{\ell-1} = x)$ can be written as a
function of the step count $\ell$, and of both the starting and ending
weights of the syndrome.
$$
p_{x,y,\ell} =
\begin{cases}
1, & \text{if } \ell=1,\, x=0,\, y=v \\
\rho(x, y, \ell), & \text{if }
\ell \geq 2, y \equiv_2 x+v, \max(0, x-v) \leq y \leq \min(x+v, r) \\
0, & \text{otherwise}
\end{cases}
$$
where, in the special case $\ell=1,x=0,y=v$, $\fliplltozero(\ell)$ is not defined, as
no flips from one to zero can be performed on a null vector.
The value $p_{0,v,1}=1$ is quantitatively justified since, during the first
step, a null syndrome (therefore with $x=0$) will deterministically turn into
a weight $v$ syndrome when a column of $\vectmat{H}$ is added.
\end{proof}

We point out that, in the derivation of $\rho(x,y,\ell)$, we are \emph{not} making use of
Assumption~\ref{assumption:indep}, since we are enforcing the total number of flipped syndrome
bits to be equal to $v$.
Indeed, the formula for $\rho(x,y,\ell)$ is equivalent to Fisher's
noncentral hypergeometric probability distribution~\cite{harkness1965extended}, modeling the
event of choosing $v$ bits to flip in the syndrome at step $\ell$ (as a consequence of adding
one column of $\vectmat{H}$ to $\vectmat{s}$) among $r-x$ clear bits and $x$ asserted bits, where the probability
of choosing asserted bits is not equal to the probability of choosing clear bits (i.e., the
extraction is \emph{not} uniform).

\begin{algorithm}[!t]
\DontPrintSemicolon
\SetInd{0.8em}{0.2em}
{\small
\KwIn{$(v, w)$-regular code parameters.\newline
$v$: column weight; $w$: row weight;\newline
$t$: error vector weight, $t\geq1$;\newline
$r$: num. of rows of the parity-check matrix;\newline
$n$: num. of columns of the parity-check matrix}
\KwOut{$\mathbf{\mathtt{wp}}_{(t)} = [\mathtt{wp}_{(t), 0}, \ldots,
\mathtt{wp}_{(t), r}]$}
$\mathbf{\mathtt{wp}} \gets [0,\ldots,0]$\;
$\mathbf{\mathtt{wp}}[v] \gets 1$\tcp*[l]{initialized as $\mathbf{\mathtt{wp}}_{(1)}$}
\For{$\ell$ \textbf{\em from} $2$ \KwTo $t$ }{
    $\mathbf{\mathtt{wp\_prev}} \gets \mathbf{\mathtt{wp}}$\;
    \For{$y$ \textbf{\em from} $0$ \KwTo $r$}{
        $\mathbf{\mathtt{wp}}[y] \gets 0$\;
        \For{$i$ \textbf{\em from} ${\max(0,y-v)}$ \KwTo ${\min(y+v, r)}$}{
            $\mathtt{p} \gets \rho(i,y,\ell)$\;
            $\mathbf{\mathtt{wp}}[y] \gets \mathbf{\mathtt{wp}}[y] + \mathtt{wp\_prev}[i]\cdot \mathtt{p}$\;
        }
    }
}
\KwRet $\mathbf{\mathtt{wp}}$\;
}
\caption{{\sc Syndrome Weight Distribution}\label{algo:distrSyndrome}}
\end{algorithm}

The pseudo-code of the procedure to derive the statistical distribution of
the weight of the syndrome of an error vector with weight $t$,
$\Pr(\mathcal{W}_t=y)$, yielding the corresponding array of probabilities
$\mathbf{\mathtt{wp}}_{(t)} = [\mathrm{wp}_{(t), 0}, \ldots,
\mathrm{wp}_{(t), r}]$ is shown by Algorithm~\ref{algo:distrSyndrome}.
The procedure start by initializing the vector containing the discrete
distribution $\mathbf{\mathtt{wp}}$ to the one of a weight $t$$=$$1$ error
vector (lines $1$--$2$).
Subsequently, the computation iterates $t$$-$$1$ times the procedure computing
the non-homogeneous Markov chain (lines $3$--$9$).
The chain is computed with two nested loops, iterating on the possible weights
(loop at $5$--$9$) and the non-null values of the transition matrix columns
$\mathbf{\mathtt{P}}_{(\ell)}$ (loop at lines $7$--$9$).

\subsection{First Iteration of a Bit Flipping Decoder}\label{subsec:first}
Given a $(v,w)$-regular binary code with a $r$$\times$$n$ parity-check matrix
$\vectmat{H} = [h_{i,j}]$ where $i$$\in$$\{0,\ldots,r$$-$$1\}$, $j$$\in$$\{0, \ldots, n$$-$$1\}$,
and the value of a syndrome $\vectmat{s}$$=$$\vectmat{H}\vectmat{e}^\mathtt{T}$ derived from an unknown error
vector $\vectmat{e}$ with weight $t$, we recall that a parity-check equation is defined as
$\bigoplus_{j=0}^{n-1} h_{i,j} e_j = s_i$, where $e_j \in \{0,1\}$ are the unknowns, $h_{i,j} \in \{0,1\}$
are the known coefficients and $s_i \in \{0,1\}$ the constant known term.
The equation is said to be {\em satisfied} if $s_i$$=$$0$, {\em unsatisfied} if
$s_i$$=$$1$.
We are going to consider the (parallel) bit flipping
decoding algorithm introduced in Section~\ref{sec:background}
that iteratively estimates the most likely value $\bar{\vectmat{e}}$ of the error
vector $\vectmat{e}$, given $\vectmat{s}$ and $\vectmat{H}$, starting from the
initial value $\bar{\vectmat{e}}=0_{1\times n}$.

Our goal is to estimate the statistical distribution of the random variable
$\mathcal{E}_{(\mathtt{iter})}$ modeling the Hamming weight of
$\bar{\vectmat{e}}\oplus \vectmat{e}$
after the $\mathtt{iter}$-th iteration of the decoding algorithm, i.e., the
count of differences between the actual error vector $\vectmat{e}$ and its current
estimate $\bar{\vectmat{e}}$.

The p.m.f.
$\Pr(\mathcal{E}_{(\mathtt{iter})} = \discre)$, $\discre$$\in$$\{0,\ldots, n\}$,
will be considered only with $\mathtt{iter} > 0$, because
$\Pr(\mathcal{E}_{(\mathtt{0})} = t)$$=$$1$,
before the beginning of the decoding algorithm, when $\bar{\vectmat{e}}$$=$$0$.
The p.m.f. $\Pr(\mathcal{E}_{(\mathtt{iter})} = \discre)$ will
be obtained, using the result in Section~\ref{subsec:synwt} about the distribution of
the weight of the syndrome of an error vector of weight $t$,
(i.e., $\mathcal{W}_t = y$), as follows:
$$
\Pr(\mathcal{E}_{(\mathtt{iter})} = \discre)=\sum_{y = 0}^{r}
\left( \condprob{\mathcal{E}_{(\mathtt{iter})} = \discre}{\mathcal{W}_t = y}
\Pr(\mathcal{W}_t = y)\right).
$$
From now on, the goal of the analysis is going to be estimating the
probability $\condprob{\mathcal{E}_{(\mathtt{iter})} = \discre}{\mathcal{W}_t = y}$,
$y$$=$$\weight{\vectmat{s}}$ considering $\mathtt{iter}$$=$$1$ in this subsection,
and $\mathtt{iter}$$=$$2$
in the next one.

In the analysis of the first iteration of the decoding algorithm,
we denote as $\mathcal{S}_{i \mid y}$, $i$$\in$$\{ 0, \ldots, r-1\}$,
the random variable modeling the value taken by any
bit of the syndrome at the beginning of the decoding process, given the total number
of asserted bits of the syndrome $y$.
Therefore, for each unsatisfied and satisfied parity-check equation, the
probability to observe a clear or asserted constant term is:
$\Pr(\mathcal{S}_{i\mid y} = 0) = \frac{r-y}{r}$, and
$\Pr(\mathcal{S}_{i\mid y} = 1) = \frac{y}{r}$, respectively.
\newline
Furthermore, we denote as $\eventij{0}$ or $\eventij{1}$ the events
describing the $j$-th error vector bit being either clear or asserted,
respectively, and such that it is also involved in the $i$-th parity-check
equation,
because $j$ equals the position of one of the $w$
asserted coefficients in the row vector $\vectmat{h}_{i,:}$, i.e.:
$\eventij{0}$$=$$\left\{h_{i,j} = 1\, \cap \, e_j = 0\right\}$, and
$\eventij{1}$$=$$\left\{h_{i,j} = 1\, \cap\, e_j = 1\right\}$, respectively.

Employing the random variable $\mathcal{F}_{t\mid y} \sim \Pr(\mathcal{F}_t = f\,|\, \mathcal{W}_t = y)$
to model the count of bit flips determining any bit of a syndrome with a given weight $y$, the
p.m.f.  of $S_{i\mid y}$ can also be described as:
$$
\Pr(\mathcal{S}_{i\mid y} = 0)=\Pr\left(\bigcup_{f = 0, \mathtt{even}}^{\min(t,w)} (\flipsy{f})\right)=
\sum_{f = 0, \mathtt{even}}^{\min(t,w)} \prob{\flipsy{f}},
$$
$$
\Pr(\mathcal{S}_{i\mid y} = 1)=\Pr\left(\bigcup_{f = 1, \mathtt{odd}}^{\min(t,w)} (\flipsy{f})\right)=
\sum_{f = 1, \mathtt{odd}}^{\min(t,w)} \prob{\flipsy{f}},
$$
where the last step is mutuated by the fact that the events $(\flipsy{f})$ are disjoint.
Note that $\flipsy{f}$ implies that there are $f$ asserted error vector bits
in the $i$-th parity-check equation that match $f$ out of $w$ asserted coefficients
in the $i$-th row of $\vectmat{H}$.
The complete derivation of the p.m.f. of the random variable
$\mathcal{F}_{t\mid y} \sim \Pr(\mathcal{F}_t = f\,|\, \mathcal{W}_t = y)$ is reported
in Appendix~\ref{app:appendix1}, as a function of the code parameters $(n,k,w,v)$,
the error weight $t$, and the syndrome weight $y$, as it follows a  line
of reasoning similar to the one employed in Section~\ref{subsec:synwt}.

We proceed by computing the probability of a parity-check being unsatisfied, given that one of the $w$ bits
included in the parity-check is correct. To proceed with our analysis,
we assume each parity-check equation in $\vectmat{H}\vectmat{e}^\mathtt{T} = \vectmat{s}$ to be identified by
the row index $i \in \{0,\ldots,r-1\}$ of the matrix $\vectmat{H}$.

\begin{proposition}
    Let $\punsatz$ be the probability $\condprob{\mathcal{S}_{i\mid y} = 1}{\eventij{0}}$ that the $i$-th
    parity-check being unsatisfied, given that one of the $w$ bits included in the parity-check, the bit in position
    $j$ is correct (i.e., $e_j=0$). Then:
    $$
    \punsatz = \frac{
    \sum_{f=1,\mathtt{odd}}^{\min(w,t)} \Pr\left(\mathcal{F}_{t|y}=f\right) \frac{w-f}{w}
    }{
    \sum_{f=0}^{\min(w,t)} \Pr\left(\mathcal{F}_{t|y}=f\right) \frac{w-f}{w}
    }
    $$
\end{proposition}

\begin{proof}
    Starting from the definition of $\punsatz$, we have:
    $$
    \punsatz = \condprob{\mathcal{S}_{i\mid y} = 1}{\eventij{0}}
    =
    \frac{
    \interprob{\mathcal{S}_{i\mid y} = 1}{\eventij{0}}
    }{
    \Pr\left(\eventij{0}\right)
    }
    $$
    The probability $\Pr\left(\eventij{0}\right)$ corresponds to the probability of choosing a correct bit among the $w$ bits in a parity-check. This probability can be expressed in terms of $\mathcal{F}_{t|y}$, since the event of selecting a correct bit is conditioned on the number of total correct bits present in the check:
    $$
    \Pr\left(\eventij{0}\right) =
    \sum_{f=0}^{\min(w,t)} \Pr\left(\mathcal{F}_{t|y}=f\right) \frac{w-f}{w}
    $$
    The probability $\interprob{\mathcal{S}_{i\mid y} = 1}{\eventij{0}}$ encodes the additional constraint on the value of $\mathcal{F}_{t|y}$, which is assumed to be odd in order for the parity-check to be unsatisfied:
    $$
    \interprob{\mathcal{S}_{i\mid y} = 1}{\eventij{0}} =
    \sum_{f=1,\mathtt{odd}}^{\min(w,t)} \Pr\left(\mathcal{F}_{t|y}=f\right) \frac{w-f}{w}
    $$
    Combining the two results, we obtain the stated formula.
\end{proof}

\begin{proposition}
    Let $\punsato$ be the probability $\condprob{\mathcal{S}_{i\mid y} = 1}{\eventij{1}}$ that
    the $i$-th parity-check being unsatisfied, given that one of the $w$ bits included in
    the parity-check, the bit in position $j$ is incorrect (i.e., $e_j=1$). Then:
    $$
    \punsato = \frac{
    \sum_{f=1,\mathtt{odd}}^{\min(w,t)} \Pr\left(\mathcal{F}_{t|y}=f\right) \frac{f}{w}
    }{
    \sum_{f=0}^{\min(w,t)} \Pr\left(\mathcal{F}_{t|y}=f\right) \frac{f}{w}
    }
    $$
\end{proposition}

\begin{proof}
    The proof is analogous to the one for $\punsato$, where $\eventij{0}$ is replaced by
    $\eventij{1}$ and the probability of choosing an incorrect bit in a parity-check,
    given $\mathcal{F}_{t|y}=f$, is $\frac{f}{w}$.
\end{proof}

We are now equipped to describe the distribution of the random variables
$\mathcal{U}_j$ modelling the integer values $\upc$ computed
during the first iteration as shown in Algorithm~\ref{algo:bitflipping}.
We are interested in modelling in particular $\condprob{\mathcal{U}_j=u}{e_j=0}$
and $\condprob{\mathcal{U}_j=u}{e_j=1}$, as they describe what is the information
available to the decoder (the upc value) to decide whether to flip or not
a given bit $\bar{e}_j$ which estimates $e_j$, depending on what the actual value
of $e_j$ is. We share the approach, described in the following proposition,
with the work of J. Chaulet~\cite{DBLP:phd/hal/Chaulet17}.

\begin{proposition}
    Let $\mathcal{U}_j$ be the random variable modelling the value of $\upc$ during the
    first iteration of Algorithm~\ref{algo:bitflipping}. Then:
$$\condprob{\mathcal{U}_j=u}{e_j=0} = \bindist{v}{u}{\punsatz}, \quad \text{and} \quad
\condprob{\mathcal{U}_j=u}{e_j=1} = \bindist{v}{u}{\punsato}.
$$
\end{proposition}

\begin{proof}
    Recalling that $\upc = \sum_{i=0}^{r-1} (s_i \cdot h_{i,j})$,
where $\vectmat{h}_{:,j}$ has only $v$ asserted elements out of $r$,
the range of possible values $u$ of $\mathcal{U}_j$ is $\{0,\ldots,v\}$.
Indeed, the value of $\upc$ in Algorithm~\ref{algo:bitflipping} results from the
sum of $v$ constant terms of the parity-check equations (i.e., $v$ bit values
$s_i$ of the syndrome out of $r$) that are selected
by the asserted elements of the $j$-th column of $\vectmat{H}$.
We denote the positions of the asserted bits in the column $\vectmat{h}_{:,j}$
as the support of $\vectmat{h}_{:,j}$,
$\support{\vectmat{h}_{:,j}}$.
We partition the support into two subsets, $\setofsat$
and $\setofunsat$, where $\setofsat$ contains all the indexes of the
satisfied parity-check equations, i.e., for all $i\in\setofsat$, we have
$s_i$$=$$0$, while $\setofunsat$ contains all the indexes of the unsatisfied
parity-check equations, i.e., for all $i\in\setofunsat$, we have $s_i$$=$$1$.

Therefore, assuming that the $j$-th error vector bit is clear
(i.e., $e_j=0$) and the syndrome value is such that $\upc=|\setofunsat| = u$,
and $|\setofsat| = v-u$, one out of the $\binom{v}{u}$ possible ways of partitioning
$\colsupport$, is captured by the following event:
$$
\left( \bigcap_{i\in\setofunsat} (\mathcal{S}_{i\mid y} = 1\, |\, \eventij{0} ) \right) \bigcap
\left(\bigcap_{i\in\setofsat} (\mathcal{S}_{i\mid y} = 0\, |\, \eventij{0} )\right).
$$
For the next step, we employ Assumption~\ref{assumption:indep}, considering
the outcomes of the parity-checks to be independent.
We thus obtain the probability of the said event as $\punsatz^u (1-\punsatz)^{v-u}$, and
consequently
$\condprob{\mathcal{U}_j=u}{e_j=0} = \bindist{v}{u}{\punsatz}$.
Following the same reasoning, we obtain
$\condprob{\mathcal{U}_j=u}{e_j=1} = \bindist{v}{u}{\punsato}$.
\end{proof}

The probability $p_{\mathtt{flip}|0} := \Pr(\upc \geq \mathtt{th} \mid e_j=0)$,
of flipping $\bar{e}_j$ given that $e_j=0$, corresponds to the likelihood of the
union of all events where the integer in $\upc$ is equal or greater than the
threshold $\mathtt{th}_{(1)}$ employed in the first iteration, while the probability $p_{\neg\mathtt{flip}|0} :=
\Pr(\upc < \mathtt{th} \mid e_j=0)$, of maintaining the value $\bar{e}_j$
given that $e_j=0$, corresponds to the likelihood of the union of all events
where the integer in $\upc$ is smaller than $\mathtt{th}_{(1)}$:
$$
p_{\mathtt{flip}|0} = \sum_{a=\mathtt{th}_{(1)}}^v \condprob{\mathcal{U}_j=a}{e_j=0}, \qquad
p_{\neg\mathtt{flip}|0} = 1 - p_{\mathtt{flip}|0} =
\sum_{a=0}^{\mathtt{th}-1} \condprob{\mathcal{U}_j=a}{e_j=0}.
$$

Through an analogous reasoning, we derive the probability
that, at the end of the first iteration, the decoder has flipped the $j$-th
bit of the error vector estimate, given that the corresponding error vector bit
is asserted, i.e., $p_{\mathtt{flip}|1}$, and the probability
that at the end of the first iteration the decoder have maintained the $j$-th
bit of the error vector estimate, given that the corresponding error vector bit
is asserted, i.e., $p_{\neg\mathtt{flip}|1}$.
$$
p_{\mathtt{flip}|1} = \sum_{a=\mathtt{th}_{(1)}}^{v}
\condprob{\mathcal{U}_j=a}{e_j=1}, \qquad
p_{\neg\mathtt{flip}|1} = 1- p_{\mathtt{flip}|1} =
\sum_{a=0}^{\mathtt{th}-1} \condprob{\mathcal{U}_j=a}{e_j=1}.
$$

Consider $\bar{\vectmat{e}}$ initialized with a null bitstring, and denote with
$\bar{\vectmat{e}}_{(i)}$ the value of $\bar{\vectmat{e}}$ after the bitflips of the $i$-th
iteration of the decoder have been applied.
We denote $\mathcal{D}_+ := \left| \Big(\{0,\ldots,n-1\}\setminus\support{\vectmat{e}}\Big) \cap
\support{\bar{\vectmat{e}}_{(1)}} \right|$ the random variable modeling the number of bitflips
applied on the $n-t$ positions of $\bar{e}$ corresponding to clear
bits in the error vector ($e_j=0$), at the end of the first decoder iteration
(i.e., they are erroneous flip-ups).
Such flips increase the amount of discrepancies between $\bar{\vectmat{e}}$ and $\vectmat{e}$.

The random variable $\mathcal{D}_- := \left|\support{\vectmat{e}} \cap \support{\bar{\vectmat{e}}_{(1)}} \right|$
models
the number of bitflips executed on the $t$ positions of the initial
values in $\bar{e}$ corresponding to asserted bits in the error vector
($e_j=1$), at the end of the first decoder iteration (i.e., they are correct
flip-ups).
Such flips decrease the number of discrepancies between $\bar{\vectmat{e}}$ and $\vectmat{e}$).

\begin{proposition}
The probability mass functions of the random variables $\mathcal{D}_+$ and
$\mathcal{D}_-$ are:
\begin{align*}
\Pr\left( \mathcal{D}_+ = d_+ \, |\, \mathcal{W}_t = y \right) :=
\bindist{n-t}{\discplus}{p_{\mathtt{flip}|0}},\ \text{with}\ d_+ \in \{0, \dots, n-t\} \\
\Pr\left( \mathcal{D}_- = d_- \, |\, \mathcal{W}_t = y \right) :=
\bindist{t}{\discminus}{p_{\mathtt{flip}|1}},\ \text{with}\ d_- \in \{0, \dots, t\}.
\phantom{MMM}
\end{align*}
\end{proposition}
\begin{proof}
    In a parallel decoder, such as the one shown by Algorithm~\ref{algo:bitflipping},
the decision to flip the $j$-th bit in the error vector estimate component (e.g.,
$\bar{e}_j$), is taken independently from the values of $\mathtt{upc}_{j'}$
with $j'\neq j$, because the $\mathtt{upc}$ values do not change until all the
flips are completed.
Therefore, the number of flips happening on the $n-t$ correct and $t$ incorrect
bit ($\discplus$ and $\discminus$) follows a binomial distribution, where each
bit has probability $p_{\mathtt{flip}|0}$ or $p_{\mathtt{flip}|1}$ of being
flipped, depending on its value.
\end{proof}

Note that, while not visually apparent in the previous formula, the
effect of conditioning on the event $(\mathcal{W}_t = y)$ is embedded in the
derivation of $\prob{\mathcal{S}_{i\mid y} = 0}$
and $\prob{\mathcal{S}_{i\mid y} = 1}$, which are needed to obtain
$p_{\mathtt{flip}|0}$ and $p_{\mathtt{flip}|1}$, which are in turn employed
to derive the p.m.f. of $\mathcal{D}_+$ and $\mathcal{D}_-$, respectively.
We are now able to obtain the p.m.f. $\condprob{\mathcal{E}_{(\mathtt{1})} = \discre}{\mathcal{W}_t = y}$,
which describes the error vector weight at the end of the first iteration, given the weight of the syndrome:

$$
\displaystyle
\condprob{\mathcal{E}_{(\mathtt{1})} = \discre}{\mathcal{W}_t = y} =
\sum_{ \substack{
\discplus,\discminus \\ t - \discminus + \discplus = \discre
} } \Pr\left( \mathcal{D}_+ = d_+ \right)\Pr\left( \mathcal{D}_- = d_- \right)
$$

Given the code parameters $n, k, v, w, t$, using the p.m.f.
$\condprob{\mathcal{E}_{(\mathtt{1})} = \discre}{\mathcal{W}_t = y}$ jointly
with the p.m.f. of the syndrome weight
$\Pr(\mathcal{W}_t = y)$ (see Algorithm~\ref{algo:distrSyndrome}) allows to compute
the statistical distribution of the weight of the error vector at the end of the
first iteration, and consequently also to assess the decoding error rate,
$\text{DFR}_{(1)}$, at the end of the first iteration as:
$$
\Pr(\mathcal{E}_{(\mathtt{1})} = \discre)=\sum_{y = 0}^{r}
\left( \condprob{\mathcal{E}_{(\mathtt{1})} = \discre}{\mathcal{W}_t = y} \cdot
\Pr(\mathcal{W}_t = y)\right), \quad \text{DFR}_{(1)} = 1 - \Pr(\mathcal{E}_{(\mathtt{1})} = 0).
$$

\subsection{Second Iteration of the Bit Flipping Decoder}\label{subsec:second}
We now tackle the modeling of the second iteration of the parallel
bit flipping decoder, with the goal of describing
$\Pr(\mathcal{E}_{(\mathtt{2})}$$=$$\discre)$.
To this end, we will derive $\Pr(\mathcal{E}_{(\mathtt{2})} = \discre|
\mathcal{W}_t = y)$, and then obtain $\Pr(\mathcal{E}_{(\mathtt{2})} =
\discre)$ using the knowledge of the distribution of
the weight of the initial syndrome, $\Pr(\mathcal{W}_t = y)$.
To derive the probabilities that each of the bits in the error vector estimate at the
end of the second iteration has been flipped with respect to their values at the end
of the first iteration, we start our analysis by partitioning the bits in
$\bar{\vectmat{e}}_{(1)}$ (and therefore their positions) into four classes.
Each class is labeled with a pair of values $(a,b)$, where, for each bit $\bar{e}_j$
belonging to the class, we have $a=e_j,\, b=e_j\oplus\bar{e}_{(1),j}$, i.e.,
the pair is defined by the value of the actual error vector bit (i.e., $e_j$)
and by the discrepancy between its value and the value assessed at the end of
the first iteration (i.e., $e_j\oplus\bar{e}_{(1),j}$).
From now on, we denote as $\mathbf{J}_{a,b}$, with $a, b$$\in$$\{0,1\}$ the sets of
positions $j \in \{0, \ldots, n-1\}$ of the bits in $\bar{\vectmat{e}}_{(1)}$ belonging to the class $(a,b)$,
i.e.:
$$
\forall (a,b) \in \{0,1\}^2, \ \mathbf{J}_{a,b} = \left\{ j \in \{0,\ldots,n-1\}\
\text{such that}\ e_j = a \wedge b = e_j\oplus\bar{e}_{(1),j} \right\}.
$$
We therefore consider the computation of the probability that a bit
in $\bar{\vectmat{e}}_{(1)}$ is flipped during the second iteration as split into four
computations, denoted as \pflipzz, \pflipzo, \pflipoz, and \pflipoo, respectively,
each one considering the position of the error estimate bit at hand to be in one of
the four $(a,b)$ classes.

In particular, \pflipzz\ denotes the probability of flipping the value of a
bit in $\bar{\vectmat{e}}_{(1)}$ having position in $\mathbf{J}_{0,0}$.
More precisely, it quantifies the likelihood of flipping the value of a bit
that, at the end of the first iteration, was (correctly) left as clear,
while the bit-value in the corresponding position of the actual error vector
is clear (i.e., $e_j = 0$ and $e_j \oplus \bar{e}_{(1), j} = 0$).
\newline
The probability \pflipoz\ quantifies the likelihood of flipping the value of a
bit in $\bar{e}_{(1)}$ having position in $\mathbf{J}_{1,0}$.
More precisely, it quantifies the likelihood of flipping the value of a bit
that, at the end of the first iteration, was correctly flipped up, while
the bit-value in the corresponding position of the actual error vector is asserted
(i.e., $e_j = 1$ and $e_j \oplus \bar{e}_{(1), j} = 0$).
\newline
The probability \pflipzo\ quantifies the likelihood of flipping the value of a
bit in $\bar{\vectmat{e}}_{(1)}$ having position in $\mathbf{J}_{0,1}$.
More precisely, it quantifies the likelihood of flipping the value of a bit
that, at the end of the first iteration, was erroneously flipped-up, while
the bit-value in the corresponding position of the actual error vector is clear
(i.e., $e_j = 0$ and $e_j \oplus \bar{e}_{(1), j} = 1$).
\newline
The probability \pflipoo\ quantifies the likelihood of flipping the value of a
bit in $\bar{\vectmat{e}}_{(1)}$ having position in $\mathbf{J}_{1,1}$.
More precisely, it quantifies the likelihood of flipping the value of a bit
that, at the end of the first iteration, was erroneusly maintained as clear
while the bit-value in the corresponding position of the actual error vector is
asserted (i.e., $e_j = 1$ and $e_j \oplus \bar{e}_{(1), j} = 1$).

We will derive \pflipzz, \pflipzo, \pflipoz, and \pflipoo\ as a function of the
cardinalities of the sets that refer to the positions of bits that were erroneously
flipped-up or erroneously maintained as clear, at the end of the first iteration
of the decoder,
i.e., $|\mathbf{J}_{0,1}|=\epsilon_{01}$ and $|\mathbf{J}_{1,1}|=\epsilon_{11}$.
These sets contain the positions of the error vector estimate $\bar{\vectmat{e}}_{(1)}$
where a discrepancy with the value stored in the corresponding position of $\vectmat{e}$
is still present after the first iteration.
\newline
Using the results in Section~\ref{subsec:first}, the
p.m.f. $\prob{|\jset{a}{b}| = m}$, for all $(a, b)$$\in$$\{0,1\}^2$,
$m$$\in$$\{0,\ldots,n\}$, can be evaluated as follows:

\begin{align*}
\epsilon_{00} = |\mathbf{J}_{0,0}|,\ & \prob{\epsilon_{00} = m \,|\, \mathcal{W}_t = y} = \Pr(\mathcal{D}_+ = n-t-m \,|\, \mathcal{W}_t = y),\\
\epsilon_{01} = |\mathbf{J}_{0,1}|,\ & \prob{\epsilon_{01} = m \,|\, \mathcal{W}_t = y} = \Pr(\mathcal{D}_+ = m \,|\, \mathcal{W}_t = y),\\
\epsilon_{10} = |\mathbf{J}_{1,0}|,\ & \prob{\epsilon_{10} = m \,|\, \mathcal{W}_t = y} = \Pr(\mathcal{D}_- = m \,|\, \mathcal{W}_t = y),\\
\epsilon_{11} = |\mathbf{J}_{1,1}|,\ & \prob{\epsilon_{11} = m \,|\, \mathcal{W}_t = y} = \Pr(\mathcal{D}_- = t-m \,|\, \mathcal{W}_t = y).
\end{align*}

Indeed, the probability that $\epsilon_{01}$ bits of the error estimate
have been erroneously flipped-up by the first iteration of the decoder, despite the bits in the
corresponding positions of the actual error vector being clear, is $\prob{\epsilon_{01} = m \,|\, \mathcal{W}_t = y} = \Pr(\mathcal{D}_+ = m \,|\, \mathcal{W}_t = y)$;
similarly, the probability that $\epsilon_{10}$ bits have been correctly flipped-up
by the first iteration of the decoder, being asserted the bits in the corresponding positions
in the actual error vector, is $\prob{\epsilon_{10} = m \,|\, \mathcal{W}_t = y} = \Pr(\mathcal{D}_- = m \,|\, \mathcal{W}_t = y)$.
Since the number $m$ of bits that are correctly maintained as clear in the error vector
estimate is $n-t-d_{+}$, it is easy to infer that
$\prob{\epsilon_{00} = m \,|\, \mathcal{W}_t = y} = \Pr(\mathcal{D}_+ = n-t-m \,|\, \mathcal{W}_t = y)$.
Similarly, in case $m$ denotes the number of bits erroneously maintained as clear in the error vector
estimate, we have that $m = t-\discminus$, $\prob{\epsilon_{11} = m \,|\, \mathcal{W}_t = y} = \Pr(\mathcal{D}_- = t-m \,|\, \mathcal{W}_t = y)$.

\textit{All the probabilities and probabilities mass functions considered in this section will be implicitly
conditioned over the event $\mathcal{W}_t = y$}, which we omit to shorten the notation.

To derive the probability of flipping a bit of the error vector estimate
during the execution of the second iteration of the decoder, we need to characterize
further the behavior of the decoding algorithm during the first iteration,
by computing the probabilities $\pflipz{\mathtt{OneEqSat}}$ and
$\pflipz{\mathtt{OneEqUnsat}}$ of observing the $j$-th bit of the error vector estimate to be
erroneously flipped-up after the first decoding iteration, knowing that the corresponding bit
in the actual error vector is clear,
i.e., $e_j = 0$, and one out of the $v$ parity-check equations in which it appears is either
satisfied or unsatisfied, respectively.
Recalling that the events $\eventij{0}$ or $\eventij{1}$
describe the $j$-th error vector bit being either clear or asserted, respectively, and involved
in the $i$-th parity-check equation,
we have that:
$$
\pflipz{\mathtt{OneEqSat}}:=\condprob{\mathcal{U}_j\geq\mathtt{th}_{(1)}}{\eventij{0}\, \cap\, s_{i}=0}
=\sum_{a=\mathtt{th}_{(1)}}^{v-1} \bindist{v-1}{a}{\punsatz}, \ \text{and}
$$
$$
\pflipz{\mathtt{OneEqUnsat}}:=\condprob{\mathcal{U}_j\geq\mathtt{th}_{(1)}}{\eventij{0}\, \cap\, s_{i}=1}
=\sum_{a=\mathtt{th}_{(1)}-1}^{v-1} \bindist{v-1}{a}{\punsatz}.
$$
The probability \pflipz{\mathtt{OneEqSat}}
captures the likelihood of the union of the events where the count of unsatisfied
parity-check equations having $e_j$ as a term is $a\geq \mathtt{th}_{(1)}$, knowing a priori
that one out of the $v$ equations is satisfied (the $i$-th one),
hence reducing the number of trials to be considered from $v$ down to $v-1$.
The probability \pflipz{\mathtt{OneEqUnsat}}
captures the likelihood of the union of the events where the count of parity-check
equations that are unsatisfied is $a$, knowing a priori
that one out of the $v$ parity-check equations is unsatisfied,
hence reducing the minimum number of remaining unsatisfied equations before
a flip takes place (successes in the binomial distribution) to $\mathtt{th}_{(1)}-1$
($a\geq \mathtt{th}_{(1)}-1$), while still also reducing the number of trials to
be considered down to $v-1$.
Analogously, the probabilities $\pnoflipo{\mathtt{OneEqSat}}$ and
$\pnoflipo{\mathtt{OneEqUnsat}}$ quantify the likelihood of erroneously maintaining as clear
(not flipping-up) the $j$-th error
vector estimate, at the end of the first iteration of the decoder,
knowing that the corresponding bit in the actual error vector is asserted, i.e., $e_j = 1$, and
one out of the $v$ parity-check equations in which it appears is either satisfied or
unsatisfied, respectively.
$$
\pnoflipo{\mathtt{OneEqSat}}:=\condprob{\mathcal{U}_j<\mathtt{th}_{(1)}}{\eventij{1}\, \cap\, s_i=0} = \sum_{a=0}^{
\mathtt{th}_{(1)}-1} \bindist{v-1}{a}{\punsato}, \ \text{and}
$$
$$
\pnoflipo{\mathtt{OneEqUnsat}}:=\condprob{\mathcal{U}_j<\mathtt{th}_{(1)}}{\eventij{0}\, \cap\, s_i=1} =
\sum_{a=0}^{\mathtt{th}_{(1)}-2} \bindist{v-1}{a}{\punsato}.
$$
The probability \pnoflipo{\mathtt{OneEqSat}}
captures the likelihood of the union of the events where the count of unsatisfied
parity-check equations having $e_j$ as a term is below the flipping threshold,
$a < \mathtt{th}_{(1)}$, knowing a priori
that one out of the $v$ equations is unsatisfied.
The probability \pnoflipo{\mathtt{OneEqUnsat}}
captures the likelihood of the union of the events where the count of unsatisfied parity-check
equations having $e_j$ as a term is below the flipping threshold $a < \mathtt{th}_{(1)}$,
knowing a priori that one out of the $v$ parity-check equations is unsatisfied,
hence reducing the number of trials to be considered in $\{0,\ldots, \mathtt{th}_{(1)}-2\}$.


\text{ }

We now derive the expression for $p_{\mathtt{flip}|ab}$, $a,b\in\{0,1\}$, that is,
the probability of flipping during the second iteration of the decoder,
a bit in $\bar{e}_{(1)}$ that has a position in $\mathbf{J}_{a,b}$.
We start by describing how to derive $\pflipzz$, that is,
of flipping a bit value that, after the first iteration of the decoder,
has been correctly maintained as clear because the bit in the
corresponding position of the actual error vector is clear.
To this end, it is instrumental to determine how many of the parity-check
equations in which the said bit is involved change their state
(from satisfied to unsatisfied and vice versa) after the first iteration
is finished.

Consider a given, say the $i$-th, parity-check equation, involving $f$
asserted error vector bits and one clear error vector bit, $e_a = 0$ for which the error
estimate was not flipped during the first round, i.e., $a \in\jset{0}{0}$.

\begin{proposition}
Let $\chi_{\updownarrow \mathtt{odd}}(f, \epsilon_{01})$, $f \in \{0,\ldots,\min(t,w)\}$ be the probability that
an odd number of erroneous flips-up, i.e., bits in $\bar{e}_{(1)}$
having positions in $\jset{0}{1}$, in the estimated error vector happen within the $i$-th parity-check
in the first decoding iteration, given that $\mathcal{F}_t=f$ and a bit in $\jset{0}{0}$ appears in the check:
$$
\chi_{\updownarrow \mathtt{odd}}(f, \epsilon_{01})  :=
\Pr\left(\left|\support{h_{i,:}} \cap \jset{0}{1}\right|\,
\text{is odd} \  \mid \ |\jset{0}{1}|=\epsilon_{01} \cap  \left( \exists\, a\in \jset{0}{0},\, h_{i,a}=1,\,
\left|\support{h_{i,:}} \cap \support{e}\right|=f \right) \right)
$$
The expression of $\chi_{\updownarrow \mathtt{odd}}(f, \epsilon_{01})$ is:
$$
\chi_{\updownarrow \mathtt{odd}}(f,\epsilon_{01})=\sum_{\ell=1, \text{odd}}^{\scriptscriptstyle
\min(\epsilon_{01}, w-f-1)} \frac{ \eta(f,\ell)\cdot \zeta(f,\ell,\epsilon_{01}) }{
\xi(f,\epsilon_{01}) },
$$
where
$$
\eta(f,\ell) :=
\begin{cases}
    \bindist{w-f-1}{\ell}{\pflipz{\mathtt{OneEqSat}}} & \quad \text{if $f$ is even}\\
    \bindist{w-f-1}{\ell}{\pflipz{\mathtt{OneEqUnsat}}} & \quad \text{if $f$ is odd}
\end{cases}$$
$$
\zeta(f,\ell,\epsilon_{01}) :=
\bindist{n-w-(t-f)}{\epsilon_{01}-\ell}{p_{\mathtt{flip}|0}}
$$
$$
\xi(f,\epsilon_{01}) := \sum_{\ell'=\max(0,\epsilon_{01}-(n-w-(t-f)))}^{\min(\epsilon_{01},
w-f-1)} \eta(f,\ell')\cdot \zeta(f,\ell',\epsilon_{01}).
$$
\end{proposition}

\begin{proof}
    The expression
of $\chi_{\updownarrow \mathtt{odd}}(f, \epsilon_{01})$ is obtained
starting from two p.m.f. s:
$\eta(f,\ell)$ and $\zeta(f,\ell,\epsilon_{01})$.
Assume, for the sake of argument, that $f$ is even.
The distribution $\eta(f,\ell)$ describes the probability
that $0 \leq \ell \leq w-f-1$ clear terms involved in the parity-check
equation (excluding $e_a$, which is assumed to be in $\jset{0}{0}$) have positions in $\jset{0}{1}$,
i.e., they are discrepancies
resulting from the erroneous flip-ups into the error vector estimate made by
the first iteration. The probability that each of such flip-ups takes place
in the first iteration is $\pflipz{\mathtt{OneEqSat}}$, as a consequence
of our assumption on $f = | \support{e} \support{\cap h_{i,:}}|$ being even,
and hence $\eta(f,\ell) = \bindist{w-f-1}{\ell}{\pflipz{\mathtt{OneEqSat}}}$.
Computing $\eta(f,\ell)$ under the (complementary) assumption of an
odd-valued $f$, only requires to replace
$\pflipz{\mathtt{OneEqSat}}$ with $\pflipz{\mathtt{OneEqUnsat}}$.

The distribution $\zeta(f,\ell,\epsilon_{01})$ models the probability that the remaining $\epsilon_{01}-\ell$
flip-ups causing a discrepancy, which took place in the first iteration,
took place on $\epsilon_{01}-\ell$ positions out of the $n-w-(t-f)$ ones,
which are not involved in the parity-check equation (hence in one of $n-w$ possible
positions),
and not containing an asserted bit (hence, we subtract the positions of asserted
bits not involved in the equation from the possible positions,
obtaining $n-w -(t-f)$).
We have that $\zeta(f,\ell,\epsilon_{01}) = \bindist{n-w-(t-f)}{\epsilon_{01}-\ell}{p_{\mathtt{flip}|0}}$
models the probability that the remaining $\epsilon_{01}-\ell$
Combining  $\eta(f,\ell)$ and $\zeta(f,\ell,\epsilon_{01})$, we obtain:
$$
\xi(f,\epsilon_{01}) := \sum_{\ell'=\max(0,\epsilon_{01}-(n-w-(t-f)))}^{\min(\epsilon_{01},
w-f-1)} \eta(f,\ell')\cdot \zeta(f,\ell',\epsilon_{01}),
$$
which represents the probability distribution of the union of the events where
the ones modeled by $\eta(f,\ell')$ and the ones modeled by
$\zeta(f,\ell',\epsilon_{01})$ take place simultaneously, for a given $\ell'$,
i.e. that, out of the $\epsilon_{01}$ indices in $\jset{0}{1}$, $\ell'$ of them
are involved in the parity-check equation at hand, while the other
$\epsilon_{01}-\ell'$ are both uninvolved and do not have an asserted error bit.
The value $\ell'$ is upper bounded by the smallest value between the amount of
flip-up affected positions $\epsilon_{01}$  and the places where they should happen
$w-f-1$, while being lower bounded by
$\epsilon_{01}-(n-w-(t-f))$, that is, the amount of flip-up affected
positions which cannot be fit among the $n-w$ not involved
in the parity-check equation, and further excluding the $t-f$ ones
where the error vector bits not involved in the parity-check equation are asserted.
Thus, informally, $\xi(f,\epsilon_{01})$ is the probability that any admissible
number of incorrect flip-ups $\ell'$, executed after the first iteration of the decoder,
are on positions involved in the $i$-th parity-check equation,
that is assumed to involve $f$ positions where the corresponding error
bit is asserted, plus a position where the corresponding error bit is clear and the
corresponding error estimate bit was not flipped by the
first iteration (i.e., a position in $\jset{0}{}$).
We therefore obtain $\chi_{\updownarrow \mathtt{odd}}(f,\epsilon_{01})$ as:
$$
\chi_{\updownarrow \mathtt{odd}}(f,\epsilon_{01})=\sum_{\ell=1, \ell\,
\text{odd}}^{\scriptscriptstyle
\min(\epsilon_{01}, w-f-1)} \frac{ \eta(f,\ell)\cdot \zeta(f,\ell,\epsilon_{01}) }{
\xi(f,\epsilon_{01}) },
$$
that is, we consider the union of all disjoint events where an odd number $\ell$
of erroneous flip-ups from the first iteration affect the positions involved in the
parity-check equation at hand, fixing the total number of erroneous flip-ups
along the $n-t$ correct bits to $\epsilon_{01}$.
\end{proof}


\begin{proposition}
Let $\chi_{\leftrightarrow \mathtt{odd}}(f, \epsilon_{11})$, $f \in \{0,\ldots,\min(t,w)\}$
be the probability that
an odd number of erroneous bits maintained as clear in the estimated error, i.e., bits in
$\bar{\vectmat{e}}_{(1)}$ with positions in $\jset{1}{1}$, appear within the
$i$-th parity-check given that $\mathcal{F}_t=f$ and a bit in $\jset{0}{0}$ appears in the check:
$$
 \chi_{\leftrightarrow \mathtt{odd}}(f, \epsilon_{11}) = \Pr\left(\left| \support{h_{i,:}} \cap \jset{1}{1}\right|
\text{is odd} \ \mid \ |\jset{1}{1}|=\epsilon_{11} \cap \left( \exists a\in \jset{0}{0},h_{i,a}=1,
\left|\support{\vectmat{h}_{i,:}} \cap \support{\vectmat{e}}\right|=f \right) \right)
$$
where we still consider the $i$-th parity-check equation composed by $w$ terms, including a bit
$a \in \jset{0}{0}$, and $f$ positions where the error vector $e$ contains asserted bits
(i.e., there are $f$ positions of bits involved in the equation which belong to $\jset{1}{0} \cup \jset{1}{1}$.
The expression of $\chi_{\leftrightarrow \mathtt{odd}}(f, \epsilon_{11})$ is:
$$
\chi_{\leftrightarrow \mathtt{odd}}(f,\epsilon_{11})= \sum_{\ell=1, \ell\, \text{odd}}^{\scriptscriptstyle
\min(\epsilon_{11}, f)} \frac{ \nu(f,\ell)\cdot \lambda(f,\ell,\epsilon_{11})
}{\theta(f,\epsilon_{11})},
$$
where
$$
\nu(f,\ell):=
\begin{cases}
    \bindist{f}{\ell}{\pnoflipo{\mathtt{OneEqSat}}} & \quad \text{if $f$ is even}\\
    \bindist{f}{\ell}{\pnoflipo{\mathtt{OneEqUnsat}}} & \quad \text{if $f$ is odd}
\end{cases}
$$
$$\lambda(f,\ell,\epsilon_{11}) :=
\bindist{t-f}{\epsilon_{11}-\ell}{p_{\neg\mathtt{flip}|1}}$$
$$
\theta(f,\epsilon_{11}):=\sum_{\ell'=\max(0,\epsilon_{11}-(t-f))}^{\min(\epsilon_{11},
{tc})} \nu(f,\ell')\cdot \lambda(f,\ell',\epsilon_{11})
$$
\end{proposition}

\begin{proof}
Along the same line of thoughts of the $\chi_{\updownarrow \mathtt{odd}}$ case,
fix the parity of $f$, this time, to be even.
Defining $\nu(f,\ell):=\bindist{f}{\ell}{\pnoflipo{\mathtt{OneEqSat}}}$,
we have the probability distribution that $\ell$ positions in $\jset{1}{1}$
are among the $f$ ones involved in the parity-check equation.
This is equivalent to saying that, among the $f$ positions involved
in the parity-check equation where the error $\vectmat{e}$ contains an asserted bit, $\ell$ are not
flipped by the first iteration in $\bar{\vectmat{e}}$, thus resulting in $\ell$ discrepancies
being maintained after the first iteration itself. Each one of such maintaining
actions takes place with probability $\pnoflipo{\mathtt{OneEqSat}}$, i.e., the
probability that a position is not flipped, given that the corresponding error
bit is asserted, and the parity-check equation in which it is involved is satisfied
(indeed, $f$ is even).
In case $f$ is even has odd parity, it is sufficient to substitute
$\pnoflipo{\mathtt{OneEqSat}}$ with $\pnoflipo{\mathtt{OneEqUnsat}}$ in
the aforementioned expression.

Defining $\lambda(f,\ell,\epsilon_{11}) :=
\bindist{t-f}{\epsilon_{11}-\ell}{p_{\neg\mathtt{flip}|1}}$
we model the probability of the remaining $\epsilon_{11}-\ell$ positions where the
first iteration did not flip the error estimate in correspondence with an asserted
error bit to be among the $t-f$ positions where the error vector is
asserted, and which are not among the ones involved in the parity-check equation.
The probability of each such position not being flipped is indeed
$p_{\neg\mathtt{flip}|1}$.
With an approach similar to what was done for $\xi(f)$, we combine
$\nu(f,\ell)$ and $\lambda(f,\ell,\epsilon_{11})$
to obtain
$$
\theta(f,\epsilon_{11}):=\sum_{\ell'=\max(0,\epsilon_{11}-(t-f))}^{\min(\epsilon_{11},
{tc})} \nu(f,\ell')\cdot \lambda(f,\ell',\epsilon_{11})
$$
We thus compose $\chi_{\leftrightarrow \mathtt{odd}}(f,\epsilon_{11})$ as:
$$
\chi_{\leftrightarrow \mathtt{odd}}(f,\epsilon_{11})= \sum_{\ell=1, \ell\, \text{odd}}^{\scriptscriptstyle
\min(\epsilon_{11}, f)} \frac{ \nu(f,\ell)\cdot \lambda(f,\ell,\epsilon_{11})
}{\theta(f,\epsilon_{11})}.
$$
that is, we consider the union of all disjoint events where an odd number $\ell$
of erroneously maintained incorrect bits from the first iteration are involved in the
parity-check equation at hand, fixing the total number of erroneously maintained bits
along the $t$ incorrect bits to $\epsilon_{11}$.
\end{proof}

The terms $\chichg$ and $\chimaint$ model the probability that an odd number of terms
in $\jset{0}{1}$ and $\jset{1}{1}$, respectively, are included in the same parity-check
including a bit in $\jset{0}{0}$.
Employing $\chichg$ and $\chimaint$, we can now model the probability
that a satisfied parity-check equation involving a position in $\jset{0}{0}$
becomes unsatisfied after the flips made by the first iteration,
$p_{00|\mathtt{BecomeUnsat}}$. This paves the way to counting how many satisfied
parity-check equations, among the ones involving an error bit with
position in $\jset{0}{0}$ are left after the first iteration, a key step in modeling
the upc of the said error bit.

The crucial observation in this modeling is that the
event of the parity-check equation at hand being unsatisfied
takes place in two cases after the flips performed in the first decoding iterations.
For any given number $f$ of asserted error bits involved in the parity-check equation, we have the following:
\begin{enumerate}
\item The first case is when an odd number of discrepancies is added, and and
even number of discrepancies is maintained, among the
bits involved in the equation. Stating so is equivalent to saying that
there is a odd number of positions of $\jset{0}{1}$, and an even number
of positions of $\jset{1}{1}$, involved in the parity-check equation.
In this case, we have that the probability of the equation becoming unsatisfied
is $\chichg(1-\chimaint)$.
\item The second case is when an even number of discrepancies is added, and
an odd number of discrepancies is maintained. Stating so is equivalent
to saying that there is an even number of positions of $\jset{0}{1}$, and
an odd number of positions of $\jset{1}{1}$, involved in the parity-check
equation.
The probability of the change in the equation satisfaction state is modeled
by $(1-\chichg)\chimaint$.
\end{enumerate}

Since the two described events are disjoint, the probability that, for a given value $f$,
a specific flip pattern from the first iteration makes a parity-check
equation containing a position in $\jset{0}{0}$ switch
from containing an even number of asserted terms to containing an odd one
is
$$\gamma_{\mathtt{UnsatPostFlips}}(f,\epsilon_{01},\epsilon_{11}):=
\chichg(1-\chimaint) +(1-\chichg)\chimaint.
$$

We are now interested in deriving the distribution of $\mathcal{F}_t$ in parity-checks
that are known to include a bit in $\jset{0}{0}$. To this end, let $\sevent{\jset{0}{0}}$
denote the event where a bit
with position in $\jset{0}{0}$ is involved in a parity-check equation.

\begin{proposition}
    The p.m.f. of $\mathcal{F}_t$ in parity-checks where a bit in $\jset{0}{0}$ is included,
    i.e., where the event $\sevent{\jset{0}{0}}$ takes place, is:
    $$
    \condprob{\flips{f}}{\sevent{\jset{0}{0}}} =
    \frac{
        \Pr(\mathcal{F}_t = f) \cdot \frac{w-f}{w} \cdot \kappa(f)
    }{
        \sum_{f'=0}^{\min(t,w)}
        \Pr(\mathcal{F}_t = f') \cdot \frac{w-f'}{w} \cdot \kappa(f')
    }
    $$
    where $\Pr(\mathcal{F}_t = f)$ is implicitly conditioned over the event $\mathcal{W}_t=y$, and
    $$
    \kappa(f) :=
\begin{cases}
    1-\pflipz{\mathtt{OneEqSat}} & \text{if $f$ is even}\\
    1-\pflipz{\mathtt{OneEqUnsat}} & \text{if $f$ is odd.}\\
\end{cases}
    $$
\end{proposition}

\begin{proof}
    Starting from the definition of $\left(\flips{f}\,|\,\sevent{\jset{0}{0}}\right)$, we have:
    $$
\condprob{\flips{f}}{\sevent{\jset{0}{0}}} =
\frac{\prob{\flips{f} \cap \sevent{\jset{0}{0}}}}{\prob{\sevent{\jset{0}{0}}}} =
\frac{\prob{\flips{f}}\condprob{\sevent{\jset{0}{0}}}{\flips{f}}}{
\sum_{f'=0}^{\min(t,w)}
\prob{\flips{f'}}\condprob{\sevent{\jset{0}{0}}}{\flips{f'}
}
},
$$
where
$\condprob{\sevent{\jset{0}{0}}}{\flips{f}}$ denotes the probability that
an error vector bit has a position in $\jset{0}{0}$ and is
also among the $w$ ones involved in the parity-check equation,
knowing that in said parity-check $\mathcal{F}_t=f$.
The probability $\Pr(\flips{f})$, that is implicitly conditioned over the event $\mathcal{W}_t=y$,
has been derived in Appendix~\ref{app:appendix1}, while
$\condprob{\sevent{\jset{0}{0}}}{\flips{f}}$ can be computed as
$
\condprob{\sevent{\jset{0}{0}}}{\flips{f}} =
\condprob{e_i=0}{\flips{f}} \cdot \condprob{\sevent{\jset{0}{0}}}{\flips{f} \cap e_i=0}.
$
Indeed, the fact that an error vector bit is in $\jset{0}{0}$
means that it is a correct bit for which the first iteration of the bit-flipping
algorithm did not flip up the corresponding estimate.
The factor $\condprob{e_i=0}{\flips{f}}$ requires to
compute the probability of choosing a clear bit involved
the parity-check equation, knowing that the number incorrect
bits (i.e., asserted bits in the actual error vector $\vectmat{e}$) included in said parity-check equation is $f$.
As a consequence, $\condprob{e_i=0}{\flips{f}} = \frac{w-f}{w}$, since the number of correct
bits involved in the equation are $w$ minus the number $f$ of
incorrect bits.
The factor $\condprob{\sevent{\jset{0}{0}}}{\flips{f} \cap e_i=0}$ requires to
compute the probability that a correct bit involved in the parity-check equation
is not flipped during the first iteration,
knowing that the number of incorrect (i.e., asserted in $\vectmat{e}$) terms of the equation is equal to $f$.
Reusing the previously introduced definitions, this probability is
computed as:
$$\condprob{\sevent{\jset{0}{0}}}{\flips{f} \cap e_i=0} =
\begin{cases}
    1-\pflipz{\mathtt{OneEqSat}} & \text{if $f$ is even}\\
    1-\pflipz{\mathtt{OneEqUnsat}} & \text{if $f$ is odd.}\\
\end{cases}
$$
Letting $\kappa(f) = \condprob{\sevent{\jset{0}{0}}}{\flips{f} \cap e_i=0}$ and
substituting the derived results, we get the stated formula.
\end{proof}

Figure~\ref{fig:paritycheck} shows the state a parity-check with $w=12$ before and
after the first decoding iteration. The white bit on the left is assumed to be in
$\jset{0}{0}$. The $w-1-f$ light blue and blue bits are correct in the error vector
(i.e., $e_j=0$), while the $f$ light red and red positions are incorrect in the error
vector (i.e., $e_j=1$). The light colored positions are guessed correctly after one iteration
(i.e., these positions are in $\jset{0}{0}$ and $\jset{1}{0}$ depending on their color), while
the dark colored positions are guessed incorrectly after one iteration (i.e., these positions
are in $\jset{0}{1}$ and $\jset{1}{1}$ depending on their color).
The p.m.f. of the number $f$ of light red and red bits included in the parity-check
is $\condprob{\flips{f}}{\sevent{\jset{0}{0}}}$.
The probability $\chichg$ corresponds to the probability that the number of dark blue terms
(i.e., terms in $\jset{0}{1}$ included in the parity-check) is odd.
The probability $\chimaint$ corresponds to the probability that the number of dark red terms
(i.e., terms in $\jset{1}{1}$ included in the parity-check) is odd.
The probability $\gamma_{\mathtt{UnsatPostFlips}}(f,\epsilon_{01},\epsilon_{11})$ corresponds
to the probability that the overall number of dark colored terms (i.e., the number of terms
in $\jset{0}{1} \cup \jset{1}{1}$ included in the parity-check) is odd, causing the
parity-check to be unsatisfied after the first iteration.

Given the previous derivations, we are now able to compute the probability
$p_{00|\mathtt{BecomeUnsat}}$ that a satisfied parity-check equation involving an
error bit with a position in $\jset{0}{0}$ becomes unsatisfied after the flips
made to the syndrome by the first iteration. We obtain it as follows:
$$
p_{00|\mathtt{BecomeUnsat}}(\epsilon_{01},\epsilon_{11}) = \frac{\sum_{f = 0,
f\text{ even}}^{\min(t,w-1)}
\condprob{\flips{f}}{\sevent{\jset{0}{0}}}
\cdot
\gamma_{\mathtt{UnsatPostFlips}}(f,\epsilon_{01},\epsilon_{11})
}
{
\sum_{f=0, f \text{ even}}^{\min(t,w-1)}
\condprob{\flips{f}}{\sevent{\jset{0}{0}}}
},
$$
that is, we add together the probabilities that a parity-check where $\mathcal{F}_t = f$, involving a
bit with position in $\jset{0}{0}$, becomes unsatisfied after the flips performed
in the first iteration (a probability depending on the cardinality of $\jset{0}{1}$ and $\jset{1}{1}$).
The existence of the denominator is justified by the fact that,
in a satisfied parity-check equation, $f$ is assumed to be even.
Informally, $p_{00|\mathtt{BecomeUnsat}}(\epsilon_{01},\epsilon_{11})$ is computed by
averaging $\gamma_{\mathtt{UnsatPostFlips}}(f,\epsilon_{01},\epsilon_{11})$ (the probability
that the parity-check equation is unsatisfied after the first iteration, given $\flips{f}$)
over all possible values of $f$.

\begin{figure}[t!]
\begin{center}
\begin{tikzpicture}[scale=0.6]


\foreach \i/\col in {
  0/white,
  1/lightblue,2/lightblue,3/lightblue,
  4/lightblue,5/lightblue,6/lightblue,
  7/darkblue,8/darkblue,
  9/lightred,10/lightred,
  11/darkred,
}{
  \fill[\col] (\i,0) rectangle (\i+1,1);
  \draw[line width=0.4pt] (\i,0) rectangle (\i+1,1);
}
\node at (0.5,0.5) {$\mathbf{1}$};
\foreach \i in {1,...,11}{
  \node at (\i+0.5,0.5) {$1$};
}

\draw[line width=0.4pt] (0,3.0) rectangle (1,4.0);
\node at (0.5,3.5) {$\mathbf{0}$};
\foreach \i in {1,...,8}{
  \draw[line width=0.4pt] (\i,3.0) rectangle (\i+1,4.0);
  \node at (\i+0.5,3.5) {$0$};
}
\foreach \i in {9,...,11}{
  \draw[line width=0.4pt] (\i,3.0) rectangle (\i+1,4.0);
  \node at (\i+0.5,3.5) {$1$};
}
\node at (12.7,3.5) {$\cdots$};
\node at (-0.7,3.5) {$\mathtt{e}$};

\draw[line width=0.4pt] (0,1.5) rectangle (1,2.5);
\node at (0.5,2.0) {$\mathbf{0}$};
\foreach \i in {1,2,3,4,5,6,11}{
  \draw[line width=0.4pt] (\i,1.5) rectangle (\i+1,2.5);
  \node at (\i+0.5,2.0) {$0$};
}
\foreach \i in {7,8,9,10}{
  \draw[line width=0.4pt] (\i,1.5) rectangle (\i+1,2.5);
  \node at (\i+0.5,2.0) {$1$};
}
\node at (12.7,2.0) {$\cdots$};
\node at (-0.7,2.0) {$\bar{\mathtt{e}}_{(1)}$};

\node at (-0.7,0.5) {$\mathtt{h}_{i,:}$};

\node[below=2pt] at (0.5,0) 
  {$\substack{\displaystyle
  \text{\rotatebox{-90}{$\in$}}
  \\ \text{ } \\
  \displaystyle \jset{0}{0}}$
  };

\draw[decorate,decoration={brace,mirror,amplitude=8pt}] 
  (1,0) -- (9,0) node[midway,below=10pt] {$w-1-f$};

\draw[decorate,decoration={brace,mirror,amplitude=8pt}] 
  (9,0) -- (12,0) node[midway,below=10pt] {$f$};

\end{tikzpicture}
\end{center}
\caption{Graphical representation of the state of a parity-check $i$, $0$$\leq$$i$$\leq r$$-$$1$,
with $w=12$ before and after the first decoding iteration. The white bit is assumed to be in
$\jset{0}{0}$. The $w$$-$$1$$-f$ light blue and blue bits are correct in the error vector
(i.e., $e_j=0$), while the $f$ light red and red positions are incorrect in the error vector
(i.e., $e_j=1$). The light coloured positions are guessed correctly after one iteration (i.e.,
these positions are in $\jset{0}{0}$ and $\jset{1}{0}$ depending on their colour), while the
dark colored positions are guessed incorrectly after one iteration (i.e., these positions are
in $\jset{0}{1}$ and $\jset{1}{1}$ depending on their color). The parity-check is unsatisfied
before the first iteration since the number $f$ of red positions is odd, and remains unsatisfied
after the first iteration since the number of dark coloured positions is odd.}
\label{fig:paritycheck}
\end{figure}

The probability $p_{00|\mathtt{StayUnsat}}(\epsilon_{01},\epsilon_{11})$,
that is the one that an unsatisfied equation, involving an error bit with position in $\jset{0}{0}$,
remains unsatisfied after the first iteration can be derived through an
analogous line of reasoning, yielding:
$$
p_{00|\mathtt{StayUnsat}}(\epsilon_{01},\epsilon_{11}) = \frac{\sum_{f = 1,
f\text{ odd}}^{\min(t,w-1)}
\condprob{\flips{f}}{\sevent{\jset{0}{0}}} \cdot
\gamma_{\mathtt{UnsatPostFlips}}(f,\epsilon_{01},\epsilon_{11})}{
\sum_{f=1, f \text{ odd}}^{\min(t,w-1)}
\condprob{\flips{f}}{\sevent{\jset{0}{0}}}
}.
$$

Having characterized the probabilistic behaviour of the changes in the
satisfaction value of a single parity-check equation involving an error bit
with position in $\jset{0}{0}$, we now move to characterizing the distribution
of the $\mathtt{upc}$ value of the said bit during the second iteration.
To this end, we turn to considering the set of $v$ parity-check equations
into which such an error bit is involved.
For any given number $a$ of parity-check equations,
among the aforementioned $v$, that are unsatisfied before the first iteration, we denote with
$\mu(\mathtt{nsat},\mathtt{nunsat},a,\epsilon_{01},\epsilon_{11})$
the probability that, as a result of the flips from the first iteration,
out of the $a$ unsatisfied ones $\mathtt{nunsat}$ stay unsatisfied,
while, out of the $v$$-$$a$ satisfied ones $\mathtt{nsat}$ become unsatisfied.
We have:
$$\mu(\mathtt{nsat},\mathtt{nunsat},a) =
\bindist{v-a}{\mathtt{nsat}}{p_{00|\mathtt{BecomeUnsat}}(\epsilon_{01},\epsilon_{11}) } \cdot
\bindist{a}{\mathtt{nunsat}}{p_{00|\mathtt{StayUnsat}}(\epsilon_{01},\epsilon_{11})}.
$$
Note that, while the expression appears to neglect the interrelations between
the $v$ parity-check equations, these are taken into account in the
$p_{00|\mathtt{BecomeUnsat}}(\epsilon_{01},\epsilon_{11})$ and
$p_{00|\mathtt{StayUnsat}}(\epsilon_{01},\epsilon_{11})$
terms.

We now have all the tools to demonstrate the following proposition.

\begin{proposition}
    Let $\pflipzz(\epsilon_{01},\epsilon_{11})$ be the probability of
    incorrectly flipping a bit in $\jset{0}{0}$ during the second decoding
    iteration, given $|\jset{0}{1}| = \epsilon_{01}$ and $|\jset{1}{1}| = \epsilon_{11}$. Then:
    $$
\begin{array}{l}
\displaystyle
\pflipzz(\epsilon_{01},\epsilon_{11})=\sum_{a=0}^{\mathtt{th}_{(1)}-1} \left(
\frac{\bindist{v}{a}{\punsatz}}{p_{\neg\mathtt{flip}|0}} \cdot
\displaystyle
\left( \sum_{\mathtt{nsat} = 0}^{v-a} \sum_{
{\substack{
\mathtt{nunsat}=\\
\max(0,\mathtt{th}_{(2)}-\mathtt{nsat})
}}
}^{a}  \mu(\mathtt{nsat},\mathtt{nunsat},a,\epsilon_{01},\epsilon_{11})\right)\right)
\end{array}
    $$
\end{proposition}

\begin{proof}
In order to
describe the probability $\pflipzz(\epsilon_{01},\epsilon_{11})$ that a bit in $\jset{0}{0}$ is actually flipped,
we employ two quantities: the probability that the said bit was not flipped
during the first iteration, due to its $\mathtt{upc}$ taking the value
$a<\mathtt{th}_{(1)}$, and the probability that, during the second iteration
the $\mathtt{upc}$ value resulting from the change in satisfaction values of
the parity equations where the bit is involved moves is above the threshold
$\mathtt{th}_{(2)}$.
The first quantity is derived by modeling
the \texttt{upc} value of a correct bit (as in Section~\ref{subsec:first}), with the additional constraint
given by the fact that such value is assumed not to
exceed the flipping threshold $\mathtt{th}_{(1)}$:
$$\Pr (\mathcal{U}_{j} = a \ |\ j\in \jset{0}{0}) =
\begin{cases}
    \frac{\bindist{v}{a}{\punsatz}}{p_{\neg\mathtt{flip}|0}} & \quad \text{if } a < \mathtt{th}_{(1)}\\
    0 & \quad \text{otherwise}\\
\end{cases}
$$
The second quantity can be obtained adding together the probability of
the (disjoint) events which result in the $\mathtt{upc}$ value for the error bit during
the second iteration to be greater or equal to the second iteration threshold.
We start by noticing that the value of the $\mathtt{upc}$ at the second iteration is
$\mathtt{nunsat}+\mathtt{nsat}$, that is the sum of the number of parity
check equations which stayed unsatisfied and the number of satisfied equations
which became unsatisfied after the first iteration.
This, in turn, implies that we should add together the probabilities
$\mu(\mathtt{nsat},\mathtt{nunsat},a,\epsilon_{01},\epsilon_{11})$, for all
the admissible numbers $\mathtt{nunsat}$ (between $0$ and $a$),
and numbers $\mathtt{nsat}$ ($0$ and $v-a$),
such that $\mathtt{nunsat}+\mathtt{nsat}\geq \mathtt{th}_{2}$.
The desired quantity is thus
$$
\sum_{\mathtt{nsat} = 0}^{v-a} \sum_{
{\tiny \begin{matrix}
\mathtt{nunsat}=\\
\max(0,\mathtt{th}_{(2)}-\mathtt{nsat})
\end{matrix}}
}^{a}  \mu(\mathtt{nsat},\mathtt{nunsat},a,\epsilon_{01},\epsilon_{11})
$$

From the two aforementioned quantities, we obtain $\pflipzz$ as a function of
$\epsilon_{01}$ and $\epsilon_{11}$, $\pflipzz(\epsilon_{01},\epsilon_{11})$ by
simply adding their products over all the values $a$ taken by the $\mathtt{upc}$ of the
bit at hand during the first iteration, i.e., $0\leq a \leq \mathtt{th}_{(1)}-1$:
$$
\begin{array}{l}
\displaystyle
\pflipzz(\epsilon_{01},\epsilon_{11})=\sum_{a=0}^{\mathtt{th}_{(1)}-1} \left(
\Pr (\mathcal{U}_{j} = a \ |\  j\in \jset{0}{0}) \cdot
\displaystyle
\left(  \sum_{\mathtt{nsat} = 0}^{v-a} \sum_{
{\tiny \begin{matrix}
\mathtt{nunsat}=\\
\max(0,\mathtt{th}_{(2)}-\mathtt{nsat})
\end{matrix}}
}^{a}  \mu(\mathtt{nsat},\mathtt{nunsat},a,\epsilon_{01},\epsilon_{11})\right)\right)
\end{array}
$$
Note that such a sum of probabilities is justified as all the events represented
by the $\mathtt{upc}$ taking a given value $a$ are disjoint, for different values of $a$.
\end{proof}

Having now completed the derivation of $\pflipzz(\epsilon_{01},\epsilon_{11})$,
we note that probabilities $\pflipzo(\epsilon_{01},\epsilon_{11}), \pflipoz(\epsilon_{01},
\epsilon_{11})$ and, $\pflipoo(\epsilon_{01},\epsilon_{11})$ are obtained through
the very same line of reasoning, taking care of considering the appropriate events.
We provide the explicit derivation of these distributions in Appendix~\ref{app:2itflippingprob}
together with the derivations of
$\condprob{\flips{f}}{\sevent{\jset{0}{1}}}$, $\condprob{\flips{f}}{\sevent{\jset{1}{0}}}$, and
$\condprob{\flips{f}}{\sevent{\jset{1}{1}}}$, which are instrumental to their
derivation.

Having obtained the probability of flipping an error estimate
bit at the second iteration, depending on which set among
$\jset{0}{0},\jset{0}{1},\jset{1}{0}$ and $\jset{1}{1}$
contains its position, and on the cardinality of $\jset{0}{1}$
and $\jset{1}{1}$, we are able to coalesce the probability
$\Pr(\mathcal{E}_{(\mathtt{2})} = 0 \,|\, \mathcal{W}_t = y, |\jset{0}{1}|=\epsilon_{01},
|\jset{1}{1}|=\epsilon_{11})$, that is the probability that no discrepancies
are left after the second iteration, i.e., the decoding algorithm
terminates with a decoding success after it.

We recall that, from the analysis of the first iteration,
we are able to compute the probability that the first iteration
introduces $\discplus$ discrepancies between the error vector estimate
and the actual value, $\prob{|\jset{0}{1}| = \discplus\,|\,\mathcal{W}_t = y}$,
as $\prob{\mathcal{D}_+ = \discplus\,|\,\mathcal{W}_t = y}$,
and the probability that it removes $\discminus$ discrepancies,
$\prob{|\jset{1}{1}| = t-\discminus\,|\,\mathcal{W}_t = y} = \prob{\mathcal{D}_- = \discminus\,|\,\mathcal{W}_t = y}$.
Thanks to this knowledge, we are able to derive the probability
of performing a correct decoding
$\Pr(\mathcal{E}_{(\mathtt{2})} = 0 |  \mathcal{W}_t = y)$,
combining together the aforementioned probabilities, combining
together the probabilities of all the events leading to zero discrepancies
after the second iteration as follows:

$$
\begin{array}{ll}
\Pr(\mathcal{E}_{(\mathtt{2})} = 0 |  \mathcal{W}_t = y) =
\displaystyle \sum_{\epsilon_{01}, \epsilon_{11} \in \{0,\ldots,n-t\}\times \{0,\ldots,t\}}
\Big( & \Pr(\mathcal{D}_+ = \epsilon_{01}\,|\,\mathcal{W}_t = y) \cdot \Pr(\mathcal{D}_- = t-\epsilon_{11}\,|\,\mathcal{W}_t = y) \cdot \\
      & \cdot (1-\pflipzz(\epsilon_{01},\epsilon_{11}))^{n-t-\epsilon_{01}} \cdot
      \pflipzo(\epsilon_{01},\epsilon_{11})^{\epsilon_{01}} \cdot \\
      & \cdot (1-\pflipoz(\epsilon_{01}))^{t-\epsilon_{11}} \cdot
      \pflipoo(\epsilon_{01},\epsilon_{11}))^{\epsilon_{11}}
\Big)
\end{array}
$$

\noindent Finally, we derive the Decoding Failure Rate (DFR) after the
second decoder iteration as $\mathrm{DFR} = 1- \Pr(\mathcal{E}_{(\mathtt{2})} = 0)$ through
an application of the total probability formula.

\begin{theorem}[Decoding Failure Rate]
	$$
	\mathrm{DFR} = 1 - \sum_{y=0}^{r} \Pr(\mathcal{E}_{(\mathtt{2})} = 0 |
	\mathcal{W}_t = y) \Pr(\mathcal{W}_t = y)
	$$
\end{theorem}

\subsection{On the Independence Assumption between Parity-checks}\label{subsec:assumption}

In this section, we discuss the impact of Assumption~\ref{assumption:indep} on
the correctness of the estimated DFR. Assumption~\ref{assumption:indep} is
employed during the study of the $\mathtt{upc}$ distributions, modeling the
counters as the sum of $v$ independent Bernoulli trials.
We start by providing a numerically backed intuition of
the impact of making Assumption~\ref{assumption:indep} during our
derivations, with respect to the interrelations among the outcomes of 
parity-check equations which all share (at least) one term.
We then proceed to prove in closed form our intuition, showing that adopting
Assumption~\ref{assumption:indep} yields a conservative DFR estimate.

\begin{figure}[t!]
    \begin{center}
        \begin{tikzpicture}
  \begin{axis}[scale=0.7,
               xmin = 7,
               xmax = 27,
               grid = major,
               ymin = 0.298,
               ymax = 0.308,
               ytick = {0.298,0.3,0.302,0.304,0.306,0.308},
               legend style={at={(1.3,0.4)},anchor=south},
               yticklabel style={/pgf/number format/.cd, fixed, precision=5},
               mark size=2pt,
               xlabel={$\#$ unsatisfied parity-checks},
               ylabel={[Probability]},
               line width=0.5pt,
               mark options=solid]
    \addplot[black,densely dotted, mark=+]
             table [x=upcm1,y expr=
             \thisrow{frac},col sep = comma]{data/upcm1_p0u_n0_2_p_4801_v_45_t_50.csv};
    \addlegendentry{Simulated}
    \addplot[blue, mark=none]
             table [x=upcm1,y expr=
             \thisrow{p0u},col sep = comma]{data/upcm1_p0u_n0_2_p_4801_v_45_t_50.csv};
    \addlegendentry{Model}

\end{axis}
\end{tikzpicture}
        \vspace{-2em}
    \end{center}
    \caption{Probability of a parity-check $i$ of being unsatisfied at the
    beginning of the decoding procedure, $0 \leq i \leq r-1$, given that one of
    the $w$ included bits is correct (i.e., $h_{i,j}=1$ and
    $e_j = 0$, $0 \leq j \leq n-1$) and given the amount of unsatisfied parity-checks
    among the other $v-1$ where the same bit is involved. Code parameters:
    $n=9602$, rate $\frac{k}{n}=\frac{1}{2}$, $v=45$, $w=90$, error weight $t=50$.}
    \label{fig:p0u_upcdep}
\end{figure}
We recall that Assumption~\ref{assumption:indep} states that the random variables 
modeling the outcomes of the $v$ parity-check equations which share a term (i.e., the
ones indexed by the support of a column of $\vectmat{H}$) are distributed as Bernoulli
variables.
In order to provide a first, visual intuition of the statistical dependence
between the outcomes of different parity-check equations, which this assumption
neglects, we report the results of a set of Monte Carlo simulations,
where we estimated the probability of a parity-check being unsatisfied, given
that one correct bit (i.e., a bit in position $j$ such that $e_j = 0$) appears
in the parity equation, and given the amount of unsatisfied parity-checks among
other $v-1$ checks where such bit is involved (i.e., the parity-checks indexed
by the support of $\vectmat{h}_{:,j}$).

Figure~\ref{fig:p0u_upcdep} shows the results of our experiment, comparing the
theoretical probability $\punsatz$ (depicted as a blue horizontal line, as the
probability is fixed, given that the parity-checks are
assumed to be independent) with the Monte Carlo simulated probability (black + signs
in Figure~\ref{fig:p0u_upcdep}).
The code parameters were chosen in the cryptographic range, corresponding
to an $80$ bit security margin~\cite{DBLP:conf/isit/MisoczkiTSB13}.
The first observation is that the values of the two probabilities are 
numerically close across all the values on the $x$-axis, with small 
fluctuations of relative magnitude $\approx 1\%$ backing the intuition that 
relying on Assumption~\ref{assumption:indep} yields values which are close to 
the actual ones.
The second observation is that the analyzed parity-check
outcomes exhibit a {\em negative correlation} with the number of unsatisfied parity-checks:
indeed, observe that the simulated probability that one parity-check is 
unsatisfied decreases if there is an increasing number of unsatisfied 
parity-checks among the other $v-1$ ones indexed by the support of $h_{:,j}$.
This implies, intuitively, that the actual $\mathtt{upc}$ distribution of the bit in common
to $v$ parity-checks is more concentrated around the average with
respect to the distribution of the modeled under Assumption~\ref{assumption:indep}
(i.e., the one of the random variable $\mathcal{U}$ defined in Section~\ref{subsec:first}).

Willing to formalize this intuition, let $\mathcal{S}_{0}, \mathcal{S}_{1}, 
\dots, \mathcal{S}_{v-1}$ be the $v$ random variables (taking values in 
$\{0,1\}$) modeling the outcome of the $v$ parity-checks where a correct bit 
is involved; we have that negative
correlation implies $\mathrm{Cov}(\mathcal{S}_a, \mathcal{S}_b) < 0$ for
$0 \leq a < b \leq v-1$, and therefore:
$$
\mathrm{Var}(\mathtt{upc}_j) = \mathrm{Var}\left(\sum_{a=0}^{v-1} \mathcal{S}_a\right)
=
\left(\sum_{0\leq a \leq v-1} \mathrm{Var}(\mathcal{S}_a)\right)
+
2\cdot \left(
\sum_{0\leq a < b \leq v-1} \mathrm{Cov}(\mathcal{S}_a, \mathcal{S}_b) \right)
<
\sum_{0\leq a \leq v-1} \mathrm{Var}(\mathcal{S}_a)
$$
The inequality $\mathrm{Var}(\mathtt{upc}_j) < \sum_{0\leq a \leq v-1}
\mathrm{Var}(\mathcal{S}_a)$ shows that the variance of $\mathtt{upc}_j$ is
strictly smaller than the variance of $v$ independent Bernoulli trials added
together.
Since $\mathbb{E}\left[\mathtt{upc}_j\right] = \sum_{0\leq a \leq v-1}
\Pr(\mathcal{S}_a = 1)$ is significantly smaller than the flipping threshold
$\mathtt{th}$ (for choices of $\mathtt{th}$ allowing effective decoding), then 
a lower variance implies that the probability of flipping a correct bit 
$\Pr(\mathtt{upc}_j \geq \mathtt{th})$ is (considering effective values of
$\mathtt{th}$) strictly smaller than the probability $\Pr(\mathcal{U} \geq 
\mathtt{th})$ where $\mathcal{U} \sim \bindist{v}{\mathtt{th}}{\punsatz}$,
which is the approximation employed in our model.
We now prove, $\mathrm{Cov}(\mathcal{S}_a,
\mathcal{S}_b) < 0$ (for $0 \leq a < b \leq v-1$), providing evidence that
Assumption~\ref{assumption:indep} yields a conservative DFR estimation.
The proof presented in the remainder of the section can be equivalently applied to
the case where the bit included in the parity-checks is incorrect (i.e.,
$e_j=1$), showing that also the probability $\Pr(\mathtt{upc}_j < \mathtt{th})$
of maintaining incorrect bits is conservatively estimated.
\begin{theorem}\label{thm:neg_cov}
 Let $\mathcal{S}_a,\mathcal{S}_b$, with $0 \leq a < b \leq v-1$, be the
random variables modeling the probabilities that two parity-check equations,
 sharing a common term are satisfied. It holds that $\mathrm{Cov}(\mathcal{S}_a,
\mathcal{S}_b) < 0$.
\end{theorem}

To the end of proving Theorem~\ref{thm:neg_cov}, we will need to prove two
instrumental lemmas.
To delineate them, we start by rewriting $\mathrm{Cov}(\mathcal{S}_a,
\mathcal{S}_b) < 0$ as follows:
$$
\mathrm{Cov}(\mathcal{S}_a, \mathcal{S}_b) =
\mathbb{E}\left[ \mathcal{S}_a \mathcal{S}_b \right]
- \mathbb{E}\left[ \mathcal{S}_a\right]
\mathbb{E}\left[ \mathcal{S}_b\right]
=
\Pr(\mathcal{S}_a = 1)\Pr(\mathcal{S}_b = 1 | \mathcal{S}_a = 1)
-
\Pr(\mathcal{S}_a = 1)\Pr(\mathcal{S}_b = 1)
=
$$
$$
=
\Pr(\mathcal{S}_a = 1) \cdot \big( \Pr(\mathcal{S}_b = 1 | \mathcal{S}_a = 1)
- \Pr(\mathcal{S}_b = 1) \big) < 0
\iff
\Pr(\mathcal{S}_b = 1 | \mathcal{S}_a = 1) < \Pr(\mathcal{S}_b = 1)
$$

In the rest of the section, we will therefore prove the last rewriting of the
statement, i.e., $\Pr(\mathcal{S}_b = 1 | \mathcal{S}_a = 1) <
\Pr(\mathcal{S}_b = 1)$. To do so, we introduce two instrumental random variables
$\widetilde{\mathcal{S}}_a$ and $\widetilde{\mathcal{S}}_b$, corresponding to
the outcome of two parity-checks sharing one common correct bit, where the
other $w-1$ terms are drawn \textit{uniformly at random} across the remaining
$n-1$ terms.
We derive an explicit formulation for $\Pr(\mathcal{S}_b = 1 | \mathcal{S}_a
= 1)$ and $\Pr(\widetilde{\mathcal{S}}_b = 1 | \widetilde{\mathcal{S}}_a = 1)$
(which is known to be $\Pr(\widetilde{\mathcal{S}}_b = 1) = \Pr(\mathcal{S}_b = 1)$, since the
two parity-checks are uncorrelated), showing that the former is strictly
smaller than the latter.

Let $\kappa(m)$ ($m \in \{0,1,\dots,w-1\}$) be the probability that two parity-checks
with a common (correct) bit share $m$ additional bit positions, assuming the parity-checks
to be drawn from a $(v,w)$-regular parity-check matrix $\vectmat{H}$
(i.e., $m = |\support{\vectmat{h}_{i_1,:}} \cap \support{h_{i_2,:}}|-1$, where
$i_1 \neq i_2$ and $\exists j$ such that $ h_{i_1,j} = h_{i_2,j} = 1$ and $e_j=0$).
Then, let $\widetilde{\kappa}(m)$ ($m \in \{0,1,\dots,w-1\}$), instead, be the 
probability that two parity-checks with a common (correct) bit share $m$ 
additional bit positions, assuming that the other $w-1$ terms in each 
parity-check are drawn \textit{uniformly at random} across the remaining $n-1$ 
terms, with no regularity constraints.
Finally, let $\nu(m)$ ($m \in \{0,1,\dots,w-1\}$) be the probability
$\Pr(\mathcal{S}_b = 1 | \mathcal{S}_a = 1)$ under the assumption that the
two parity-checks, share one correct bit position and $m$ additional bit positions.
Then, by definition:
$$
\Pr(\mathcal{S}_b = 1 | \mathcal{S}_a = 1) = \sum_{m=0}^{w-1} \kappa(m) \nu(m)
\qquad \text{and} \qquad
\Pr(\mathcal{S}_b = 1) = \Pr(\widetilde{\mathcal{S}}_b = 1 |
\widetilde{\mathcal{S}}_a = 1) = \sum_{m=0}^{w-1} \widetilde{\kappa}(m) \nu(m).
$$
We now prove the first technical lemma:

\begin{lemma}\label{prop:stochdom}
    Let $\mathcal{M}$ be a random variable bound to the p.m.f. $\Pr(\mathcal{M} = m) = \kappa(m)$,
    and $\widetilde{\mathcal{M}}$ be a random variable bound to the p.m.f.
    $\Pr(\widetilde{\mathcal{M}} = m) = \widetilde{\kappa}(m)$.
    Then $\widetilde{\mathcal{M}}$ first-order stochastically dominates $\mathcal{M}$, i.e.,
    $$
    \widetilde{\mathcal{M}} \geq_{\textrm{st}} \mathcal{M}
    $$
\end{lemma}
\begin{proof}
    We begin by deriving $\kappa(m)$ and $\widetilde{\kappa}(m)$ explicitly.
    The quantity $\kappa(m)$ is the probability that two parity-checks with a
    common bit share $m$ additional bit positions, assuming the parity-checks
    to be drawn from a $(v,w)$-regular parity-check matrix $\vectmat{H}$.
    Consider two binary vectors $\vectmat{h}_{i_1,:}$ (two rows of $\vectmat{H}$) 
    and $\vectmat{h}_{i_2,:}$ ($0 \leq i_1 , i_2 \leq n-1$ and $i_1 \neq i_2$) 
    sharing a common position $j \in \{0, 1, \dots, n-1\}$ with a set bit, i.e., $\vectmat{h}_{i_1,j}=\vectmat{h}_{i_2,j}=1$, such that $e_j=0$, and consider any other 
    bit position $\ell \neq j$.
    Since each column of the parity-check matrix $\vectmat{H}$ has exactly $v$ 
    asserted bits, we have, by counting argument, that:
    $$
    \rho_{1|1} := \Pr(h_{i_2,\ell} = 1 | h_{i_1,\ell} = 1) = \frac{v-1}{r-1}
    $$
    $$
    \rho_{1|0} := \Pr(h_{i_2,\ell} = 1 | h_{i_1,\ell} = 0) = \frac{v}{r-1}
    $$
    Moreover, since the total number of bit positions $\ell \neq j$
    where $h_{i_2,\ell} = 1$ is $w-1$, we have that the probability $\kappa(m)$
    of having $m$ positions where $h_{i_1,\ell} = 1 \cap h_{i_2,\ell} = 1$ and
    $w-1-m$ bit positions where $h_{i_1,\ell} = 0 \cap h_{i_2,\ell} = 1$ is:
    $$
    \kappa(m) = \frac{
    \bindist{w-1}{m}{\rho_{1|1}}
    \cdot
    \bindist{n-w}{w-1-m}{\rho_{1|0}}
    }{
    \sum_{m'=0}^{w-1}
    \bindist{w-1}{m'}{\rho_{1|1}}
    \cdot
    \bindist{n-w}{w-1-m'}{\rho_{1|0}}
    }
    $$
    This p.m.f. corresponds to Fisher's noncentral hypergeometric
    distribution~\cite{harkness1965extended}, modeling the selection of the $w-1$
    bit positions where $h_{i_2,\ell} = 1$, knowing that the probability of choosing
    a certain bit position is $\rho_{1|1}$ if $h_{i_1,\ell} = 1$ and $\rho_{1|0}$
    if $h_{i_1,\ell} = 0$. Denoting as $\rho_{\texttt{ratio}}$ the following, $\rho_{\texttt{ratio}} =\frac{\rho_{1|1}(1-\rho_{1|0})}{(1-\rho_{1|1})
    \rho_{1|0}}$, we get:
    $$
    \kappa(m) = \frac{
    \binom{w-1}{m} (\rho_{1|1})^m (1-\rho_{1|1})^{w-1-m}
    \cdot
    \binom{n-w}{w-1-m} (\rho_{1|0})^{w-1-m} (1-\rho_{1|0})^{n-2w+m+1}
    }{
    \sum_{m'=0}^{w-1}
    \binom{w-1}{m'} (\rho_{1|1})^{m'} (1-\rho_{1|1})^{w-1-m'}
    \cdot
    \binom{n-w}{w-1-m'} (\rho_{1|0})^{w-1-m'} (1-\rho_{1|0})^{n-2w+m'+1}
    } =
    $$
    $$
    =
    \frac{
    \binom{w-1}{m} \left(\frac{\rho_{1|1}}{1-\rho_{1|1}}\right)^m (1-\rho_{1|1})^{w-1}
    \cdot
    \binom{n-w}{w-1-m} \left(\frac{\rho_{1|0}}{1-\rho_{1|0}}\right)^{w-1-m} (1-\rho_{1|0})^{n-w}
    }{
    \sum_{m'=0}^{w-1}
    \binom{w-1}{m'} \left(\frac{\rho_{1|1}}{1-\rho_{1|1}}\right)^{m'} (1-\rho_{1|1})^{w-1}
    \cdot
    \binom{n-w}{w-1-m'} \left(\frac{\rho_{1|0}}{1-\rho_{1|0}}\right)^{w-1-m'} (1-\rho_{1|0})^{n-w}
    }=
    $$
    $$
    =
    \frac{
    \binom{w-1}{m}\binom{n-w}{w-1-m} \left(\rho_{\texttt{ratio}}\right)^m
    \cdot
    (1-\rho_{1|1})^{w-1}\left(\frac{\rho_{1|0}}{1-\rho_{1|0}}\right)^{w-1} (1-\rho_{1|0})^{n-w}
    }{
    \sum_{m'=0}^{w-1}
    \binom{w-1}{m'}\binom{n-w}{w-1-m'} \left(\rho_{\texttt{ratio}}\right)^{m'}
    \cdot
    (1-\rho_{1|1})^{w-1} \left(\frac{\rho_{1|0}}{1-\rho_{1|0}}\right)^{w-1} (1-\rho_{1|0})^{n-w}
    }=
    $$
    $$
    =
    \frac{
    \binom{w-1}{m}\binom{n-w}{w-1-m} \left(\rho_{\texttt{ratio}}\right)^m
    }{
    \sum_{m'=0}^{w-1}
    \binom{w-1}{m'}\binom{n-w}{w-1-m'} \left(\rho_{\texttt{ratio}}\right)^{m'}
    }
    $$
    In particular, we have that $\kappa(m)$ is proportional to
    $\binom{w-1}{m}\binom{n-w}{w-1-m}\left(\rho_{\texttt{ratio}}\right)^m$, i.e.,
    $\kappa(m) \propto \binom{w-1}{m}\binom{n-w}{w-1-m}
    \left(\rho_{\texttt{ratio}}\right)^m$, since the denominator is fixed w.r.t. $m$, and
    $$
    \rho_{\texttt{ratio}} =
    \frac{\rho_{1|1}(1-\rho_{1|0})}{(1-\rho_{1|1})\rho_{1|0}}
    =
    \frac{
    \frac{v-1}{r-1}(1-\frac{v}{r-1})
    }{
    (1-\frac{v-1}{r-1})\frac{v}{r-1}
    } =
    \frac{
    (v-1)(r-1-v)
    }{
    v(r-v)
    }=
    \frac{
    rv - v^2 - (r-1)
    }{
    rv - v^2
    } < 1
    $$
    Moving on to $\widetilde{\kappa}(m)$, we have that the probability of having $m$
    common asserted positions between two binary vectors of length $n-1$ and weight $w-1$,
    when these are drawn uniformly at random, is:
    $$
    \widetilde{\kappa}(m) = \hygdist{n-1}{w-1}{w-1}{m} = \frac{\binom{w-1}{m}\binom{n-w}{w-1-m}}{\binom{n-1}{w-1}}
    $$
    This implies that $\widetilde{\kappa}(m)$ is proportional to $\binom{w-1}{m}\binom{n-w}{w-1-m}$, i.e.,
    $\widetilde{\kappa}(m) \propto \binom{w-1}{m}\binom{n-w}{w-1-m}$.
    Therefore:
    $$
    \frac{
    \Pr(\widetilde{\mathcal{M}} = m)
    }{
    \Pr(\mathcal{M} = m)
    } =
    \frac{\widetilde{\kappa}(m)}{\kappa(m)} \propto \left( \frac{1}{\rho_{\texttt{ratio}}} \right)^m
    $$
    The term $\left( \frac{1}{\rho_{\texttt{ratio}}} \right)^m$ is strictly
    increasing in $m$ (since $\rho_{\texttt{ratio}} < 1$), meaning that the two
    p.m.f. of $\widetilde{\mathcal{M}}$ and $\mathcal{M}$ enjoy the
    \textit{monotone likelihood ratio} property, which in turn implies the
    first-order stochastic dominance of $\widetilde{\mathcal{M}}$ over
    $\mathcal{M}$. Indeed, for every $m_0, m_1$, $0 \leq m_0 < m_1 \leq w-1$,
    we have that
    $$
    \frac{\widetilde{\kappa}(m_0)}{\kappa(m_0)}
    =
    \frac{
    \Pr(\widetilde{\mathcal{M}} = m_0)
    }{
    \Pr(\mathcal{M} = m_0)
    } < \frac{
    \Pr(\widetilde{\mathcal{M}} = m_1)
    }{
    \Pr(\mathcal{M} = m_1)
    } = \frac{\widetilde{\kappa}(m_1)}{\kappa(m_1)}
    $$
    from which we have that $\widetilde{k}(m_0)\kappa(m_1) < \widetilde{k}(m_1)\kappa(m_0) $.
    From this relation, we obtain the following two inequalities for every value of $m \in \{0, \dots, w-2\}$:
    $$
    \Pr(\widetilde{\mathcal{M}} \leq m)\Pr(\mathcal{M} = m)
    =
    \sum_{m_0=0}^{m}
    \widetilde{\kappa}(m_0)\kappa(m)
    <
    \sum_{m_0=0}^{m}
    \widetilde{\kappa}(m)\kappa(m_0)
    =
    \Pr(\widetilde{\mathcal{M}} = m)\Pr(\mathcal{M} \leq m)
    $$
    $$
    \Pr(\widetilde{\mathcal{M}} = m)\Pr(\mathcal{M} > m)
    =
    \sum_{m_1=m+1}^{w-1}
    \widetilde{\kappa}(m_0)\kappa(m_1)
    <
    \sum_{m_1=m+1}^{w-1}
    \widetilde{\kappa}(m_1)\kappa(m)
    =
    \Pr(\widetilde{\mathcal{M}} > m)\Pr(\mathcal{M} = m)
    $$
    Which imply, respectively:
    $$
    \frac{
    1 - \Pr(\widetilde{\mathcal{M}} > m)
    }{
    1 - \Pr(\mathcal{M} > m)
    } <
    \frac{
    \Pr(\widetilde{\mathcal{M}} = m)
    }{
    \Pr(\mathcal{M} = m)
    }
    $$
    $$
    \frac{
    \Pr(\widetilde{\mathcal{M}} = m)
    }{
    \Pr(\mathcal{M} = m)
    }
    <
    \frac{
    \Pr(\widetilde{\mathcal{M}} > m)
    }{
    \Pr(\mathcal{M} > m)
    }
    $$
    Merging the two inequalities, we get:
    $$
    \frac{
    1 - \Pr(\widetilde{\mathcal{M}} > m)
    }{
    1 - \Pr(\mathcal{M} > m)
    } <
    \frac{
    \Pr(\widetilde{\mathcal{M}} > m)
    }{
    \Pr(\mathcal{M} > m)
    }
    \quad
    \implies
    \quad
    $$
    $$\quad
    \implies
    \quad
    \Pr(\mathcal{M} > m) - \Pr(\mathcal{M} > m)\Pr(\widetilde{\mathcal{M}} > m) <
    \Pr(\widetilde{\mathcal{M}} > m) - \Pr(\mathcal{M} > m)\Pr(\widetilde{\mathcal{M}} > m)
    \quad
    \implies
    \quad
    $$
    $$
    \quad
    \implies
    \quad
    \Pr(\mathcal{M} > m) < \Pr(\widetilde{\mathcal{M}} > m) \quad \forall m \in \{0, \dots, w-2\}
    $$
    Which implies $\widetilde{\mathcal{M}} \geq_{\textrm{st}} \mathcal{M}$.
\end{proof}

We now move on to the analysis of $\nu(m)$, the probability that a parity-check 
equation $b$ (whose outcome is modeled by $\mathcal{S}_b$) 
involving a correct bit in position $j \in \{0, \dots, n-1\}$  is unsatisfied, 
given that one of the $v-1$ other parity-checks which also involve the said bit, say, the one 
of index $a$ (whose outcome is modeled by $\mathcal{S}_a$), is also unsatisfied, 
and given that the two parity-checks intersect in $m$ positions different from $j$.

\begin{figure*}[!t]
	\centering
	\begin{minipage}{0.45\textwidth}
		\centering
		\subfloat[$m=0$\label{fig:vm_0}]{
			\begin{tikzpicture}[scale=0.6]
	
	\node at (0-0.8,0.5) {$\mathtt{h}_{b,:}$};
	\fill[darkblue] (0,0) rectangle (0+1,1);
	\node at (0+0.5,0.5) {$\mathbf{1}$};
	\draw[line width=0.4pt] (0,0) rectangle (0+1,1);
	\foreach \i/\col/\num in {
		1/white/0,2/white/0,3/lightblue/1,
		4/lightblue/1,
	}{
		\fill[\col] (\i,0) rectangle (\i+1,1);
		\node at (\i+0.5,0.5) {$\num$};
		\draw[line width=0.4pt] (\i,0) rectangle (\i+1,1);
	}
	\node at (7-0.8,0.5) {$s_{b}$};
	\fill[lightred] (7,0) rectangle (7+1,1);
	\node at (7+0.5,0.5) {$1$};
	\draw[line width=0.4pt] (7,0) rectangle (7+1,1);
	
	\node at (0-0.8,1.8) {$\mathtt{h}_{a,:}$};
	\fill[darkblue] (0,1.3) rectangle (0+1,2.3);
	\node at (0+0.5,1.8) {$\mathbf{1}$};
	\draw[line width=0.4pt] (0,1.3) rectangle (0+1,2.3);
	\foreach \i/\col/\num in {
		1/lightblue/1,2/lightblue/1,3/white/0,
		4/white/0,
	}{
		\fill[\col] (\i,1.3) rectangle (\i+1,2.3);
		\node at (\i+0.5,1.8) {$\num$};
		\draw[line width=0.4pt] (\i,1.3) rectangle (\i+1,2.3);
	}
	\node at (7-0.8,1.8) {$s_{a}$};
	\fill[lightred] (7,1.3) rectangle (7+1,2.3);
	\node at (7+0.5,1.8) {$1$};
	\draw[line width=0.4pt] (7,1.3) rectangle (7+1,2.3);
	
	\node at (0-0.8,3.5) {$\mathtt{e}$};
	\fill[darkblue] (0,3) rectangle (0+1,4);
	\node at (0+0.5,3.5) {$\mathbf{0}$};
	\draw[line width=0.4pt] (0,3) rectangle (0+1,4);
	\foreach \i/\col/\num in {
		1/white/1,2/white/0,3/white/0,
		4/white/1,
	}{
		\fill[\col] (\i,3) rectangle (\i+1,4);
		\node at (\i+0.5,3.5) {$\num$};
		\draw[line width=0.4pt] (\i,3) rectangle (\i+1,4);
	}
	
	\fill[white] (0,-0.5) rectangle (1,-0.5);
\end{tikzpicture}
		}
	\end{minipage}
	\begin{minipage}{0.45\textwidth}
		\centering
		\subfloat[$m=1$\label{fig:vm_1}]{
			\begin{tikzpicture}[scale=0.6]
	
	\node at (0-0.8,0.5) {$\mathtt{h}_{b,:}$};
	\fill[darkblue] (0,0) rectangle (0+1,1);
	\node at (0+0.5,0.5) {$\mathbf{1}$};
	\draw[line width=0.4pt] (0,0) rectangle (0+1,1);
	\foreach \i/\col/\num in {
		1/white/0,2/lightblue/1,3/lightblue/1,
		4/white/0,
	}{
		\fill[\col] (\i,0) rectangle (\i+1,1);
		\node at (\i+0.5,0.5) {$\num$};
		\draw[line width=0.4pt] (\i,0) rectangle (\i+1,1);
	}
	\node at (7-0.8,0.5) {$s_{b}$};
	\fill[lightblue] (7,0) rectangle (7+1,1);
	\node at (7+0.5,0.5) {$0$};
	\draw[line width=0.4pt] (7,0) rectangle (7+1,1);
	
	\node at (0-0.8,1.8) {$\mathtt{h}_{a,:}$};
	\fill[darkblue] (0,1.3) rectangle (0+1,2.3);
	\node at (0+0.5,1.8) {$\mathbf{1}$};
	\draw[line width=0.4pt] (0,1.3) rectangle (0+1,2.3);
	\foreach \i/\col/\num in {
		1/lightblue/1,2/lightblue/1,3/white/0,
		4/white/0,
	}{
		\fill[\col] (\i,1.3) rectangle (\i+1,2.3);
		\node at (\i+0.5,1.8) {$\num$};
		\draw[line width=0.4pt] (\i,1.3) rectangle (\i+1,2.3);
	}
	\node at (7-0.8,1.8) {$s_{a}$};
	\fill[lightred] (7,1.3) rectangle (7+1,2.3);
	\node at (7+0.5,1.8) {$1$};
	\draw[line width=0.4pt] (7,1.3) rectangle (7+1,2.3);
	
	\node at (0-0.8,3.5) {$\mathtt{e}$};
	\fill[darkblue] (0,3) rectangle (0+1,4);
	\node at (0+0.5,3.5) {$\mathbf{0}$};
	\draw[line width=0.4pt] (0,3) rectangle (0+1,4);
	\foreach \i/\col/\num in {
		1/white/1,2/white/0,3/white/0,
		4/white/1,
	}{
		\fill[\col] (\i,3) rectangle (\i+1,4);
		\node at (\i+0.5,3.5) {$\num$};
		\draw[line width=0.4pt] (\i,3) rectangle (\i+1,4);
	}
	
	\fill[white] (0,-0.5) rectangle (1,-0.5);
\end{tikzpicture}
		}
	\end{minipage}
	\caption{Graphical representation of two
		parity-check equations ($\mathtt{h}_{a,:}$ and $\mathtt{h}_{b,:}$)
		sharing a correct bit (highlighted in dark blue),
		along with their corresponding outcomes ($s_{a}$ and $s_{b}$),
		given an error vector $\mathtt{e}$ with $\weight{\mathtt{e}}=2$.
		In Figure~\ref{fig:vm_0}, the parity checks share $m=0$ additional bits.
		In Figure~\ref{fig:vm_1}, the parity checks share $m=1$ additional bit.
		\label{fig:vm_example}}
\end{figure*}

We start by an intuitive argument backed by a running example,
shown in Figure~\ref{fig:vm_example}, where $n=5$, $w=3$. We know that 
$\nu(w-1) = 1$, since if the two equations fully overlap then their
outcome is equal, meaning that $\mathcal{S}_a = 1 \implies \mathcal{S}_b = 1$ with
probability $1$. 
Moreover, we can state that in general $\nu(m)$ is \textit{not}
strictly increasing in $m$. Indeed, consider two parity-checks ($\mathtt{h}_{a,:}$ and 
$\mathtt{h}_{b,:}$ in Figure~\ref{fig:vm_example}), and an error vector with 
$\weight{e}=2$. If $m=0$, then the two parity-checks
share one position (the one assumed to be correct, dark blue in 
Figure~\ref{fig:vm_example}) and, out of the $n-1=4$ remaining
ones, $w-1=2$ are included in one check and $2$ are included in the other
(Figure~\ref{fig:vm_0}).
If the parity-check $a$ is unsatisfied, then one incorrect bit is in check $a$,
forcing the other incorrect bit to be included in parity-check $b$, meaning that
parity-check $b$ is unsatisfied as well.
If, however, checks $a$ and $b$ share $m=1$ additional term, then the probability of
parity-check $b$ being unsatisfied is not equal to $1$ (Figure~\ref{fig:vm_1}).
With these parameters, we have $\nu(0)=1$ and $\nu(1)<1$, meaning that $\nu(m)$ is
not necessarily increasing in $m$.

This example, however, does not fully capture a realistic parameter set employed either
in communication-grade or cryptographic-grade LDPC/MDPC codes, where instead $n$
is large (i.e., $n \geq 100$) and $t$ and $w$ are a small fraction of $n$.
We now prove that, under these assumptions on the code parameters,
$\nu(m)$ is strictly increasing in $m$.

\begin{lemma}\label{prop:increasing}
    Let $\nu(m)$ be the probability $\Pr(\mathcal{S}_b = 1 | \mathcal{S}_a = 1)$
    assuming that the two parity-checks share $m$ ``additional'' bit positions.
    If $n$ is sufficiently large (i.e., $n \geq 100$) and $t,w \ll n$, then $\nu(m)$ is strictly increasing in $m$.
\end{lemma}

\begin{proof}
    The proof is structured as follows: we first derive the exact
    formula for $\Pr(\mathcal{S}_b = 1 | \mathcal{S}_a = 1)$, and
    then perform a series of substitutions 
    that yield to a formulation which is strictly increasing in $m$.
    Define $q(x)$, $x \in \{0, \dots, n-1\}$, to be the following quantity:
    $$
    q(x) := \sum_{i=1, \text{odd}}^{\min(x,t)} \hygdist{n-1}{x}{t}{i} =
    \sum_{i=1, \text{odd}}^{\min(x,t)} \frac{\binom{x}{i}\binom{n-1-x}{t-i}}{\binom{n-1}{t}}
    $$
    The function $q(x)$ describes the probability that, for any pick of a 
    subset of size $x$ of positions in the error vector out of the $n-1$ positions
    in $\vectmat{e}$ different from the one which is assumed to be correct, 
    there is an odd amount of selected positions $\ell \leq x$ such that $e_\ell=1$.
    We immediately have the following:
    $$
    \Pr(\mathcal{S}_a = 1) = \Pr(\mathcal{S}_b = 1) = q(w-1)
    $$
    We can also compute the probability $\Pr(\mathcal{S}_a \neq \mathcal{S}_b)$ in terms of $q(x)$.
    Consider the number of erroneous bits included in one of the two checks,
    but not in the other. If $\mathcal{S}_a \neq \mathcal{S}_b$, then this quantity
    must be odd for one check and even for the other.
    Indeed, since the erroneous bits among the
    $m$ common positions are shared between the two checks, if the parity of 
    the two quantities were
    to be the same, then the outcome of both parity-checks would be the same, which contradicts the
    assumption that $\mathcal{S}_a \neq \mathcal{S}_b$. Given that the total number of terms included
    in one of the two parity-checks but not in both is $2(w-1-m)$, we have that the probability of
    the event $\mathcal{S}_a \neq \mathcal{S}_b$ is equal to the probability that the overall number
    of erroneous bits included in the $2(w-1-m)$ terms  belonging to a single
    parity check equation is odd:
    $$
    \Pr(\mathcal{S}_a \neq \mathcal{S}_b) = q\left(2(w-1-m)\right)
    $$
    The probability $\Pr(\mathcal{S}_a \neq \mathcal{S}_b)$ can also be written as:
    \begin{equation}
    \Pr(\mathcal{S}_a \neq \mathcal{S}_b) = \Pr(\mathcal{S}_a = 1)\big(1-\Pr(\mathcal{S}_b = 1
    | \mathcal{S}_a = 1) \big) + \big(1-\Pr(\mathcal{S}_a = 1)\big)\Pr(\mathcal{S}_b = 1 | \mathcal{S}_a = 0) \label{prob:sa_neq_sb}
    \end{equation}
    Moreover:
    $$
    \Pr(\mathcal{S}_b = 1) = \Pr(\mathcal{S}_b = 1 \,\cap\, \mathcal{S}_a = 0)
    + \Pr(\mathcal{S}_b = 1 \,\cap\, \mathcal{S}_a = 1)
    =
    $$
    $$
    =
    \big(1-\Pr(\mathcal{S}_a = 1)\big)\Pr(\mathcal{S}_b = 1 \,|\, \mathcal{S}_a = 0)
    + \Pr(\mathcal{S}_a = 1)\Pr(\mathcal{S}_b = 1 \,|\, \mathcal{S}_a = 1)
    $$
    From which we get:
    $$\Pr(\mathcal{S}_b = 1 \,|\, \mathcal{S}_a = 0)
    = \frac{\Pr(\mathcal{S}_b = 1) - \Pr(\mathcal{S}_a = 1)\Pr(\mathcal{S}_b = 1 \,
    |\, \mathcal{S}_a = 1)}{1-\Pr(\mathcal{S}_a = 1)}
    $$
    Substituting $\Pr(\mathcal{S}_b = 1 \,|\, \mathcal{S}_a = 0)$ in Equation~\ref{prob:sa_neq_sb}  and solving for $\Pr(\mathcal{S}_b = 1 \,|\, \mathcal{S}_a = 1)$, we get:
    \begin{equation}\label{eq:nu_exact}
    \nu(m) = \Pr(\mathcal{S}_b = 1 \,|\, \mathcal{S}_a = 1) =
    \frac{
    \Pr(\mathcal{S}_a = 1) + \Pr(\mathcal{S}_b = 1) - \Pr(\mathcal{S}_a \neq \mathcal{S}_b)
    }{2\Pr(\mathcal{S}_a = 1)} = 1 - \frac{q\left(2(w-1-m)\right)}{2 q(w-1)}
	\end{equation}
    As previously stated, $\nu(m)$ is not always increasing in $m$. However,
    if we assume $n$ to be large and $w$ and $t$ to be $\ll n$, then we can simplify
    the expression for $q(x)$.
    Indeed $q(x)$ measures the probability that a random variable
    $\chi \sim \Pr(\chi = i)$ with hypergeometric distribution, i.e.,
    $ \Pr(\chi = i) = \hygdist{n-1}{t}{x}{i}$,  takes odd $i$ values.
    For large $n$ and $x \ll n$, this distribution converges to
    $\Pr(\chi = i) = \bindist{x}{i}{\frac{t}{n-1}}$, corresponding to
    the distribution of the sum of $x$ independent Bernoulli trials with success
    probability $\frac{t}{n-1}$.
    Denoting with $\text{\sc exp}\left(\cdot \right)$ the formula of the exponential
    function having the base equal to the Euler's number (a.k.a. Napier's constant),
    the probability that a odd number of successes happens within the $x$ trials
    has been derived in~\cite{DBLP:journals/tit/Gallager62}:
    $
    \frac{1}{2}\left(1 - \prod_{i=1}^{x}\left(1-2\frac{t}{n-1}\right)\right)
    $.
    Consequently, keeping into account that $x, t \ll n$:
    $$
    q(x)
    \simeq \frac{1 - \left(1-2\frac{t}{n-1}\right)^x}{2}
    \simeq \frac{1 - \text{\sc exp}\left({-\frac{2tx}{n-1}}\right)}{2}
    $$
    Therefore:
    \begin{equation}\label{eq:nu_approx}
    \nu(m) \simeq 1 - \frac{
    1-\text{\sc exp}\left({-\frac{4t(w-1-m)}{n-1}}\right)
    }{
    2\left(1 - \text{\sc exp}\left({-\frac{2t(w-1)}{n-1}}\right)\right)
    } =
    1 +
    \frac{
    \text{\sc exp}\left({\frac{4t(m+1-w)}{n-1}}\right) - 1
    }{
    2\left(1 - \text{\sc exp}\left({-\frac{2t(w-1)}{n-1}}\right)\right)
    }
	\end{equation}
    where the only term depending on $m$ is $\text{\sc exp}\left({\frac{4t(m+1-w)}{n-1}}\right)$,
    which is strictly increasing in $m$. Since the denominator is positive,
    this implies that $\nu(m)$ is strictly increasing in $m$.
\end{proof}

\begin{figure}[t!]
    \centering
\begin{tikzpicture}
  \begin{axis}[scale=0.7,
                width = 22cm,
                height = 7.5cm,
               xmin = 0,
               xmax = 89,
               grid = major,
               ymin = 0.2,
               ymax = 1,
               legend style={at={(1.1,0.4)},anchor=south},
               yticklabel style={/pgf/number format/.cd, fixed, precision=5},
               mark size=2pt,
               xlabel={$m$},
               ylabel={$\nu(m)$},
               line width=0.5pt,
               mark options=solid]
    
    \addplot[red, mark=none]
             table [x=m,y expr=
             \thisrow{approx_nu_m},col sep = comma]{data/nu_9602_90_50.csv};
    \addlegendentry{Approx.}
    \addplot[black,only marks, mark=x]
             table [x=m,y expr=
             \thisrow{nu_m},col sep = comma]{data/nu_9602_90_50.csv};
    \addlegendentry{Exact}

\end{axis}
\end{tikzpicture}
	\vspace{-2.5em}
    \caption{Values of $\nu(m)$ using the exact and approximated formulas.
    Code parameters: $n=9602$, rate $\frac{k}{n}=\frac{1}{2}$, $v=45$, $w=90$, error weight $t=50$.}
    \label{fig:nuvalues}
\end{figure}
To provide a visual intuition of the goodness of fit between the exact formula
for $\nu(m)$ (Equation~\ref{eq:nu_exact}) and the approximated one shown at the end of the previous proof (Equation~\ref{eq:nu_approx})
for cryptographic grade parameters, Figure~\ref{fig:nuvalues} shows a
comparison between them using the same parameter set of Figure~\ref{fig:p0u_upcdep}
(which can be thought of as the smallest one for cryptographic purposes, 
as they provide a $2^{80}$ security margin~\cite{DBLP:conf/isit/MisoczkiTSB13}).
From this plot, we can see that the two functions match almost exactly, and
that $\nu(m)$ increases with $m$, as proven for large $n$ and $t,v \ll n$.

We now have all the tools needed to prove the Theorem~\ref{thm:neg_cov},
i.e.:
    Let $\mathcal{S}_{0}, \mathcal{S}_{1}, \dots, \mathcal{S}_{v-1}$ be the $v$
    random variables (taking values in $\{0,1\}$) modeling the outcome of the
    $v$ parity-checks of a correct bit (i.e., a bit $j$ where $e_j=0$).
    Then $\mathrm{Cov}(\mathcal{S}_a, \mathcal{S}_b) < 0$ for $0 \leq a < b \leq v-1$.

\begin{proof}
    As previously stated, $\mathrm{Cov}(\mathcal{S}_a, \mathcal{S}_b) < 0$
    is equivalent to $\Pr(\mathcal{S}_b = 1 | \mathcal{S}_a = 1) < \Pr(\mathcal{S}_b = 1)$.
    Moreover, we expressed the two terms of the inequality as:
    $$
    \Pr(\mathcal{S}_b = 1 | \mathcal{S}_a = 1) = \sum_{m=0}^{w-1} \kappa(m) \nu(m)
    $$
    $$
    \Pr(\mathcal{S}_b = 1) = \Pr(\widetilde{\mathcal{S}}_b = 1 |
    \widetilde{\mathcal{S}}_a = 1) = \sum_{m=0}^{w-1} \widetilde{\kappa}(m) \nu(m)
    $$
    Let $\mathcal{M}$ be a random variable bound to the p.m.f. $\Pr(\mathcal{M} = m) =
    \kappa(m)$, and $\widetilde{\mathcal{M}}$ be a random variable bound to the p.m.f.
    $\Pr(\widetilde{\mathcal{M}} = m) = \widetilde{\kappa}(m)$. Then:
    $$
    \Pr(\mathcal{S}_b = 1 | \mathcal{S}_a = 1)
    = \sum_{m=0}^{w-1} \kappa(m) \nu(m)
     =
     \sum_{m=0}^{w-1} \Pr(\mathcal{M} = m) \nu(m)
     =
    \mathbb{E}\left[ \nu(\mathcal{M}) \right]
    $$
    $$
    \Pr(\mathcal{S}_b = 1)
     =
    \sum_{m=0}^{w-1} \widetilde{\kappa}(m) \nu(m) =
     \sum_{m=0}^{w-1} \Pr(\widetilde{\mathcal{M}} = m) \nu(m)
     =
    \mathbb{E}\left[ \nu(\widetilde{\mathcal{M}}) \right]
    $$
    From Lemma~\ref{prop:increasing} we have that $\nu(m)$ is strictly increasing
    (for reasonable code parameters).
    From Lemma~\ref{prop:stochdom} we have that $\widetilde{\mathcal{M}} \geq_{\textrm{st}}
    \mathcal{M}$, which implies that $\mathbb{E}\left[f(\widetilde{\mathcal{M}})\right] >
    \mathbb{E}\left[f(\mathcal{M})\right]$ for every increasing function $f(m)$.
    Letting $f(m) = \nu(m)$ proves that $\Pr(\mathcal{S}_b = 1) > \Pr(\mathcal{S}_b = 1 |
    \mathcal{S}_a = 1)$, and therefore that $\mathrm{Cov}(\mathcal{S}_a, \mathcal{S}_b) < 0$.
\end{proof}

%
%
\section{Efficient Estimation Engineering and Numerical Validation}

In this section, we describe a set of numerical techniques and approximations
that can be employed to speed-up the computation of the DFR estimates without
compromising the accuracy in a noticeable way, and report the results of
the numerical validations of our theoretical results against
Monte Carlo simulations.

\subsection{Enhancing Performance of the DFR Estimate Computation}

While the mathematical model we propose can be computed
\textit{as-is} for transmission-grade codes (i.e., the ones with code lengths
in the low thousands, and weights in the tens), obtaining DFR estimates for
cryptographic-grade codes
(i.e., the ones with code lenghts in the tens of thousands, and weights in the hundreds)
becomes computationally expensive.
To allow an efficient estimation of the decoding capabilities in these cases,
useful for the extensive testing and parameter tuning typical of cryptosystem
design, we employ a series of approximations to skip unnecessary calculations
without compromising the final results.
Employing the following techniques, the computation time for our estimates
is reduced by an order of magnitude or more for all code parameters of practical
interest (both for transmission and cryptographic purposes).

These approximations have been introduced and validated incrementally against
the corresponding exact methods, in order to ensure correct results.
Additionally, all the calculations  performed employing
double-precision floating-point arithmetic have been validated through the
use of an additional implementation which is employing arbitrary precision mantissas
and $64$-bit wide exponents to validate their numerical stability and
correctness.

The first optimization we report aims to minimize the dependence between the
failure probability after the second iteration and the weight of the syndrome
vector before the first iteration.
To achieve this, we average the probabilities needed for calculating the DFR
(as reported in~\ref{subsec:second}) over the syndrome weight.
For example, the average distribution (over all syndrome weights) of flips
among correct bits, $\Pr(\mathcal{D}_+ = \discplus)$, can be computed as:
$$
\Pr(\mathcal{D}_+ = \discplus) = \sum_y \Pr(\mathcal{D}_+ = \discplus|
\mathcal{W}_t = y) \Pr(\mathcal{W}_t = y)
$$
All the probabilities and probability mass functions derived in
Section~\ref{subsec:first} and Section~\ref{subsec:second} can be averaged
in a similar way. This modification
eliminates the need to recompute the second iteration DFR for every possible
syndrome weight, speeding up calculations by a factor of
$\mathcal{O}(\min(v \cdot t,r))$.
The final result is not altered significantly with respect to its exact value,
since the knowledge of the
syndrome weight impacts the distributions $\mathcal{D}_+$ and
$\mathcal{D}_-$, but it does not directly affect the final DFR, once the values
$\mathcal{D}_+$ and $\mathcal{D}_-$ are assumed to be known as well ($\epsilon_{01}$ and $t-\epsilon_{11}$, respectively).

The second optimization we realized splits the DFR calculations in
two steps.
The first step computes the distribution of the syndrome weight and the
necessary probabilities averaged over such distribution. The second step
derives the final decoding failure rate based on $\epsilon_{01} = \discplus$
and $\epsilon_{11} = t-\discminus$. The number of admissible values for
$(\epsilon_{01},\epsilon_{11}) \in \{0,\ldots,n-t\}\times \{0,\ldots,t\}$ is
$(n-t+1)(t+1) = nt-t^2+1$, which results in a large number of calculations when
$n$ and $t$ are large.
To mitigate this we employ a cutoff strategy, deriving $\Pr(\mathcal{E}_{(2)} = 0)$
incrementally with the formula presented in Section~\ref{subsec:second} avoiding the
computation of terms on the sum (iterating over values of $\epsilon_{01}$
and $\epsilon_{11}$), which do not impact the final result numerically. We
identify such instances of $\epsilon_{01}$ and $\epsilon_{11}$ by noting that,
if the coefficient $\Pr(\mathcal{D}_+ = \epsilon_{01}) \cdot \Pr(\mathcal{D}_- = t-\epsilon_{11})$
is several orders of magnitude smaller than the partial sum computed over
previous terms, the contribution of the corresponding term can be safely omitted.
This strategy is justified by the fact that, using double precision floating point
variables, the \textit{machine epsilon} (i.e., an upper bound on the relative
approximation error due to rounding) is in the order of $10^{-16}$, meaning
that a larger difference in the order of magnitude between two addends causes
the binary representation of the result to be \textit{equal} to the one of
the larger term in the worst case.
More formally,  given a partial sum $S_I$ computed over a set
$I \subset \{0,\ldots,n-t\}\times \{0,\ldots,t\}$ of values
for $\epsilon_{01}$ and $\epsilon_{11}$:
$$
\begin{array}{ll}
S_I =
\displaystyle \sum_{\epsilon_{01}, \epsilon_{11} \in I}
\Big( & \Pr(\mathcal{D}_+ = \epsilon_{01}) \cdot \Pr(\mathcal{D}_- = t-\epsilon_{11}) \cdot \\
      & \cdot (1-\pflipzz(\epsilon_{01},\epsilon_{11}))^{n-t-\epsilon{01}} \cdot
      \pflipzo(\epsilon_{01},\epsilon_{11})^{\epsilon_{01}} \cdot \\
      & \cdot (1-\pflipoz(\epsilon_{01},\epsilon_{11}))^{t-\epsilon_{11}} \cdot
      \pflipoo(\epsilon_{01},\epsilon_{11}))^{\epsilon_{11}}
\Big)
\end{array}
$$
we can omit the contribution to the sum given by values of $\epsilon_{01},\epsilon_{11} \notin I$ where $\Pr(\mathcal{D}_+ = \epsilon_{01}) \cdot \Pr(\mathcal{D}_- = t-\epsilon_{11}) < S_I \cdot 10^{-16}$ since in such cases, due to finite precision, we would have the following:
$$
S_I + \Pr(\mathcal{D}_+ = \epsilon_{01}) \cdot \Pr(\mathcal{D}_- = t-\epsilon_{11}) \cdot \alpha < S_I \cdot (1 + 10^{-16}) \overset{\text{fp}}{=} S_I
$$
where $\alpha$ can be any number in $[0,1]$ (i.e., a decoding failure probability),
and $\overset{\text{fp}}{=}$ denotes that the elements appearing at its left
and right yield the same number when numerically computed with double precision
floating point numbers.

Finally, regarding the numerical stability in the computation of the inner
terms of the sum:
$$
(1-\pflipzz)^{n-t-\epsilon{01}} \cdot \pflipzo^{\epsilon_{01}} \cdot (1-\pflipoz)^{t-\epsilon_{11}} \cdot \pflipoo)^{\epsilon_{11}}
$$
we note that $(1-\pflipzz) \overset{\text{fp}}{=} 1$ and $(1-\pflipoz) \overset{\text{fp}}{=} 1$ if $\pflipzz$ and $\pflipoz$ are smaller than
$10^{-16}$, for the same reasons regarding finite precision stated above.
Similarly, $\pflipzo$ and $\pflipoo$ prove to be numerically unstable as they
approach $1$. To address this, we compute $p_{\neg \mathtt{flip}|01}$ and
$p_{\neg \mathtt{flip}|11}$ rather than $\pflipzo$ and $\pflipoo$, and we
derive $\Pr(\mathcal{E}_{(2)} \geq 1)$ instead of $\Pr(\mathcal{E}_{(2)} = 0)$
employing the following (conservative) approximation:
$$
\Pr(\mathcal{E}_{(2)} \geq 1) = \sum_{\epsilon \geq 1} \Pr(\mathcal{E}_{(2)} = \epsilon)
\leq
\sum_{\epsilon \geq 1} \epsilon \cdot \Pr(\mathcal{E}_{(2)} = \epsilon) = \mathbb{E}\left[\mathcal{E}_{(2)}\right]
$$
where $\mathbb{E}\left[\mathcal{E}_{(2)}\right]$ can be efficiently
(and safely) computed as:
$$
\mathbb{E}[\mathcal{E}_{(2)}] = \pflipzz \cdot (n-t-\epsilon_{01}) + p_{\neg \mathtt{flip}|01} \cdot \epsilon_{01} + \pflipoz \cdot (t-\epsilon_{11}) + p_{\neg \mathtt{flip}|11} \cdot \epsilon_{11}
$$

These optimizations ensure an efficient and numerically stable computation of
the DFR, making the model practical for cryptographic-grade parameter sizes.

\subsection{Numerical validations}
In the following, we provide numerical validations of \emph{i)} our syndrome weight
estimation technique, \emph{ii)} the probability distributions
of the discrepancies (i.e., differences between the error vector estimate and
the actual error vector) after the first iteration,
$\Pr(\mathcal{D}_+ = \discplus)$ and $\Pr(\mathcal{D}_- = \discminus)$, \emph{iii)} numerical validation
 of the goodness of fit for the flipping probabilities at the second iteration
$\pflipzz,\pflipzo,\pflipoz,\pflipoo$; \emph{iv)} the two-iterations
decoding failure rate of a parallel bit-flipping decoder, while varying the code
density and the number of errors; \emph{v)} we report the result of applying
our two-iterations DFR estimation technique to re-evaluate the DFR of the
code parameters proposed in the LEDAcrypt specification~\cite{LEDA} providing
tighter, yet conservative estimates.

The numeric simulations were run on two Dell PowerEdge R$630$ nodes,
each one endowed with two Intel Xeon CPU E$5$-$2698$ v$4$
($20$ cores/$40$ threads each), and a Dell PowerEdge R$7425$ equipped with two AMD Epyc
$7551$ ($32$ cores/$64$ threads each), taking around $50$k core-hours.
The memory footprint of the simulations was small ($<$$200$MiB).
The model results were obtained on a desktop equipped with an Intel Core
i$7$-$12700$, and took approximately half a core-hour overall.

\begin{figure*}[!t]
\begin{center}
    \subfloat[$n=4400, k=2200,v=11,t=18$\label{fig:sweight_toy}]{
      \begin{tikzpicture}
  \begin{axis}[scale=0.72,
               xmax = 200,
               grid = major,
               ymode= log,
               ymax = 1,
               ymin = 1e-8,
               ytick={1,1e-2,1e-4,1e-6,1e-8},
               legend style={at={(1.3,0.35)},anchor=south},
               mark size=2pt,
               xlabel={$y:\weight{s}$},
               ylabel={$\Pr(\mathcal{W}_t=y)$},
               line width=0.5pt,
               mark options=solid]
    \addplot[red,dotted, mark=x]
             table [x=w_init_val,y expr=
             \thisrow{prob},col sep = comma]{data/model_sw_2_11_18_0.csv};
    \addplot[black,only marks, mark=+]
             table [x=w_init_val,y expr=
             \thisrow{prob},col sep = comma]{data/sw_2_11_18_0.csv};
\end{axis}
\end{tikzpicture}
}
    \subfloat[$n_0=4, p=13397, v=83$, $t=66$\label{fig:sweight_leda}]{
     \begin{tikzpicture}
  \begin{axis}[scale=0.72,
               xmin = 3500,
               xmax = 4000,
               grid = major,
               ymode= log,
               ymax = 1,
               ymin = 1e-8,
               ytick={1,1e-2,1e-4,1e-6,1e-8},
               legend style={at={(1.3,0.35)},anchor=south},
               mark size=2pt,
               xlabel={$y:\weight{s}$},
               ylabel={$\Pr(\mathcal{W}_t=y)$},
               line width=0.2pt,
               mark options=solid]
    \addplot[red,dotted, mark=x]
             table [x = w_init_val,
                    y expr = \thisrow{prob}, 
                    col sep = comma]{data/sw_4_13397_83_66_model.csv};
    \addlegendentry{Model};
    \addplot[black,only marks, mark=+]
             table [x = w_init_val,
                    y expr = \thisrow{prob}, 
                    col sep = comma]{data/sw_4_13397_83_66_10e9.csv};
    \addlegendentry{Numerical};
\end{axis}
\end{tikzpicture}
}
\end{center}
\caption{Numerical validation of the model of syndrome weight distribution,
simulation on a $(v,w)$-regular code parity-check matrix, picking a communications-grade
code parameter set (left) and a cryptography grade code parameter set (right). Numerical
results obtained with $10^9$ random syndrome samples\label{fig:sweight}}
\end{figure*}
Figure~\ref{fig:sweight} reports the comparison between our modeled (red) and
the numerically estimated (black) distribution of syndrome weights
$\Pr(\mathcal{W}_t=y)$ for a small $(11-22)$ regular LDPC
code with code length $n=4400$ and error weight $t=18$ (Figure~\ref{fig:sweight_toy}),
and for the code with rate $\frac{3}{4}$ employed in the LEDAcrypt
specification~\cite{LEDA} (Figure~\ref{fig:sweight_leda}),
for NIST security category $1$.
In both cases, our estimation technique provides a very good match for
the numerically simulated probabilities.
We note that the asymmetric shape of Figure~\ref{fig:sweight_toy} is justified
by the code being rather sparse, in turn causing little interaction among
the bits of the columns of the parity-check matrix being added to the syndrome.
Indeed, a non-negligible amount of syndromes have the maximum weight $v\cdot t= 198$.

\begin{figure*}[!t]
\begin{tikzpicture}
  \begin{axis}[scale=0.52,
               xmin = 0,
               xmax = 35,
               grid = major,
               ymode= log,
               ymin = 0.00001,
               ymax = 1,
               ytick={1,1e-1,1e-2,1e-3,1e-4,1e-5},
               legend style={at={(1.55,0.35)},anchor=south},
               mark size=2pt,
               xlabel={$\discplus$},
               ylabel={$\Pr(\mathcal{D}_+ = \discplus)$ [Probability]},
               line width=0.5pt,
               xlabel style={font=\footnotesize},
               ylabel style={font=\footnotesize},
               xticklabel style = {font=\footnotesize},
               yticklabel style = {font=\footnotesize},
               mark options=solid]
    \addplot[blue, mark=x]
             table [x=err,y expr=
             \thisrow{leda},col sep = comma]{data/nd_4_13397_83_95_45.csv};
    \addplot[red, mark=x]
             table [x=err,y expr=
             \thisrow{mod},col sep = comma]{data/nd_4_13397_83_95_45.csv};
    \addplot[black,only marks, mark=+]
             table [x=err,y expr=
             \thisrow{sim},col sep = comma]{data/nd_4_13397_83_95_45.csv};

\end{axis}
\end{tikzpicture}
\begin{tikzpicture}
  \begin{axis}[scale=0.52,
                xtick={80,85,...,95},
               xmin = 80,
               xmax = 95,
               grid = major,
               ymode= log,
               ymin = 0.00001,
               ymax = 1,
               ytick={1,1e-1,1e-2,1e-3,1e-4,1e-5},
               legend style={at={(1.55,0.35)},anchor=south},
               mark size=2pt,
               xlabel={$\discminus$},
               ylabel={$\Pr(\mathcal{D}_- = \discminus)$ [Probability]},
               line width=0.5pt,
               xlabel style={font=\footnotesize},
               ylabel style={font=\footnotesize},
               xticklabel style = {font=\footnotesize},
               yticklabel style = {font=\footnotesize},               
               mark options=solid]
    \addplot[blue, mark=x]
             table [x expr = 95-\thisrow{err},y expr=
             \thisrow{leda},col sep = comma]{data/od_4_13397_83_95_47.csv};
    \addplot[red, mark=x]
             table [x expr = 95-\thisrow{err},y expr=
             \thisrow{mod},col sep = comma]{data/od_4_13397_83_95_47.csv};
    \addplot[black,only marks, mark=+]
             table [x expr = 95-\thisrow{err},y expr=
             \thisrow{sim},col sep = comma]{data/od_4_13397_83_95_47.csv};

\end{axis}
\end{tikzpicture}
\begin{tikzpicture}
  \begin{axis}[scale=0.52,
               xtick={0,10,...,40},
               xmin = 0,
               xmax = 40,
               grid = major,
               ymode= log,
               ymin = 1e-5,
               ymax = 1,
               ytick={1,1e-1,1e-2,1e-3,1e-4,1e-5},
               legend style={at={(1.45,0.35)},anchor=south,font=\footnotesize},
               mark size=2pt,
               xlabel={$\discre$},
               ylabel={$\prob{\mathcal{E}_{(\mathtt{1})} = \discre}$ [Probability]},
               line width=0.5pt,
               xlabel style={font=\footnotesize},
               ylabel style={font=\footnotesize},
               xticklabel style = {font=\footnotesize},
               yticklabel style = {font=\footnotesize},
               mark options=solid]
    \addplot[blue, mark=x]
             table [x expr = \thisrow{err},
                    y expr=\thisrow{leda},col sep = comma]{data/d_4_13397_83_95_45.csv};
    \addlegendentry{LEDAcrypt};
    \addplot[red, mark=x]
             table [x expr = \thisrow{err},
                    y expr=\thisrow{mod},col sep = comma]{data/d_4_13397_83_95_45.csv};
    \addlegendentry{Model};
    
    \addplot[black,only marks, mark=+]
             table [x expr = \thisrow{err},
                    y expr=\thisrow{sim},col sep = comma]{data/d_4_13397_83_95_45.csv};
    \addlegendentry{Simulated};
\end{axis}
\end{tikzpicture}
\caption{Number flips on $\bar{e}_j$ which took place when $e_j=0$ ($\discplus$)
and number of flips not made on $\bar{e}_j$ when $e_j=1$ ($t-\discminus$)
after the first iteration for the LEDAcrypt code with parameters $n_0=4, p=13397,
n=n_0p, k=(n_0-1)p, v=83$, results obtained with $10^5$ randomly generated error
vectors of weight $t=95$ for each point.\label{fig:discrepancies}}
\end{figure*}
Figure~\ref{fig:discrepancies} reports the distribution of $\discre$, $\discplus$
and $\discminus$. We recall that the aforementioned values are the number
of discrepancies between the error vector and its estimate in the decoder
after the first iteration ($\discre$), the number of discrepancies introduced
by the first iteration flipping up positions where the error vector does not
have a set bit ($\discplus$), and the number of discrepancies removed by
the first iteration not flipping a position of the estimate where the error
vector has a set bit ($\discminus$).
In all three cases, our model yields a distribution that depends on the syndrome
weight; the depicted values are thus obtained as the weighted average
over all syndrome weights.
As it can be seen, our model provides a closer fit to the sample distribution
of $\discre$, $\discplus$ and $\discminus$, w.r.t. the one employed
in~\cite{LEDA}.

\begin{figure*}[!t]
\begin{center}
    \subfloat[$\mathtt{th1}= \mathtt{th2}= 25$]{
     \begin{tikzpicture}
  \begin{axis}[scale=0.65,
               xtick={50,54,...,70},
							 ytick={1,0.1,0.001,0.00001,0.0000001},
               xmin = 54,
               xmax = 70,
               grid = major,
               ymode= log,
               ymin = 0.00001,
               ymax = 1,
               log basis y=10,
               mark size=2pt,
               xlabel={$t$},
               ylabel={Probability},
               line width=0.5pt,
               xlabel style={font=\footnotesize},
               ylabel style={font=\footnotesize},
               xticklabel style = {font=\footnotesize},
               yticklabel style = {font=\footnotesize},
               mark options=solid]
    \addplot[red]
             table [x=t,y expr=
             \thisrow{m00},col sep = comma]{data/cat_2_4801_45_25_25.csv};
    \addplot[red,dotted, mark=+]
             table [x=t,y expr=
             \thisrow{s00},col sep = comma]{data/cat_2_4801_45_25_25.csv};

    \addplot[orange]
             table [x=t,y expr=
             \thisrow{m01},col sep = comma]{data/cat_2_4801_45_25_25.csv};
    \addplot[orange,dotted, mark=+]
             table [x=t,y expr=
             \thisrow{s01},col sep = comma]{data/cat_2_4801_45_25_25.csv};

    \addplot[green]
             table [x=t,y expr=
             \thisrow{m10},col sep = comma]{data/cat_2_4801_45_25_25.csv};
    \addplot[green,dotted, mark=+]
             table [x=t,y expr=
             \thisrow{s10},col sep = comma]{data/cat_2_4801_45_25_25.csv};

    \addplot[blue]
             table [x=t,y expr=
             \thisrow{m11},col sep = comma]{data/cat_2_4801_45_25_25.csv};
    \addplot[blue,dotted, mark=+]
             table [x=t,y expr=
             \thisrow{s11},col sep = comma]{data/cat_2_4801_45_25_25.csv};
\end{axis}
\end{tikzpicture} \label{fig:Jset_th_values25}
}
    \hfill
    \subfloat[$\mathtt{th1}= \mathtt{th2}= 27$\ \ \ \ \ \ \ \ \ \ \ \
		\ \ \ \ \ \ \ \ \ \ \ \ \ \ \ \ \ \ \ \ \ \ \ \ \ \ \ \ \ \ \ \ \
		\ \ \ \ \ \ \ \ ]{
      \begin{tikzpicture}
  \begin{axis}[scale=0.65,
               xtick={60,64,...,80},
							 ytick={1,0.1,0.001,0.00001,0.0000001},
               legend columns=1,
               xmin = 60,
               xmax = 80,
               grid = major,
               ymode= log,
               ymin = 0.00001,
               ymax = 1,
               log basis y=10,
							 legend style={at={(1.65,0.02)},anchor=south},
               mark size=2pt,
               xlabel={$t$},
               ylabel={Probability},
               line width=0.5pt,
               xlabel style={font=\footnotesize},
               ylabel style={font=\footnotesize},
               xticklabel style = {font=\footnotesize},
               yticklabel style = {font=\footnotesize},
               mark options=solid]
    \addplot[red]
             table [x=t,y expr=
             \thisrow{m00},col sep = comma]{data/cat_2_4801_45_27_27.csv};
    \addlegendentry{$p_{\mathtt{flip}|00}$ (Model)};
    \addplot[red,dotted, mark=+]
             table [x=t,y expr=
             \thisrow{s00},col sep = comma]{data/cat_2_4801_45_27_27.csv};
    \addlegendentry{$p_{\mathtt{flip}|00}$ (Simulated)};

    \addplot[orange]
             table [x=t,y expr=
             \thisrow{m01},col sep = comma]{data/cat_2_4801_45_27_27.csv};
    \addlegendentry{$p_{\neg \mathtt{flip}|01}$ (Model)};
    \addplot[orange,dotted, mark=+]
             table [x=t,y expr=
             \thisrow{s01},col sep = comma]{data/cat_2_4801_45_27_27.csv};
    \addlegendentry{$p_{\neg \mathtt{flip}|01}$ (Simulated)};

    \addplot[green]
             table [x=t,y expr=
             \thisrow{m10},col sep = comma]{data/cat_2_4801_45_27_27.csv};
    \addlegendentry{$p_{\mathtt{flip}|10}$ (Model)};
    \addplot[green,dotted, mark=+]
             table [x=t,y expr=
             \thisrow{s10},col sep = comma]{data/cat_2_4801_45_27_27.csv};
    \addlegendentry{$p_{\mathtt{flip}|10}$ (Simulated)};

    \addplot[blue]
             table [x=t,y expr=
             \thisrow{m11},col sep = comma]{data/cat_2_4801_45_27_27.csv};
    \addlegendentry{$p_{\neg \mathtt{flip}|11}$ (Model)};
    \addplot[blue,dotted, mark=+]
             table [x=t,y expr=
             \thisrow{s11},col sep = comma]{data/cat_2_4801_45_27_27.csv};
    \addlegendentry{$p_{\neg \mathtt{flip}|11}$ (Simulated)};
\end{axis}
\end{tikzpicture}\label{fig:Jset_th_values27}
}
\end{center}
  \caption{Flipping probabilities for bits in
  $\mathbf{J}_{0,0}$, $\mathbf{J}_{0,1}$, $\mathbf{J}_{1,0}$ and $\mathbf{J}_{1,1}$
  during the second iteration of a parallel decoder, with two different choices for
  $\mathtt{th1}$ and $\mathtt{th2}$. Code parameters matching
the ones in the LEDAcrypt specifications~\cite{LEDA}, Section $4$.$1$,
Figure $4$.$1$: $n_0=2, p= 4801, n=n_0p, k=p, v=45$.
Simulation data obtained from $10^5$ randomly generated error vectors of weight $t$.}
\label{fig:categories}
\end{figure*}
Figure~\ref{fig:categories} reports the probability of flipping or maintaining
correct and incorrect bits during the second iteration of the decoding procedure
(averaged over all admissible values for the syndrome weight, $\epsilon_{01}$
and $\epsilon_{11}$), distinguishing the class in which they belong based on the
partition defined in Section~\ref{subsec:second}. The difference between the
probability of flipping correct bits ($\mathbf{J}_{0,0}$ and $\mathbf{J}_{1,0}$)
and maintaining incorrect bits ($\mathbf{J}_{0,1}$ and $\mathbf{J}_{1,1}$)
based on their correctness \textit{before} the first iteration is evident,
and our model characterizes the behavior of each set of bits accurately.

\begin{figure*}[!t]
\begin{center}
    \subfloat[Waterfall and floor regions.]{
     \pgfkeys{/pgf/number format/.cd,1000 sep={}}
\begin{tikzpicture}
  \begin{axis}[scale=0.75,
               legend columns=2,
               xmin = 800,
               xmax = 12000,
               grid = major,
               ymode= log,
               log basis y=10,
               ymin = 2e-8,
               ymax = 1,
               ytick = {1,1e-1,1e-3,1e-5,1e-7},
               legend style={at={(1.55,0.35)},anchor=south},
               mark size=1.5pt,
               xlabel={code length $n$},
               ylabel={DFR},
               line width=0.5pt,
               legend columns=1, 
               cycle list name=waterfall-floor,
               xticklabel style={
                 /pgf/number format/fixed,
                 /pgf/number format/precision=5
               },
scaled x ticks=false]
   \foreach \vvalue in {9,11,13,...,17} {
     \addplot table [x expr = \thisrow{p}*2,
                     y expr= \thisrow{v\vvalue},
                     restrict y to domain=-18.1:1, 
                     col sep = comma] {data/wef_2_18.csv};
     \addplot table [x expr = \thisrowno{2}*2,
                     y expr = \thisrowno{9}, 
                     restrict y to domain=-18.3:1, 
                     col sep = comma] {data/10m8_density_sweep_LEDA_spec_regular_codes/regular_n0_2_v_\vvalue_t_18.csv};
   }

\end{axis}
\end{tikzpicture} \label{fig:vsweep_full}
}
    \hfill
    \subfloat[Focus on the beginning of the waterfall region.]{
      \begin{tikzpicture}
  \begin{axis}[scale=0.75,
               xmin = 1200,
               xmax = 3600,
               grid = major,
               ymode= log,
               ymin = 1e-3,
               ymax = 1,
               legend style={at={(1.4,0.01)},anchor=south, font=\footnotesize},
               mark size=1.5pt,
               xlabel={code length $n$},
               ylabel={DFR},
               line width=0.5pt,
               legend columns=1,                
               mark options=solid,
               cycle list name=waterfall-floor]
   \foreach \vvalue in {9,11,...,17} {
     \addplot table [x expr = \thisrow{p}*2,y expr= \thisrow{v\vvalue},col sep = comma] {data/wef_2_18.csv};
     \addlegendentryexpanded{$v=\vvalue$\ Mod. };     
     \addplot table [x expr=\thisrowno{2}*2,y expr = \thisrowno{9}, col sep = comma] {data/10m8_density_sweep_LEDA_spec_regular_codes/regular_n0_2_v_\vvalue_t_18.csv};
     \addlegendentryexpanded{$v=\vvalue$\ Sim. };
   }
\end{axis}
\end{tikzpicture}\label{fig:vsweep_waterfall}
}
\end{center}
  \caption{Two iterations DFR values for $(v,2v)$-regular LDPC codes,
  $v \in\{9,11,13,15,17\}$, with rate $\frac{k}{n}=\frac{1}{2}$, $t=18$,
  parallel decoder employing majority thresholds, i.e.,
  $\mathtt{th1}= \mathtt{th2}=\lceil \frac{v+1}{2}\rceil$.\label{fig:vsweep1} Each data point
  was obtained performing $10^8$ decoding actions, or a sufficient number
  to obtain $100$ decoding failures.}
\end{figure*}
We now validate the goodness of fit of our two-iterations DFR predictions
against numerical simulations. To this end, we chose to sweep over the code
length $n$, column weight $v$ and error weight $t$ parameters.
All numerical simulations are done performing $10^8$ decoding actions, or a
sufficient number to obtain $100$ decoding failures, whichever takes place
first. Due to computation time constraints, we restricted our systematic
numerical validation for the two-iterations DFR to $(v,2v)$ regular codes
with rate $\frac{1}{2}$.

We present our results examining first the effects of changing code length $n$
and column weight $v$, as depicted in Figure~\ref{fig:vsweep1}. We chose
the parameter regime in such a fashion that the code with the smallest column
weight $v=9$ achieves a floor regime well within the explored range of code lengths
($n\in\{1200,\ldots,12000\}$), and visible failure rates. 
Figure~\ref{fig:vsweep_full} shows how our two-iterations DFR prediction provides
a conservative estimate for the decoder behaviour when after the floor regime
has been reached, while providing a remarkably good fit for the waterfall region.
This is visible in a clearer fashion in Figure~\ref{fig:vsweep_waterfall}, which
provides a zoom on the waterfall region of all codes, showing the closeness
between our model and the numerical simulations.

\begin{figure*}[t]
\begin{center}
    \subfloat[Waterfall and floor regions.]{
    \begin{tikzpicture}
  \begin{axis}[scale=0.72,
               xmin = 800,
               xmax = 11000,
               grid = major,
               ymode= log,
               log basis y=10,
               ymin = 1e-8,
               ymax = 1,
               mark size=1.5pt,
               xlabel={code length $n$},
               ylabel={DFR},
               line width=0.5pt,
               mark options=solid,
               cycle list name=waterfall-floor,
               xticklabel style={
                 /pgf/number format/fixed,
                 /pgf/number format/precision=5
               },
scaled x ticks=false
               ]
   \foreach \tvalue in {39,36,...,12,10} {
     \addplot table [x expr= \thisrowno{2}*2,
                     y expr = \thisrowno{7},
                     col sep = comma] {data/model_error_sweep/model_QC_0_v_11_t_\tvalue_th1_6_th2_6.csv};
     \addplot table [x expr= \thisrowno{2}*2,
                     y expr = \thisrowno{9},
                     restrict y to domain=-18.1:1,
                     col sep = comma] {data/10m8_error_sweep_LEDA_spec_regular_codes/results_QC_0_v_11_t_\tvalue_th1_6_th2_6.csv};
   }
\end{axis}
\end{tikzpicture} \label{fig:tsweep_full}
}
    \hfill
    \subfloat[Focus on the beginning of the waterfall region.]{
      \begin{tikzpicture}
  \begin{axis}[scale=0.72,
               legend columns=2,
               xmin = 800,
               xmax = 4200,
               grid = major,
               ymode= log,
               ymin = 1e-5,
               ymax = 1,
               legend style={at={(1.50,0)},anchor=south, font=\footnotesize},
               mark size=1.5pt,
               xlabel={code length $n$},
               ylabel={DFR},
               line width=0.5pt,
               mark options=solid,
               cycle list name=waterfall-floor]
   \foreach \tvalue in {39,36,...,12,10} {
     \addplot table [x expr=\thisrowno{2}*2,
                     y expr = \thisrowno{7},
                     col sep = comma] {data/model_error_sweep/model_QC_0_v_11_t_\tvalue_th1_6_th2_6.csv};
     \addlegendentryexpanded{$t=\tvalue$\ Mod. };
     \addplot table [x expr = \thisrowno{2}*2,
                     y expr = \thisrowno{9},
                     restrict y to domain=-18.1:1,
                     col sep = comma] {data/10m8_error_sweep_LEDA_spec_regular_codes/results_QC_0_v_11_t_\tvalue_th1_6_th2_6.csv};
     \addlegendentryexpanded{Sim.};
   }
\end{axis}
\end{tikzpicture}\label{fig:tsweep_waterfall}
      }
\end{center}
  \caption{
  Two iterations DFR values for $(v,2v)$-regular LDPC codes,
  $t \in\{10,\ldots,39\}$, with rate $\frac{k}{n}=\frac{1}{2}$, $v=11$,
  parallel decoder employing majority thresholds, i.e.,
  $\mathtt{th1}=\mathtt{th2}=\lceil \frac{v+1}{2}\rceil$.\label{fig:tsweep} Each data point
  was obtained performing $10^8$ decoding actions, or a sufficient number
  to obtain $100$ decoding failures.}
\end{figure*}

We move onto the analysis of robustness of our estimates, when changing the
number of errors $t\in\{10,13,\ldots,37\}$ while retaining the same column
weight $v$ and exploring the code length range $\{800,\ldots,11000\}$. The
code length range to be explored, and the error weight were chosen with the same
intent as the previous exploration, i.e., an attempt at covering as much as
possible the variety of behaviours when transitioning from a waterfall to a floor
regime.

Figure~\ref{fig:tsweep} depicts the results of the validation campaign,
providing the overview of the entire parameter sweep in Figure~\ref{fig:tsweep_full},
and a zoom on the smaller code lengths in Figure~\ref{fig:tsweep_waterfall}.
As it can be seen, our DFR estimation technique still provides consistent and accurate
estimates for the waterfall region of the examined codes, while providing a
conservative (i.e., higher) estimated DFR value in the floor regime of the codes.

\begin{figure}[t]
\begin{center}
\begin{tikzpicture}
  \begin{axis}[scale=0.7,
               xtick={40,50,...,80},
               xmin = 40,
               xmax = 80,
               grid = major,
               ymode= log,
               ymin = 0.000002,
               ymax = 1,
               ytick={1,1e-1,1e-2,1e-3,1e-4,1e-5},
               legend style={at={(1.3,0.35)},anchor=south,font=\footnotesize},
               mark size=2pt,
               xlabel={$t$},
               ylabel={DFR},
               line width=0.5pt,
               mark options=solid]
    \addplot[blue,densely dotted, mark=+]
             table [x=err_t,y expr=
             \thisrow{leda_dfr},col sep = comma]{data/dfr_2_4801_45_25_25.csv};
    \addlegendentry{LEDAcrypt};
    \addplot[red,densely dotted, mark=+]
             table [x=err_t,y expr=
             \thisrow{new_dfr},col sep = comma]{data/dfr_2_4801_45_25_25.csv};
    \addlegendentry{New model};
    \addplot[black,only marks, mark=x]
             table [x=err_t,y expr=
             \thisrow{sim_dfr},col sep = comma]{data/dfr_2_4801_45_25_25.csv};
    \addlegendentry{Simulated};

\end{axis}
\end{tikzpicture}
\end{center}
\caption{Comparison of our two-iterations DFR estimation technique with the one
currently employed in the LEDAcrypt parameter design. Code parameters matching
the ones in the LEDAcrypt specifications~\cite{LEDA}, Section $4$.$1$,
Figure $4$.$1$:
$n_0=2, p= 4801, n=n_0p, k=p, v=45, \texttt{th1}=25, \texttt{th2}=25$.
Simulation data obtained from $10^6$ randomly generated error vectors of weight
$t$, decoded with a $2$-iterations parallel (\textit{Out-of-place} in LEDAcrypt specification)
decoder.\label{fig:leda_oldvsnew}}
\end{figure}

Finally, we move onto the last step of our numerical validation, i.e., the use
of the proposed two-iterations DFR estimate to analyze the current parameter
sets from LEDAcrypt.
First of all, we quantify the improvement in the estimate of two-iterations
DFR, replicating the numeric experiment reported in the LEDAcrypt specification,
Section $4$.$1$.
Figure~\ref{fig:leda_oldvsnew} reports the result of numerically obtained
DFR values on a relatively small code $n_0=2, p= 4801, n=n_0p, k=p, v=45$,
obtained performing $10^6$ decoding actions, while sweeping the range of
error weight $\{40,\ldots,80\}$.
As it can be seen, we effectively improve the tightness of the DFR estimate
by $\approx 10^5$, in the visible waterfall region of the code.
\begin{table}[!t]
{\small
\begin{center}
    \caption{Estimates for the DFR of a two-iterations out-of-order decoding from this work, compared to the
    ones of the LEDAcrypt~\cite{LEDA} specifications. Codes are $[n_0p,(n_0-1)p]$ QC-LDPC codes,
    hence $(v,n_0v)$ regular\label{tab:lc1}}
    \begin{tabular}{ccccc|c|c}
        \toprule
 $\mathbf{NIST}$ & \multirow{2}{*}{$\mathbf{n}_{\mathbf{0}}$} & \multirow{2}{*}{$\mathbf{p}$} & \multirow{2}{*}{$\mathbf{v}$} & \multirow{2}{*}{$\mathbf{t}$} & \textbf{LEDAcrypt} & \textbf{Our} \\
 $\mathbf{Cat.}$ &                                            &                               &                               &                               & \textbf{DFR}       & \textbf{DFR} \\
                      \midrule
\multirow{6}{*}{$1$}  & $2$ &  $23,371$ & $71$ & $130$ &  $2^{-64}$ & $2^{-140}$\\
                      & $3$ &  $16,067$ & $79$ & $83$  &  $2^{-64}$ & $2^{-135}$\\
                      & $4$ &  $13,397$ & $83$ & $66$  &  $2^{-64}$ & $2^{-131}$\\
                      \cmidrule{2-7}
                      & $2$ &  $28,277$ & $69$ & $129$ & $2^{-128}$ & $2^{-169}$\\
                      & $3$ &  $19,709$ & $79$ &  $82$ & $2^{-128}$ & $2^{-170}$\\
                      & $4$ &  $16,229$ & $83$ &  $65$ & $2^{-128}$ & $2^{-166}$\\
                      \midrule
\multirow{6}{*}{$3$}  &$2$ &  $40,787$ & $103$ & $195$ & $2^{-64}$  & $2^{-189}$\\
                      &$3$ &  $28,411$ & $117$ & $124$ & $2^{-64}$  & $2^{-183}$\\
                      &$4$ &  $22,901$ & $123$ &  $98$ & $2^{-64}$  & $2^{-168}$\\
                      \cmidrule{2-7}
                      &$2$ &  $52,667$ & $103$ & $195$ & $2^{-192}$ & $2^{-258}$\\
                      &$3$ &  $36,629$ & $115$ & $123$ & $2^{-192}$ & $2^{-256}$\\
                      &$4$ &  $30,803$ & $123$ &  $98$ & $2^{-192}$ & $2^{-256}$\\
                      \midrule
\multirow{6}{*}{$5$}  & $2$ &  $61,717$ & $137$ & $261$ & $2^{-64}$ & $2^{-224}$\\
                      & $3$ &  $42,677$ & $153$ & $165$ & $2^{-64}$ & $2^{-216}$\\
                      & $4$ &  $35,507$ & $163$ & $131$ & $2^{-64}$ & $2^{-207}$\\
                      \cmidrule{2-7}
                      & $2$ &  $83,579$ & $135$ & $260$ & $2^{-256}$ & $2^{-350}$\\
                      & $3$ &  $58,171$ & $153$ & $165$ & $2^{-256}$ & $2^{-347}$\\
                      & $4$ &  $48,371$ & $161$ & $131$ & $2^{-256}$ & $2^{-343}$\\
                      \bottomrule

    \end{tabular}
\end{center}
}
\end{table}

This significant improvement in the tightness of the estimate allows us to
re-analyze the parameters proposed in the LEDAcrypt specification, as reported
in Table~\ref{tab:lc1}.
LEDAcrypt employs quasi-cyclic codes, with rates among
$\frac{1}{2},\frac{2}{3},\frac{3}{4}$, providing security-equivalent parameter
sets for all the rates.
For each rate and security level, the LEDAcrypt specification proposed two
sets of parameters, one with DFR low enough to formally guarantee the
resistance against an active attacker (IND-CCA$2$ security), and one
with a practically low enough DFR $2^{-64}$ so that the event of an attacker
succeeding in causing a failure in decryption is extremely unlikely.
Applying our estimate techniques, we observe that the parameter sets which
were proposed with DFR $\leq 2^{-64}$ for NIST security category $1$ actually
guarantee a DFR $< 2^{-128}$, in turn fully meeting the requirement for IND-CCA2
security ($\leq 2^{-128}$, for category $1$), and indeed leaving further margin
for the reduction of the code size.
The parameters proposed with DFR $\leq 2^{-64}$ for NIST security category $3$
come relatively close to the requirement for IND-CCA2 security ($\leq 2^{-192}$,
for category $3$), albeit not meeting it already, while their analogues for
NIST category $5$ have DFRs decidedly lower than the required $\leq 2^{-256}$.
These results point to the possibility a further reduction of the code sizes
for LEDAcrypt at NIST security category $1$ by more than $20$\% with respect
to the currently proposed ones, while a significant reduction is expected
also for category $3$ and $5$.

As a final note, we point out that quasi-cyclic codes, 
under specific decoding techniques,
are known to suffer from high error floors due to the so called \textit{near codewords}.
Nevertheless, the large size of LEDAcrypt codes and the small number of iterations allows
the probability of converging to a near codeword to be considered negligible. Indeed, for
$n_0=2$, $v=71$ and $t=134$ (almost equal to the parameters of LEDAcrypt for NIST category $1$),
the probability of converging to a near codeword has been shown to be $\leq 2^{-128}$ for
$p \geq 13477$ \cite{DBLP:journals/iacr/ArpinLPRTV25}, meaning that for codes with a
circulant block twice the size (e.g., $p=23371$ and $p=28277$) this probability can
be considered negligible with respect to the target DFR. For $n_0=3$ and $n_0=4$,
this phenomenon is even less severe due to larger code lengths and near codewords
with higher weights.


%
%
\section{Concluding Remarks}
In this work, we presented a new technique to estimate the decoding failure
rate of two-iterations parallel bit flipping decoders. Our technique
relies on a sound model for the syndrome weight distribution, and
on a novel approach to determining if a bit of the error vector estimation
is flipped at the second iteration of the parallel bit flipping decoder.
We validated numerically our approach showing a good fit of our prediction
to the actual DFRs.
In light of our results, we analyzed the DFR estimates made by LEDAcrypt in
their parameter design, and have shown that it is possible to reduce
the chosen code lengths by $\approx 20$\% for NIST category $1$, while
retaining a suitable DFR.

%
%
%
%
%
\bibliographystyle{IEEEtran}
\bibliography{biblio.bib}
%
%
%
\appendices
\section{Derivation of the Probability Mass Function
\texorpdfstring{$\Pr(\mathcal{F}_t = f\,|\, \mathcal{W}_t = y)$}{}}
\label{app:appendix1}
\noindent In this section, we derive an explicit formulation for the distribution of
the number of bit positions in each parity-check equation taking as a value an asserted
bit before the execution
of the first iteration of the decoding algorithm, given the syndrome weight.
In other words, we provide a complete derivation of the probability mass function
of the number of bitflips $\mathcal{F}_t$ on any parity-check equation induced during
the sequential computation of the syndrome by the actual and unknown value of the error vector $\vectmat{e}$
having weight $t$ (i.e. during the sequential additions of the columns of $\vectmat{H}$ indexed
by asserted bits in $\vectmat{e}$), given the weight of the syndrome $\mathcal{W}_t$, i.e.,
$\mathcal{F}_{t|y} \sim \condprob{\flips{f}}{\mathcal{W}_t=y}$.

To calculate $\condprob{\flips{f}}{\mathcal{W}_t=y}$, we first formulate the problem in
the following way:
$$
\condprob{\flips{f}}{\mathcal{W}_t=y} = \frac{\prob{\flips{f} \cap \mathcal{W}_t=y}}{\prob{\mathcal{W}_t=y}} =
\condprob{\mathcal{W}_t=y}{\flips{f}} \frac{\prob{\flips{f}}}{\prob{\mathcal{W}_t=y}}.
$$
While the two quantities on the right of the expression have been already calculated,
the probability $\condprob{\mathcal{W}_t=y}{\flips{f}}$ expresses the probability that
the syndrome weight is equal to $y$, given the amount of flips applied to one of the
syndrome bits.
To derive this probability, we undergo a process similar to the one employed for the
calculation of the syndrome weight distribution. We begin by defining a state vector
$\mathtt{wp}_{(\ell)}=[\mathtt{wp}_{(\ell),0}, \ldots, \mathtt{wp}_{(\ell),r-1}]$ modeling the number
of asserted bits in the $r-1$ syndrome bits different from the selected one.
As before, the starting state is the one corresponding to a null syndrome,
thus $\mathtt{wp}_{(0)}=[1,0,\ldots,0]$.
Given the $t$ total bit flips applied to the error vector, $f$ of these will cause a
flip in the selected bit and $v-1$ flips in the other $r-1$ syndrome bits, while the
remaining $t-f$ error bits will not cause any flip in the selected parity-check, while
causing instead $v$ flips in the $r-1$ other positions.
To complete the definition of this modified discrete-time non-homogeneous Markov chain,
we need to define the transition matrices
$\mathbf{\mathtt{P}}_{(\ell)} =[p_{x,y,\ell}]_{\scriptstyle x,y \in \{0,\ldots,r\}}$.
We can assume, without loss of generality, that the $f$ flips affecting the selected
syndrome bit happen after the other $t-f$ flips. The distribution of the number of
flips applied during the process to a generic syndrome bit:
$$
\phi_\ell(f) = \Pr(\mathcal{F}_\ell = f) =
\frac{\binom{w}{f}\binom{n-w}{\ell-f}}{\binom{n}{\ell}}
$$
\noindent remains unchanged.
Subsequently, the definitions of $\pi_{\stackrel{\ell-1 \to \ell}{\mathtt{\scriptscriptstyle flip}\, 0 \to 1}}(\ell)$ and
$\pi_{\stackrel{\ell-1 \to \ell}{\mathtt{\scriptscriptstyle flip}\, 1 \to 0}}(\ell)$, corresponding to the probability of
flipping up or down a syndrome bit during step $\ell$, do not change either.
The probabilities we need to modify are $\omega_{\mathtt{up}}(a)$,
corresponding to the probability of flipping up $a$ bits, and $\omega_{\mathtt{down}}(a)$,
corresponding to the probability of flipping down $v-a$ (or $v-1-a$, depending on $\ell$) bits,
where $x$ is the number of asserted bits and $\ell$ is the current step.
The following modification is justified by the fact that only $r-1$ positions can be modified
instead of $r$, and the number of bit flips happening at each step depends on the value of $\ell$.
For $1 \leq \ell \leq t-f$, we have that $v$ flips take place at each step, so the formulation of
$\omega_{\mathtt{up}}(a)$ and $\omega_{\mathtt{down}}(a)$ becomes:

$$
\omega_{\mathtt{up}}(a) = \bindist{r-1-x}{a}{\pi_{\stackrel{\ell-1 \to \ell}{\, \mathtt{\scriptscriptstyle flip}\, 0\to 1}}(\ell)}
= {\displaystyle\binom{r-1-x}{a}}
\left(\pi_{\stackrel{\ell-1 \to \ell}{\, \mathtt{\scriptscriptstyle flip}\, 0\to 1}}(\ell)\right)^{a}
\left(1-\pi_{\stackrel{\ell-1 \to \ell}{\, \mathtt{\scriptscriptstyle flip}\, 0\to 1}}(\ell)\right)^{(r-1-x)-a}
$$
$$
\omega_{\mathtt{down}}(a) = \bindist{x}{v-a}{\pi_{\stackrel{\ell-1 \to \ell}{\, \mathtt{\scriptscriptstyle flip}\, 1\to 0}}(\ell)}
= {\displaystyle\binom{x}{v-a}}
\left(\pi_{\stackrel{\ell-1 \to \ell}{\, \mathtt{\scriptscriptstyle flip}\, 1\to 0}}(\ell)\right)^{v-a}
\left(1-\pi_{\stackrel{\ell-1 \to \ell}{\, \mathtt{\scriptscriptstyle flip}\, 1\to 0}}(\ell)\right)^{x-(v-a)}
$$

\noindent The probability of moving to any admissible syndrome weight has to be modified in a similar way:
$$
\displaystyle
\sum_{i=\max\{0,v-x\}}^{\min\{r-1-x, v\}}
\left(
 \omega_{\mathtt{up}}(i)
 \cdot \omega_{\mathtt{down}}(i)
\right).
$$
\noindent Finally, the elements of the transition matrix $p_{x,y,\ell}$ indicating the probability of changing the number of
asserted bits in the syndrome from $x$ to $y$ can be defined as:
$$
p_{x,y,\ell} =
\begin{cases}
1, & {\ell=1,x=0,y=v} \\
\rho(x, y, \ell),  &\
{\scriptstyle\hspace{-1em}
\begin{array}{l}
\ell \geq 2 \\
\max(0, x-v) \leq y \leq \min(x+v, r-1)\\
y \equiv_2 (x+v)
\end{array}}\\
0,
&
{\mathrm{otherwise}}
\end{cases}
$$
Where:
$$
\rho(x,y,\ell) = \frac{
\omega_{\mathtt{up}}(\frac{y-x+v}{2})
\cdot
\omega_{\mathtt{down}}(\frac{y-x+v}{2}) 
}{
\displaystyle
\sum_{i=\max\{0,v-x\}}^{\min\{r-1-x, v\}}
\left(
 \omega_{\mathtt{up}}(i)
 \cdot \omega_{\mathtt{down}}(i)
\right)
}
$$

For $t-f+1\leq \ell \leq t$, we apply a similar modification to the previous formulas, noting that in the last $f$
steps only $v-1$ flips take place on the $r-1$ syndrome bits under analysis:
$$
\omega_{\mathtt{up}}(a) =
\bindist{r-1-x}{a}{\pi_{\stackrel{\ell-1 \to \ell}{\, \mathtt{\scriptscriptstyle flip}\, 0\to 1}}(\ell)},
\quad \text{and} \quad
\omega_{\mathtt{down}}(a) =
\bindist{x}{v-1-a}{\pi_{\stackrel{\ell-1 \to \ell}{\, \mathtt{\scriptscriptstyle flip}\, 1\to 0}}(\ell)}
$$

$$
\rho(x,y,\ell)=
\frac{
\omega_{\mathtt{up}}(\frac{y-x+v-1}{2})
\cdot
\omega_{\mathtt{down}}(\frac{y-x+v-1}{2})
}{
\sum_{i=\max\{0,v-1-x\}}^{\min\{r-1-x, v-1\}}
\left(
 \omega_{\mathtt{up}}(i) 
 \cdot \omega_{\mathtt{down}}(i)
\right)
}
$$

\noindent The definition of the transition probability for the last $f$ rounds is the following:
$$
p_{x,y,\ell} =
\begin{cases}
1, & {\ell=1,x=0,y=v-1} \\
\rho(x, y, \ell),  &\
{\scriptstyle\hspace{-1em}
\begin{array}{l}
\ell \geq 2 \\
\max(0, x-v+1) \leq y \leq \min(x+v-1, r-1)\\
y \equiv_2 (x+v-1)
\end{array}}\\
0,
&
{\mathrm{otherwise}}
\end{cases}
$$

Once the transition matrices have been derived, the Markov chain is well defined and the final state vector
$\mathtt{wp}_{(t)}$ can be obtained. The resulting vector contains the probability distribution of the number
of asserted bits in the $r-1$ bits different from the selected one.
The conditioned probability $\condprob{\mathcal{W}_t=y}{\flips{f}}$ can then be obtained by adding the
selected bit to the count:
$$
\condprob{\mathcal{W}_t=y}{\flips{f}} =
\begin{cases}
 \mathrm{wp}_{(t),y} & \text{if $f$ is even}\\
 \mathrm{wp}_{(t),y-1} & \text{if $f$ is odd}\\
\end{cases}
$$

%
%
\section{Derivation of the Second Iteration Bit-flipping Probabilities}
\label{app:2itflippingprob}
\noindent As shown in Section~\ref{subsec:second}, the formula for computing the DFR estimation of the
(parallel) bit-flipping decoder, at the end of its second iteration,
as a function of the probability of performing correct and incorrect
bit-flips of the values stored in the error vector estimate
requires the evaluation of the following
quantities $p_{\mathtt{flip}|00}, p_{\mathtt{flip}|01}, p_{\mathtt{flip}|10}, p_{\mathtt{flip}|11}$.
The deductions underlying the writing of the formula corresponding to
$p_{\mathtt{flip}|00}$, i.e., the probability of performing
a bit-flip in a position of the error vector estimate that correctly
stores a clear bit are shown in Section~\ref{subsec:second}.
In the following, we report the derivations needed to explicitly
compute also the remaining probabilities and highligh in blue the
differences w.r.t. the formulas related to $p_{\mathtt{flip}|00}$.

\subsection{Modeling the probability of flipping a bit in
\texorpdfstring{$\bar{\vectmat{e}}_{(1)}$}{}, with position in
\texorpdfstring{$\mathbf{J}_{0,1}$}{}}
Maintaining the definitions and the notation, introduced in Section~\ref{subsec:second}, we model
the probability of performing, during the second iteration of the decoding algorithm,
a bit-flip in a position categorized to be in
$\jset{0}{1}$ at end the end of the first decoding iteration.

\noindent The expression of $\chi_{\updownarrow \mathtt{odd}}(f,\epsilon_{01})$ is:
$$
\chi_{\updownarrow \mathtt{odd}}(f,\epsilon_{01})=\sum_{\ell=1, \ell\,
\text{odd}}^{\scriptscriptstyle
\min(\mathalert{\epsilon_{01}-1}, w-f-1)} \frac{ \eta(f,\ell)\cdot
\zeta(f,\ell,\mathalert{\epsilon_{01}-1}) }{ \xi(f,\mathalert{\epsilon_{01}-1}) },
$$
changing from the one when the examined position is in $\jset{0}{0}$ only in 
replacing the occurrences of $\epsilon_{01}$ with $\epsilon_{01}-1$:
indeed, one of the positions $j\in\jset{0}{1}$ involved in the equation is actually the variable being considered.
The expression of $\chi_{\leftrightarrow \mathtt{odd}}(f)$ does not change, as the fact
that $a$ is in $\jset{0}{1}$ instead of
$\jset{0}{0}$ does not have any effect on its formulation.

The probability $p_{01|\mathtt{BecomeUnsat}}$ is obtained as (changes highlighted in blue)
$$
\frac{\sum_{f = 0, f\text{ even}}^{\min(t,w-1)}
\condprob{\flips{f}}{\mathalert{\sevent{\jset{0}{1}}}}
\cdot (
\mathalert{1-}
\gamma_{\mathtt{UnsatPostFlips}}(f,\epsilon_{01},\epsilon_{11})
)
}
{
\sum_{f = 0, f\text{ even}}^{\min(t,w-1)}
\condprob{\flips{\mathtt{f}}}{\mathalert{\sevent{\jset{0}{1}}}}
}
$$
Indeed, the two changes from $p_{00|\mathtt{BecomeUnsat}}$ involve the fact
that the position being considered is in $\jset{0}{1}$ instead of $\jset{0}{0}$
and the fact that the contribution to the parity taking place on the $w-1$ 
positions of the equation excluding the one being considered should be even in order to have an unsatisfied check.
Symmetrically, $p_{01|\mathtt{StayUnsat}}$ is obtained as (changes highlighted in blue)
$$
\frac{\sum_{f = 1, f\text{ odd}}^{\min(t,w-1)}
\condprob{\flips{f}}{\mathalert{\sevent{\jset{0}{1}}}} \cdot
(
\mathalert{1-}
\gamma_{\mathtt{UnsatPostFlips}}
(f,\epsilon_{01},\epsilon_{11})
)}{
\sum_{f = 1, f\text{ odd}}^{\min(t,w-1)}
\condprob{\flips{f}}{\mathalert{\sevent{\jset{0}{1}}}}
}.
$$
We now derive $\condprob{\flips{f}}{\sevent{\jset{0}{1}}}$ as:
$$
\condprob{\flips{f}}{\sevent{\jset{0}{1}}} =
\frac{\prob{\flips{f} \cap \mathalert{\sevent{\jset{0}{1}}}}}{\prob{\mathalert{\sevent{\jset{0}{1}}}}} =
\frac{\condprob{\mathalert{\sevent{\jset{0}{1}}}}{\flips{f}}\prob{\flips{f}}}{
\sum_{f=0}^{\min(t,w)}
\condprob{\mathalert{\sevent{\jset{0}{1}}}}{\flips{f}}
\prob{\flips{f}}
},
$$
Where 
$
\condprob{\sevent{\jset{0}{1}}}{\flips{f}} = 
\condprob{e_i=0}{\flips{f}} \cdot \condprob{\mathalert{\sevent{\jset{0}{1}}}}{\flips{f} \cap e_i=0}
$. We have that $\condprob{e_i=0}{\flips{f}} = \frac{w-f}{w}$ as in the case of $\jset{0}{0}$, since
in both cases the bit in the actual error vector is clear. On the other hand,
$\condprob{\sevent{\jset{0}{1}}}{\flips{f} \cap e_i=0}$, the probability of a clear
bit to be erroneously flipped, given the number $f$ of asserted bits involved in the parity-check, is computed as:
$$\condprob{\sevent{\jset{0}{1}}}{\flips{f} \cap e_i=0} =
\begin{cases}
    \mathalert{\pflipz{\mathtt{OneEqSat}}} & \text{if $f$ is even}\\
    \mathalert{\pflipz{\mathtt{OneEqUnsat}}} & \text{if $f$ is odd.}\\
\end{cases}
$$

The probability distribution of the $j$-th upc being valued $a$, given that $j\in\jset{0}{1}$ is
(changes w.r.t $\Pr (\mathcal{U}_{j} = a  \ |\  j\in \jset{0}{0})$ highlighted in blue):
$$\Pr (\mathcal{U}_{j} = a \ |\ j\in \jset{0}{1}) =
\begin{cases}
    \frac{\bindist{v}{a}{\punsatz}}{p_{\mathalert{\mathtt{flip}|0}}} &
    \quad \text{if } x\ \mathalert{\geq}\ \mathtt{th}_{(1)}\\
    0 & \quad \text{otherwise}\\
\end{cases}
$$
Since the expression of $\mu$ does not change, we obtain 
$\pflipzo(\epsilon_{01},\epsilon_{11})$ as
$$
\begin{array}{l}
\displaystyle
\pflipzo(\epsilon_{01},\epsilon_{11})=\sum_{\mathalert{a=\mathtt{th}_{(1)}}}^{\mathalert{v}}
\Pr (\mathcal{U}_{j} = a  \ |\  \mathalert{j\in \jset{0}{1}}) \cdot
\displaystyle
\left(  \sum_{\mathtt{nsat} = 0}^{v-a} \sum_{
{\scriptscriptstyle \begin{matrix}
\mathtt{nunsat}=\\
\max(0,\mathtt{th}_{(2)}-\mathtt{nsat})
\end{matrix}}
}^{a}  \mu(\mathtt{nsat},\mathtt{nunsat},a,\epsilon_{01},\epsilon_{11})\right)
\end{array}
$$
since we are interested in the probability of performing a flip
at the second iteration (which does not change from the computation
of $\pflipzz$), but the first iteration performed a flip, therefore
the first-iteration upc value $a$ should be above the first iteration
threshold $\mathtt{th}_{(1)}$.

\subsection{Modeling the probability of flipping a bit in
\texorpdfstring{$\bar{e}_{(1)}$}{}, with position in
\texorpdfstring{$\mathbf{J}_{1,0}$}{}}
Maintaining the definitions and the notation, introduced in Section~\ref{subsec:second}, we model
the probability of performing, during the second iteration of the decoding algorithm,
a bit-flip in a position categorized to be in
$\jset{1}{0}$ at end the end of the first decoding iteration.

\noindent The expression of $\chi_{\updownarrow \mathtt{odd}}(f,\epsilon_{01})$ is:
$$
\eta(f,\ell) = \bindist{\mathalert{w-f}}{\ell}{\pflipz{\mathtt{OneEqSat}}}
$$
$$
\chi_{\updownarrow \mathtt{odd}}(f,\epsilon_{01})=\sum_{\ell=1, \ell\, \text{odd}}^{\scriptscriptstyle
\min(\epsilon_{01}, \mathalert{w-f})} \frac{ \eta(f,\ell)\cdot
\zeta(f,\ell,\epsilon_{01}) }{ \xi(f,\epsilon_{01}) },
$$
where the modified indexed are justified by the fact that the bit we are analyzing is not in
the set of initially clear bits, so we do not need to subtract it in the calculation of
$\chi_{\uparrow \mathtt{odd}}(f,\epsilon_{01})$.
For the same reason, $\chi_{\leftrightarrow \mathtt{odd}}(f,\epsilon_{11})$ becomes:
$$\nu(f,\ell)=\bindist{\mathalert{f-1}}{\ell}{\pnoflipo{\mathtt{OneEqSat}}}$$
$$\chi_{\leftrightarrow \mathtt{odd}}(f,\epsilon_{11})= \sum_{\ell=1, \ell\, \text{odd}}^{\scriptscriptstyle
\min(\epsilon_{11}, \mathalert{f-1})}\frac{ \nu(f,\ell)\cdot \lambda(f,\ell,\epsilon_{11})
}{\theta(f,\epsilon_{11})}$$
Where $f-1$ is the number of asserted positions in the error vector where a flip may happen.
$p_{10|\mathtt{BecomeUnsat}}$ is obtained as

$$
\frac{\sum_{f = \mathalert{2}, f\text{ even}}^{\min(t,\mathalert{w})}
\condprob{\flips{f}}{\mathalert{\sevent{\jset{1}{0}}}}
\cdot
\gamma_{\mathtt{UnsatPostFlips}}(f,\epsilon_{01},\epsilon_{11})
}
{
\sum_{f = \mathalert{2}, f\text{ even}}^{\min(t,\mathalert{w})}
\condprob{\flips{\mathtt{f}}}{\mathalert{\sevent{\jset{1}{0}}}}
}
$$
substituting $w-1$ with $w$ (since a clear bit in the check is not guaranteed) and
$f = 0$ with $f = 2$ (since an asserted bit is now guaranteed).
$p_{10|\mathtt{StayUnsat}}$ is obtained as
$$
\frac{\sum_{f = 1, f\text{ odd}}^{\min(t,\mathalert{w})}
\condprob{\flips{f}}{\mathalert{\sevent{\jset{1}{0}}}} \cdot
\gamma_{\mathtt{UnsatPostFlips}}(f,\epsilon_{01},\epsilon_{11})}{
\sum_{f = 1, f\text{ odd}}^{\min(t,\mathalert{w})}
\condprob{\flips{\mathtt{f}}}{\mathalert{\sevent{\jset{1}{0}}}}
}.
$$
We now derive $\condprob{\flips{f}}{\sevent{\jset{1}{0}}}$ as:
$$
\condprob{\flips{f}}{\sevent{\jset{1}{0}}} =
\frac{\prob{\flips{f} \cap \mathalert{\sevent{\jset{1}{0}}}}}{\prob{\mathalert{\sevent{\jset{1}{0}}}}} =
\frac{\condprob{\mathalert{\sevent{\jset{1}{0}}}}{\flips{f}}\prob{\flips{f}}}{
\sum_{f=0}^{\min(t,w)}
\condprob{\mathalert{\sevent{\jset{1}{0}}}}{\flips{f}}
\prob{\flips{f}}
},
$$
Where 
$
\condprob{\sevent{\jset{1}{0}}}{\flips{f}} = 
\condprob{\mathalert{e_i=1}}{\flips{f}} \cdot \condprob{\mathalert{\sevent{\jset{1}{0}}}}{\flips{f} \cap \mathalert{e_i=1}}
$.
We have $\condprob{e_i=1}{\flips{f}} = \frac{f}{w}$, since
the analyzed bit is now assumed to be asserted.
On the other hand, $\condprob{\sevent{\jset{1}{0}}}{\flips{f} \cap e_i=1}$,
the probability of a set bit to be correctly flipped, given the number $f$ of
asserted bits involved in the parity-check, is computed as:
$$\condprob{\sevent{\jset{1}{0}}}{\flips{f} \cap e_i=1} =
\begin{cases}
    \mathalert{1-\pnoflipo{\mathtt{OneEqSat}}} & \text{if $f$ is even}\\
    \mathalert{1-\pnoflipo{\mathtt{OneEqUnsat}}} & \text{if $f$ is odd.}\\
\end{cases}
$$

\noindent The probability distribution of the $j$-th upc being valued $a$, given that $j\in\jset{1}{0}$
is (changes w.r.t $\Pr (\mathcal{U}_{j} = a  \ |\  j\in \jset{0}{0})$ highlighted in blue):
$$\Pr (\mathcal{U}_{j} = a  \ |\  j\in \jset{1}{0}) =
\begin{cases}
    \frac{\bindist{v}{a}{\mathalert{\punsato}}}{p_{\mathalert{\mathtt{flip}|1}}} &
    \quad \text{if } x\ \mathalert{\geq}\ \mathtt{th}_{(1)}\\
    0 & \quad \text{otherwise}\\
\end{cases}
$$
\noindent
Since the expression of $\mu$ does not change, we obtain 
$\pflipoz(\epsilon_{01},\epsilon_{11}):$
$$
\begin{array}{l}
\displaystyle
\pflipoz(\epsilon_{01},\epsilon_{11})=\sum_{\mathalert{a=\mathtt{th}_{(1)}}}^{\mathalert{v}}
\Pr (\mathcal{U}_{j} = a  \ |\  \mathalert{j\in \jset{1}{0}})
\displaystyle
\cdot\left(  \sum_{\mathtt{nsat} = 0}^{v-a} \sum_{
{\scriptscriptstyle \begin{matrix}
\mathtt{nunsat}=\\
\max(0,\mathtt{th}_{(2)}-\mathtt{nsat})
\end{matrix}}
}^{a}  \mu(\mathtt{nsat},\mathtt{nunsat},a,\epsilon_{01},\epsilon_{11})\right)
\end{array}
$$

\subsection{Modeling the probability of flipping a bit in
\texorpdfstring{$\bar{\vectmat{e}}_{(1)}$}{}, with position in
\texorpdfstring{$\mathbf{J}_{1,1}$}{}}
Maintaining the definitions and the notation, introduced in Section~\ref{subsec:second}, we model
the probability of performing, during the second iteration of the decoding algorithm,
a bit-flip in a position categorized to be in
$\jset{1}{1}$ at end the end of the first decoding iteration.

\noindent Switching from $\jset{1}{0}$ to $\jset{1}{1}$ does not change the definition of
$\chi_{\updownarrow \mathtt{odd}}(f,\epsilon_{01})$, while
$\chi_{\leftrightarrow \mathtt{odd}}(f,\epsilon_{11})$ becomes:
$$
\chi_{\leftrightarrow \mathtt{odd}}(f,\epsilon_{11})= \sum_{\ell=1, \ell\, \text{odd}}^{\scriptscriptstyle
\min(\mathalert{\epsilon_{11}-1}, \mathalert{f-1})}\frac{ \nu(f,\ell)\cdot
\lambda(f,\ell,\mathalert{\epsilon_{11}-1})
}{\theta(f,\mathalert{\epsilon_{11}-1})}
$$
\noindent where, with the same line of reasoning as the $\jset{0}{1}$ case, we exclude one of
the bits in $\jset{1}{1}$ being it the one under consideration.
$p_{11|\mathtt{BecomeUnsat}}$
is obtained as
$$
\frac{\sum_{f = \mathalert{2}, f\text{ even}}^{\min(t,\mathalert{w})}
\condprob{\flips{f}}{\mathalert{\sevent{\jset{1}{1}}}}
\cdot
(
\mathalert{1-}
\gamma_{\mathtt{UnsatPostFlips}}(f,\epsilon_{01},\epsilon_{11})
)}
{
\sum_{f = \mathalert{2}, f\text{ even}}^{\min(t,\mathalert{w})}
\condprob{\flips{f}}{\mathalert{\sevent{\jset{1}{1}}}}
}
$$
noting that, for the parity-check to be unsatisfied, an even number of asserted bits must result
from the other $w-1$ bits.
$p_{11|\mathtt{StayUnsat}}$ is obtained as
$$
\frac{\sum_{f = 1, f\text{ odd}}^{\min(t,\mathalert{w})}
\condprob{\flips{f}}{\mathalert{\sevent{\jset{1}{1}}}} \cdot
(
\mathalert{1-}
\gamma_{\mathtt{UnsatPostFlips}}(f,\epsilon_{01},\epsilon_{11})
)}{
\sum_{f = 1, f\text{ odd}}^{\min(t,\mathalert{w})}
\condprob{\flips{\mathtt{f}}}{\mathalert{\sevent{\jset{1}{1}}}}
}.
$$
We now derive $\condprob{\flips{f}}{\sevent{\jset{1}{1}}}$ as:
$$
\condprob{\flips{f}}{\mathalert{\sevent{\jset{1}{1}}}} =
\frac{\prob{\flips{f} \cap \mathalert{\sevent{\jset{1}{1}}}}}{\prob{\mathalert{\sevent{\jset{1}{1}}}}} =
\frac{\condprob{\mathalert{\sevent{\jset{1}{1}}}}{\flips{f}}\prob{\flips{f}}}{
\sum_{f=0}^{\min(t,w)}
\condprob{\mathalert{\sevent{\jset{1}{1}}}}{\flips{f}}
\prob{\flips{f}}
},
$$
Where 
$
\condprob{\sevent{\jset{1}{1}}}{\flips{f}} = 
\condprob{\mathalert{e_i=1}}{\flips{f}} \cdot \condprob{\mathalert{\sevent{\jset{1}{1}}}}{\flips{f} \cap \mathalert{e_i=1}}
$.
We have $\condprob{e_i=1}{\flips{f}} = \frac{f}{w}$, since the analyzed bit
is assumed to be asserted in the actual error vector, as in the case of $\jset{1}{0}$.
On the other hand, $\condprob{\sevent{\jset{1}{1}}}{\flips{f} \cap e_i=1}$, the probability of an asserted
bit to be incorrectly maintained, given the number $f$ of asserted bits involved in the parity-check, is computed as:
$$\condprob{\sevent{\jset{1}{1}}}{\flips{f} \cap e_i=1} =
\begin{cases}
    \mathalert{\pnoflipo{\mathtt{OneEqSat}}} & \text{if $f$ is even}\\
    \mathalert{\pnoflipo{\mathtt{OneEqUnsat}}} & \text{if $f$ is odd.}\\
\end{cases}
$$
\noindent
The probability distribution of the $j$-th upc being valued $a$, given that $j\in\jset{1}{1}$ is
(changes w.r.t $\Pr (\mathcal{U}_{j} = a  \ |\  j\in \jset{0}{0})$ highlighted in blue):
$$\Pr (\mathcal{U}_{j} = a  \ |\  j\in \jset{1}{1}) =
\begin{cases}
    \frac{\bindist{v}{a}{\mathalert{\punsato}}}{p_{\mathalert{\mathtt{flip}|1}}} &
    \quad \text{if } x< \mathtt{th}_{(1)}\\
    0 & \quad \text{otherwise}\\
\end{cases}
$$
Since the expression of $\mu$ does not change, we obtain 
$\pflipoo(\epsilon_{01},\epsilon_{11}):$
$$
\begin{array}{l}
\displaystyle
\pflipoo(\epsilon_{01},\epsilon_{11})=\sum_{a=0}^{\mathtt{th}_{(1)}-1}
\Pr (\mathcal{U}_{j} = a  \ |\  \mathalert{j\in \jset{1}{1}})
\displaystyle
\cdot\left(  \sum_{\mathtt{nsat} = 0}^{v-a} \sum_{
{\scriptscriptstyle \begin{matrix}
\mathtt{nunsat}=\\
\max(0,\mathtt{th}_{(2)}-\mathtt{nsat})
\end{matrix}}
}^{a}  \mu(\mathtt{nsat},\mathtt{nunsat},a,\epsilon_{01},\epsilon_{11})\right)
\end{array}
$$

%
%
%
%
%
\vspace{11pt}
%
%
\begin{IEEEbiographynophoto}{Alessandro Annechini}
TbW
\end{IEEEbiographynophoto}
%
%
\begin{IEEEbiographynophoto}{Alessandro Barenghi}
TbW
\end{IEEEbiographynophoto}
%
%
\begin{IEEEbiographynophoto}{Gerardo Pelosi}
TbW
\end{IEEEbiographynophoto}
%
%
\vfill
%
%
\end{document}